\documentclass[aps,pra,twocolumn,notitlepage,10pt,superscriptaddress,longbibliography]{revtex4-1}

\usepackage{graphicx}
\usepackage{times,bbm,amsmath,amssymb}

\usepackage{hyperref}
\usepackage{xcolor}
\usepackage{cleveref}
\crefname{figure}{Fig.}{Figs.}
\Crefname{figure}{Figure}{Figures}

\usepackage{graphicx}
\usepackage{dcolumn}
\usepackage{bm}
\usepackage{soul}

\usepackage{nicefrac, xfrac}
\usepackage{pifont}
\usepackage{bbold}
\usepackage{physics}
\usepackage{xr}
\usepackage{lineno}

\usepackage{easyReview}
\usepackage{parskip}
\usepackage{libertine}[newtxmath]

\usepackage{comment}
\usepackage{xcolor}

\begin{document}
%\linenumbers
\title{Time-series forecasting with multiphoton quantum states and integrated photonics}

\author{Rosario Di Bartolo}
\affiliation{Dipartimento di Fisica, Sapienza Universit\`{a} di Roma, Piazzale Aldo Moro 5, I-00185 Roma, Italy}

\author{Simone Piacentini}
\thanks{Current affiliation: Quandela, 7 Rue Léonard de Vinci, 91300 Massy, France}
\affiliation{Istituto di Fotonica e Nanotecnologie, Consiglio Nazionale delle Ricerche (IFN-CNR), Piazza Leonardo da Vinci, 32, I-20133 Milano, Italy}

\author{Francesco Ceccarelli}
\affiliation{Istituto di Fotonica e Nanotecnologie, Consiglio Nazionale delle Ricerche (IFN-CNR), Piazza Leonardo da Vinci, 32, I-20133 Milano, Italy}

\author{Giacomo Corrielli}
\affiliation{Istituto di Fotonica e Nanotecnologie, Consiglio Nazionale delle Ricerche (IFN-CNR), Piazza Leonardo da Vinci, 32, I-20133 Milano, Italy}

\author{Roberto Osellame}
\affiliation{Istituto di Fotonica e Nanotecnologie, Consiglio Nazionale delle Ricerche (IFN-CNR), Piazza Leonardo da Vinci, 32, I-20133 Milano, Italy}

\author{Valeria Cimini}
\email{valeria.cimini@uniroma1.it}
\affiliation{Dipartimento di Fisica, Sapienza Universit\`{a} di Roma, Piazzale Aldo Moro 5, I-00185 Roma, Italy}

\author{Fabio Sciarrino}
\affiliation{Dipartimento di Fisica, Sapienza Universit\`{a} di Roma, Piazzale Aldo Moro 5, I-00185 Roma, Italy}

\begin{abstract}

Quantum machine learning algorithms have very recently attracted significant attention in photonic platforms. In particular, reconfigurable integrated photonic circuits offer a promising route, thanks to the possibility of implementing adaptive feedback loops, which is an essential ingredient for achieving the necessary nonlinear behavior characteristic of neural networks. 
Here, we implement a quantum reservoir computing protocol in which information is processed through a reconfigurable linear optical integrated photonic circuit and measured using single-photon detectors. We exploit a multiphoton-based setup for time-series forecasting tasks in a variety of scenarios, where the input signal is encoded in one of the circuit's optical phases, thus modulating the quantum reservoir state. The resulting output probabilities are used to set the feedback phases and, at the end of the computation, are fed to a classical digital layer trained via linear regression to perform predictions. We then focus on the investigation of the role of input photon indistinguishability in the reservoir’s capabilities of predicting time-series. We experimentally demonstrate that two-photon indistinguishable input states lead to significantly better performance compared to distinguishable ones. This enhancement arises from the quantum correlations present in indistinguishable states, which enable the system to approximate higher-order nonlinear functions when using comparable physical resources, highlighting the importance of quantum interference and indistinguishability as a resource in photonic quantum reservoir computing. 

\end{abstract}

\maketitle

\section{Introduction}\label{sec:introduction}

The recent surge of artificial intelligence, relying on huge neural network (NN) architectures that require high-speed and high-dimensional data processing, is driving renewed interest in alternative hardware platforms beyond conventional microelectronic processors. 
The core of the NN computation consists of two operations: (1) matrix–vector multiplication, which enables the propagation of information across the network layers, and (2) the application of nonlinear activation functions at each node. Scaling these operations to process increasingly high-dimensional data and complex tasks demands deeper and more expressive models \cite{raghu2017expressive}. Yet, traditional digital processors face two major challenges in meeting these demands. The first is computational: the time and memory requirements scale unfavorably with network depth and size, imposing practical limits on achievable performance. The second challenge is linked to the energy consumption and the associated carbon footprint of training and running such large-scale models, making standard processors unsustainable \cite{patterson2021carbon}. 
This is the reason why we are now at a point, often referred to as the electronic bottleneck, which constrains the scalability, speed, and energy efficiency of deep learning models. 

In response to these challenges, one promising alternative is neuromorphic computing \cite{markovic2020physics,schuman2022opportunities,mehonic2024roadmap}, a paradigm introduced in the 1990s by Carver Mead \cite{mead1990neuromorphic} that proposed to emulate the structure and function of the brain using analog circuits to enhance computation efficiency. Although it represented a visionary strategy at the time, it experienced a long period of limited progress after its proposal, largely due to the absence of compelling real-world applications and the lack of a clear incentive to deviate from digital computing, whose circuit components continued to grow following Moore's law.
Today, the context has fundamentally changed, catalyzed by the urgent need for novel energy-efficient hardware solutions \cite{pal2024ultra, goltz2021fast,xu2023low} capable of sustaining the rapid evolution of deep NN models. 

Two properties are fundamental for these models: \emph{expressivity}, consisting of the ability to approximate complex nonlinear functions, and \emph{memory capacity} to retain and process information across multiple layers or time steps. These two ingredients are critical for solving a wide range of machine learning (ML) tasks, including time-series forecasting \cite{du2017reservoir}.
In this path, a new frontier is to explore the adoption of quantum systems to develop more efficient neuromorphic architectures \cite{spagnolo2022experimental}, which have access to an enlarged Hilbert space. This expanded state space leads to higher expressivity \cite{schuld2019quantum}, allowing quantum systems to represent and manipulate complex data structures with fewer physical resources. At the same time, quantum coherence and entanglement can offer mechanisms for encoding correlations and memory in ways that are not readily achievable classically. 
However, this requires the challenging task of optimizing the quantum system, which presents several practical problems.

Quantum reservoir computing (QRC) offers a practical and promising alternative in the near/medium term \cite{fujii2021quantum, mujal2021opportunities, ghosh2021quantum, innocenti2023potential,xiong2025fundamental} by restricting the training to a classical readout layer while keeping fixed the quantum system dynamics, thus avoiding the complexities of directly optimizing quantum parameters. This framework has already demonstrated strong potential for addressing nonlinear and temporal learning problems \cite{fujii2017harnessing}, thus holding significant promise for advancing time series forecasting, which is fundamental in fields ranging from finance and climate modeling to natural language analysis.
Several physical platforms, including superconducting circuits \cite{senanian2024microwave,yasuda2023quantum} and  Rydberg atom arrays \cite{bravo2022quantum,kornjavca2024large}, have already been employed to demonstrate the effectiveness of the QRC paradigm. 
In particular, photonics systems have been widely adopted lately for QRC, leveraging both continuous-variable squeezed states \cite{paparelle2025experimental, cimini2025large, garcia2023scalable, garcia2024squeezing} and discrete-variable encodings\cite{zia2025quantum, nerenberg2025photon, suprano2024experimental}. Recent advances have shown that nonlinearity and memory can be introduced through adaptive feedback schemes \cite{kobayashi2024feedback}, where quantum circuits are dynamically modified based on intermediate measurement outcomes, making integrated photonic a strong candidate for further investigation.

In this work, we demonstrate the implementation of a quantum reservoir architecture based on a photonic integrated circuit (PIC) injected with different optical input states. While previous feedback-driven approaches based on integrated circuits have been restricted to the single-photon regime \cite{selimovic2025experimental}, here we investigate time-series forecasting with multiphoton inputs. We focus on assessing the role of quantum correlation in the system's computational capabilities by comparing the performance achieved when the system involves indistinguishable two-photon states and distinguishable photon inputs under otherwise identical conditions. Specifically, we fix the number of physical resources and employ the same classical optimization protocols across all configurations to ensure a fair comparison. The achieved experimental results show that the use of indistinguishable photon states, combined with active feedback dynamics, significantly enhances the expressivity of the reservoir, enabling it to accurately reconstruct higher-order nonlinear functions. As a result, this enhanced nonlinear processing capability does not compromise the fading memory property of the system, which remains essential for effective temporal information processing. Therefore, this improvement translates into superior performance on time-dependent benchmark tasks, including temporal XOR, NARMA sequences, and the prediction of Mackey-Glass chaotic series. On the other hand, we prove that classical correlations alone, present in distinguishable two-photon states, provide little to no advantage compared to quantum correlations.

\section{Photonic Quantum Reservoir Computing}\label{sec:photonic}
\subsection{Protocol}\label{sec:protocol}

\begin{figure*}[ht]
    \centering \includegraphics[width=0.99\textwidth]{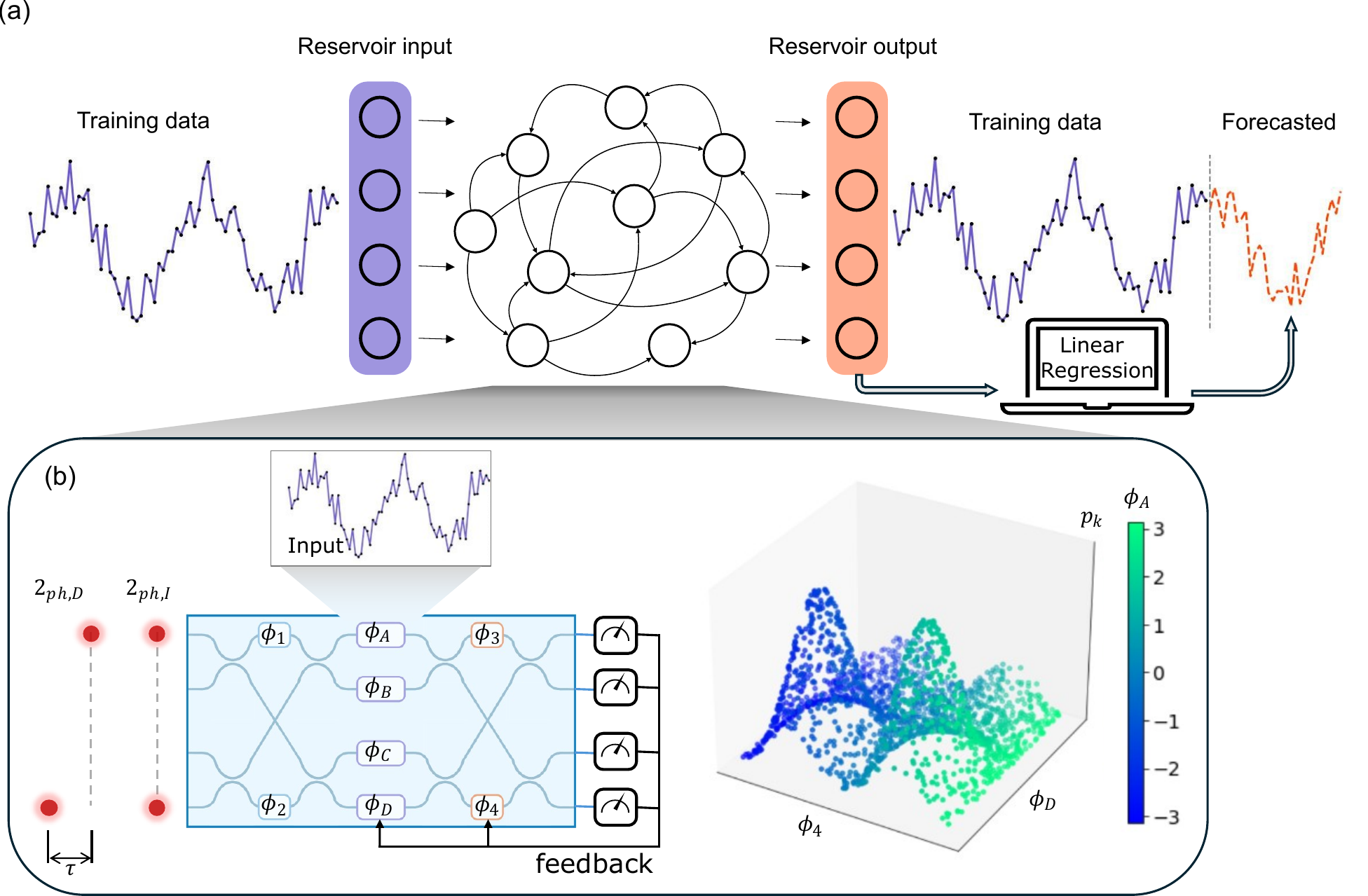}
    \caption{\textbf{Schematic overview of the realized photonic quantum reservoir computing.} \textbf{(a)} Scheme of a classical echo state network (ESN) for time-series forecasting \cite{jaeger2001echo}. The input data is injected into a fixed, recurrent reservoir of nonlinear nodes. The reservoir output is linearly combined and trained via linear regression to predict future values. 
    \textbf{(b)} Photonic implementation of a quantum reservoir computing scheme. Two-photon input states, either indistinguishable ($2_{ph,I}$) or distinguishable ($2_{ph,D}$) with a control on the delay $\tau$, are injected into a four-mode reconfigurable photonic integrated platform. This includes the tunable phases $\phi_i$ (with $i=1,2,A,B,C,D,3,4$). The input data is encoded into one phase of the central layer. The output statistics are reconstructed using single-photon detectors. A feedback-loop mechanism is based on the reinjection of the past measurement outcomes and is implemented via phase reconfiguration of $\phi_D$ and $\phi_4$. On the right, the plot of an example of a component of the output probability distribution as a function of the internal phases $\phi_A$, $\phi_D$ and $\phi_4$.}
    \label{fig:Fig_scheme}
\end{figure*}

Reservoir computing is a computational framework inspired by recurrent neural networks (RNNs) in which information is encoded into the dynamics of a fixed nonlinear system, the reservoir. The reservoir maps the inputs into a high-dimensional space, thus enabling complex learning tasks to be solved efficiently simply via linear regression. The reservoir can be a physical system exhibiting quantum behavior, in this case, the approach is known as QRC. 

A schematic representation of the protocol we implement for temporal signal processing is illustrated in Fig. \ref{fig:Fig_scheme}. In this approach, the training data are sequentially fed into the reservoir as input. The reservoir, treated as a \textit{black box} with random internal connections, dynamically evolves and nonlinearly maps the input features into a higher-dimensional space. The resulting outputs encode temporal patterns and are subsequently processed by a digital linear readout layer, which enables forecasting of future values of the signal.

Our quantum reservoir architecture is based on an integrated quantum photonic platform, injected with either single or multiphoton states, as depicted in Fig. \ref{fig:Fig_scheme}b. The time series is mapped into the reservoir dynamics by modulating an optical phase governing the transformation implemented by the circuit on the input photons. In particular, we focus on evaluating the performance of the system by comparing the results obtained when using distinguishable and indistinguishable photon inputs, aiming to assess the impact of quantum correlations on the system's computational capabilities, while the training procedure remains entirely classical.

Whether exploiting photon indistinguishability as a resource for classical ML tasks provides a consistent advantage over classical systems has remained an open question. Recent works have highlighted the potential of indistinguishable photons to enrich the feature space of photonic learning systems and increase the system's computational expressivity \cite{joly2025harnessing,yin2025quantum}. However, experimental investigations have been mostly limited to photonic extreme learning machines without feedback, and have reported little or no measurable advantage from photon indistinguishability in low-dimensional systems with only two photons \cite{joly2025harnessing}.

Given a classical input sequence $\{s_k\}$, we want to predict the time series $\{y_k\}$ generated by the input sequence through an unknown nonlinear or time-varying function. The protocol we implement consists of four key steps: encoding the input into the quantum system, evolving the quantum reservoir, measuring the reservoir output states, and applying a feedback operation conditioned on the measurement outcome. 

The first step is to encode the input value $s_k \in [0,1]$ into the quantum system by modulating the optical phase as $\phi_B = a_{\text{in}} \cdot s_k$, where $a_{\text{in}} \in [-\pi, \pi]$ is an input scaling factor. This phase modulation acts on photons via a phase shifter on the selected optical mode, described by the unitary matrix: 

\begin{equation}
    U(\phi_B) = \begin{pmatrix}
    1 & 0 & 0 & 0 \\
    0 & e^{i\phi_B} & 0 & 0 \\
    0 & 0 & 1 & 0 \\
    0 & 0 & 0 & 1
\end{pmatrix}.
\end{equation}
At each time step $k$, the reservoir evolves in response to the encoded input $s_k$. 
After the evolution corresponding to each element of the input sequence, the resulting output state is measured using a positive operator-valued measure (POVM) ${\Pi_\mu}$, where $\mu$ is the number of possible measurement outcomes. To accurately reconstruct the output probabilities, each input in the sequence is sampled multiple times, collecting sufficient statistics over repeated measurements. This process yields an output probability distribution $\mathbf{p}_k(\vec{\mu}) = [p_{k,1}, p_{k,2}, \ldots, p_{k,\mu}]$ associated to the $k$ input.

Finally, the application of the feedback loop plays a crucial role in enriching the computational dynamics of the reservoir by introducing both memory and nonlinearity into the dynamics, allowing the system to adapt based on past outputs. In our implementation, the information derived from selected measurement outcomes $\mathbf{p}_k$ is reinjected into the system during the next cycle $k+1$. This feedback operation is performed by encoding the selected components of $\mathbf{p}_k$ in the circuit parameters as phase shifts of the form $a_{\text{fb}} \cdot p_{k}(\mu')$, where $a_{\text{fb}} \in [-\pi,\pi]$ is the feedback scaling weight. This mechanism breaks the unitary evolution typically associated with quantum mechanics and allows the system to process temporal patterns and nonlinear input-output dependencies, while preserving its internal architecture.

For each input $s_k$ of the sequence, the vector of output probabilities $\mathbf{p}_k$ is reconstructed from photon-coincidence events detected between the different outputs of the circuit. These vectors act as high-dimensional feature representations of the system’s response to each input of the sequence. Once the probability vectors for each element of the sequence have been collected, they are associated with the corresponding target $y_k$. At this point part of the collected data $\{(\mathbf{p}_1, y_1), \ldots, (\mathbf{p}_K, y_K)\}$ of total length $K$ is split into a training set including a sequence of size $K_\text{tr}$, used to train the linear regression model, while the remaining $K_\text{ts} = K - K_\text{tr}$ data, is reserved to evaluate its performance on unseen data. We denote model predictions with a tilde. The goal is to learn a function that, given unseen inputs $\mathbf{p}_\text{ts}$ from the test set, can accurately predict their corresponding target values $\tilde{\mathbf{y}}_\text{ts}$.

We train the digital linear model using \textit{Ridge} regression (see Methods in Sec. \ref{sec:methods}), and evaluate its performance by comparing the predicted output vector $\tilde{\mathbf{y}}_\text{ts}$ to the true target vector $\mathbf{y}_\text{ts}$ of the test dataset. In particular, the determination coefficient $R^2$ is a figure of merit which quantifies their similarity, and is defined as:

\begin{equation}
    R^2(\mathbf{\Tilde{y}}_\text{ts}, \mathbf{y}_\text{ts}) = \frac{\text{cov}^2(\mathbf{\Tilde{y}}_\text{ts}, \mathbf{y}_\text{ts})}{\sigma^2(\mathbf{\Tilde{y}}_\text{ts}) \sigma^2(\mathbf{y}_\text{ts})},
\end{equation}

where cov and $\sigma^2$ are the covariance and variance, respectively. This measures the strength of a linear relationship between the two variables. It ranges from $0$ to $1$, where $0$ indicates no linear correlation, while $1$ indicates a perfect correlation.

Another metric is the mean-squared error (MSE):

\begin{equation}
    \text{MSE} = \frac{1}{K_{ts}} \sum_{k=1}^{K_{ts}} \left( \tilde{y}_{\text{ts},k} - y_{\text{ts},k} \right)^2,
\end{equation}

which quantifies how closely the predicted values match the target ones.

\subsection{Task-Independent Characterization}\label{sec:characterization}

To evaluate the computational capabilities of the QRC model, we adopt a set of standard benchmarks and performance metrics from classical ML. These tasks enable both to evaluate the performance of the protocol and to characterize the underlying physical system used as the computational substrate. The two main properties typically investigated are the \textit{short-term memory} and the \textit{expressive power}. The former quantifies the ability of the system to retain and recall information about past inputs, while the latter evaluates the range and complexity of input-output transformations the reservoir can implement. High expressivity indicates a strong capacity to approximate a broad class of nonlinear functions of the inputs.

The analysis of the memory capacity of a ML model involves processing an input sequence randomly sampled from a uniform distribution $s_k \in [0, 1]$, and the target is to reproduce the input value from $d$ cycles earlier, i.e., $y_k = s_{k-d}$. The performance of the model at each delay $d$ is measured using the determination coefficient $R^2_d$. The total memory capacity is then defined as the sum of these values: $C_\Sigma = \sum_{d=0}^{d_{\max}} R^2_d$, where $d_{\max}$ is an integer such that $R^2_{d_{\max}}$ is approximately zero.

We evaluate the expressivity of the system by feeding an ordered input sequence $s_k \in [0,1]$ into the reservoir, where each value is mapped to a corresponding optical phase $\phi_B$. The readout layer is then trained to approximate nonlinear target functions such as monomials or more complex polynomials. Specifically, the choice of input encoding and the effective dimensionality explored by the reservoir determine its expressive capacity, limiting the set of functions it can represent, typically characterized by a finite set of Fourier frequencies \cite{schuld2021effect}.

A common approach to forecasting time series data relies on RNN architectures, since these tasks require both high memory capacity and nonlinear processing capabilities. However, training RNNs can be challenging, especially in neuromorphic implementations. A simplified alternative to address such problems is represented by echo state networks (ESNs)  \cite{jaeger2001echo} or, more broadly, by reservoir computing. In an ESN, the recurrent component is a fixed, sparsely connected network of nonlinear nodes left untrained, as shown in Fig. \ref{fig:Fig_scheme}a. 
This architecture significantly reduces training complexity while preserving the ability to perform complex time-dependent tasks. However, this is contingent on the optimal exploitation of the system’s available degrees of freedom. 
In particular, the total memory capacity of these architectures is upper-bounded by the number of reservoir nodes $N_r$, i.e., $C_\Sigma \leq N_r$, and more generally by the rank of the reservoir’s correlation matrix $\mathbf{R}$ \cite{jaeger2002short}. 

In QRC, dimensional limitations are particularly critical because the available degrees of freedom are fundamentally constrained by physical resources, such as the number of modes, the number of photons, and the set of accessible measurement outcomes. Furthermore, practical implementations are affected by real-world imperfections, including statistical noise\cite{khan2021physical,hu2024generalization}, losses, and imperfect detection. These factors can further reduce the effective dimensionality of the computational space explored by the system.
In this context, we characterize the dimension of the explored space via the Gram matrix of the reservoir state $\mathbf{X}$, where each row represents the state at a given time step. It is computed as $\mathbf{G} = \mathbf{X}^T \mathbf{X}$, and encodes the pairwise inner products between reservoir states across time, offering a quantitative measure of the effective dimensionality explored by the observables \cite{nerenberg2025photon}.

\section{Experimental protocol}

%\subsection{Architecture of the Photonic Quantum Reservoir}
\label{sec:architecture}

Our photonic quantum reservoir consists of a hybrid platform made up of a single-photon source, a reconfigurable four-arm integrated interferometer, fabricated via femtosecond laser waveguide writing in glass \cite{corrielli2021femtosecond}, illustrated in all its details in \cite{valeri2023experimental}, and single-photon detectors. The source is a BBO crystal designed for Type-II Spontaneous Parametric Down Conversion (SPDC), which is used to generate photon pairs that are subsequently coupled into single-mode fibers (SMFs). The photons are spatially separated, and their indistinguishability, in both time and polarization, is controlled using delay lines and fiber polarization controllers, respectively. 

The PIC architecture is shown in Fig. \ref{fig:Fig_scheme}b, the device includes only linear optical elements, such as beam splitters (BSs), thermal phase shifters (PSs), and SWAP gates. This architecture allows us to implement the quantum reservoir protocol.  In the following, we detail each of the four main steps: state preparation and input encoding, evolution, measurement, and feedback operation.

\emph{Preparation and encoding--}
We evaluate the system performance by exploiting the path degree of freedom for three different input configurations:  single-photon ($n_{ph} = 1$), two distinguishable photons ($n_{ph,D} = 2$), and two indistinguishable photons ($n_{ph,I} = 2$) inputs. 
In the single-photon configuration, the input state $\vert 1 0 0 0 \rangle$ corresponds to injecting a single photon into the first mode of the PIC. The second photon here is used solely as a trigger for post-selected coincidence measurements. The temporal input sequence is encoded into the phase parameter $\phi_B$, which modulates the single-photon state as follows:
\begin{equation}
    \begin{aligned}
        \vert \psi_B \rangle_{1} = \frac{1}{2} (\vert 1\,0\,0\,0\rangle & - i e^{i\phi_B} \vert 0\,1\,0\,0\rangle \\ & - i \vert 0\,0\,1\,0\rangle - \vert 0\,0\,0\,1\rangle)
    \end{aligned}
\end{equation}

By also injecting the second photon into the last mode of the device, we initialize the input state in $\vert 1 0 0 1 \rangle$. In this scenario, we have access to a larger computational space, therefore, we can assess whether additional dimensionality enhances the system’s ability to solve the target tasks.
In the case of distinguishable photons, the measurement outcomes provide direct access to classical joint probability distributions over pairs of output modes, reflecting correlations that arise from the co-propagation of distinguishable photons through the circuit. 
These distributions contain additional information compared to single-photon inputs, reflecting classical correlations among the photons' paths. In contrast, when indistinguishable photons are injected, quantum interference leads to measurement statistics that reflect non-classical correlations. These can be exploited to realize richer transformations of the input sequence, thus serving as an additional resource to increase the expressivity and computational capabilities of the reservoir. When injecting indistinguishable photons into the two modes of the PIC, the evolution through the preparation layer allows us to generate two-photon entangled states by leveraging the effect of quantum interference at the BS level. Thus, we investigate the scenarios where the photons are injected either indistinguishably 
in time or with a controlled temporal delay $\tau$ as represented in Fig.\ref{fig:Fig_scheme}b, allowing us to prepare in a controlled way separable or entangled two-photon states.

In the encoding layer, the input sequence value $s_k$ is encoded via a phase shifter acting on the phase $\phi_B$. Consequently, once the state is prepared and the phase is encoded, the entangled two-photon state becomes:

\begin{equation}
    \begin{aligned}
    \vert \psi_B \rangle_{2, I} = & -\frac{1}{2 \sqrt{2}} ( \vert 2\,0\,0\,0\rangle + e^{2i\phi_B} \vert 0\,2\,0\,0\rangle +  \vert 0\,0\,2\,0\rangle \\
    &+ \vert 0\,0\,0\,2\rangle) + \frac{1}{2} (\vert 1\,0\,0\,1\rangle - e^{i\phi_B} \vert 0\,1\,1\,0\rangle ).
    \end{aligned}
\end{equation}

On the other hand, the two-photon distinguishable state takes the form:

\begin{equation}
    \begin{aligned}
    \vert \psi_B \rangle_{2, D} = & \frac{1}{4} \bigl[(\vert 1\,0\,0\,0\rangle_1 + i e^{i\phi_B} \vert 0\,1\,0\,0\rangle_1 + i  \vert 0\,0\,1\,0\rangle_1\\
    &- \vert 0\,0\,0\,1\rangle_1) \otimes (\vert 1\,0\,0\,0\rangle_2 + i  e^{i\phi_B}  \vert 0\,1\,0\,0\rangle_2\\
    &+ \vert 0\,0\,1\,0\rangle_2 - \vert 0\,0\,0\,1\rangle_2)\bigr],
    \end{aligned}
\end{equation}

where the indices 1 and 2 refer to the two distinguishable photons, which evolve independently across the four spatial modes.

\emph{Evolution--}
At this stage, the system evolves according to a unitary evolution applied to the input modes of the injected photons, depending on the feedback applied to the phases $\phi_D$ and $\phi_4$. This evolution is realized by the PIC and is given by the unitary operator:
\begin{equation}
    \mathcal{U}_{\phi_4} \otimes \mathcal{U}_{\phi_D}=
    \frac{1}{2}
    \begin{pmatrix}
    e^{i (\phi_D + \phi_4)} & i e^{i \phi_4} & i & -1 \\
    i e^{i (\phi_D + \phi_4)} & -e^{i \phi_4} & 1 & i \\
    i e^{i \phi_D} & 1 & -1 & i \\
    -e^{i \phi_D} & i & i & 1
    
    \end{pmatrix}.
\end{equation}

This unitary dynamics, which incorporates the input-dependent feedback encoded in the phases, drives the internal quantum state of the reservoir and encodes temporal correlations across input sequences. Further details about the evolution of the states injected into the PIC are provided in the Supplementary Information S1.

\emph{Measurement--}
The probability distribution of the circuit’s output states is measured using avalanche photodiodes (APDs). For each experimental configuration, we collect an average total number $N_{tot}$ of coincidences counts, which are used to reconstruct the output probability vector $\mathbf{p}_k(\Vec{\mu})$. 
Specifically, the same average counts have been chosen for all the different optical inputs to ensure consistent statistical resolution and a fair comparison, thus granting that differences in performance arise from the system’s dynamics rather than variations in measurement statistics.
To reliably reconstruct all the possible output configurations and distinguish the events in which the two photons exit through different modes from those in which both photons are detected in the same mode, we employ fiber beam splitters at each output port. This detection scheme yields 10 distinct outcomes when two-photon input states are injected, and 4 possible outcomes when single-photon inputs are used.

\emph{Feedback operation--}
The measurement performed at the time step $k$ will condition the encoding and the measurement of the next step $k+1$ since a part of the distribution $\mathbf{p}_k(\vec{\mu})$ is fed back into the reservoir in order to implement the temporal feedback loop. 
Specifically, the feedback applied to the phase $\phi_D$ during the step $k$ is determined by the measurement outcome corresponding to a selected component $\mu'$ of the probability reconstructed at the previous time step i.e., $p_{k-1}(\mu')$. On the other hand, the feedback applied to $\phi_4$ is set based on a probability component that incorporates the system evolution in two steps earlier i.e., $p_{k-2}(\mu'')$.
After acquiring the entire sequence of measurements, the model is trained using Ridge regression, which maps the recorded output probabilities to the target values of the computational task.

\section{Experimental results}\label{sec:experimental_results}

\begin{figure*}[ht]
    \centering \includegraphics[width=0.99\textwidth]{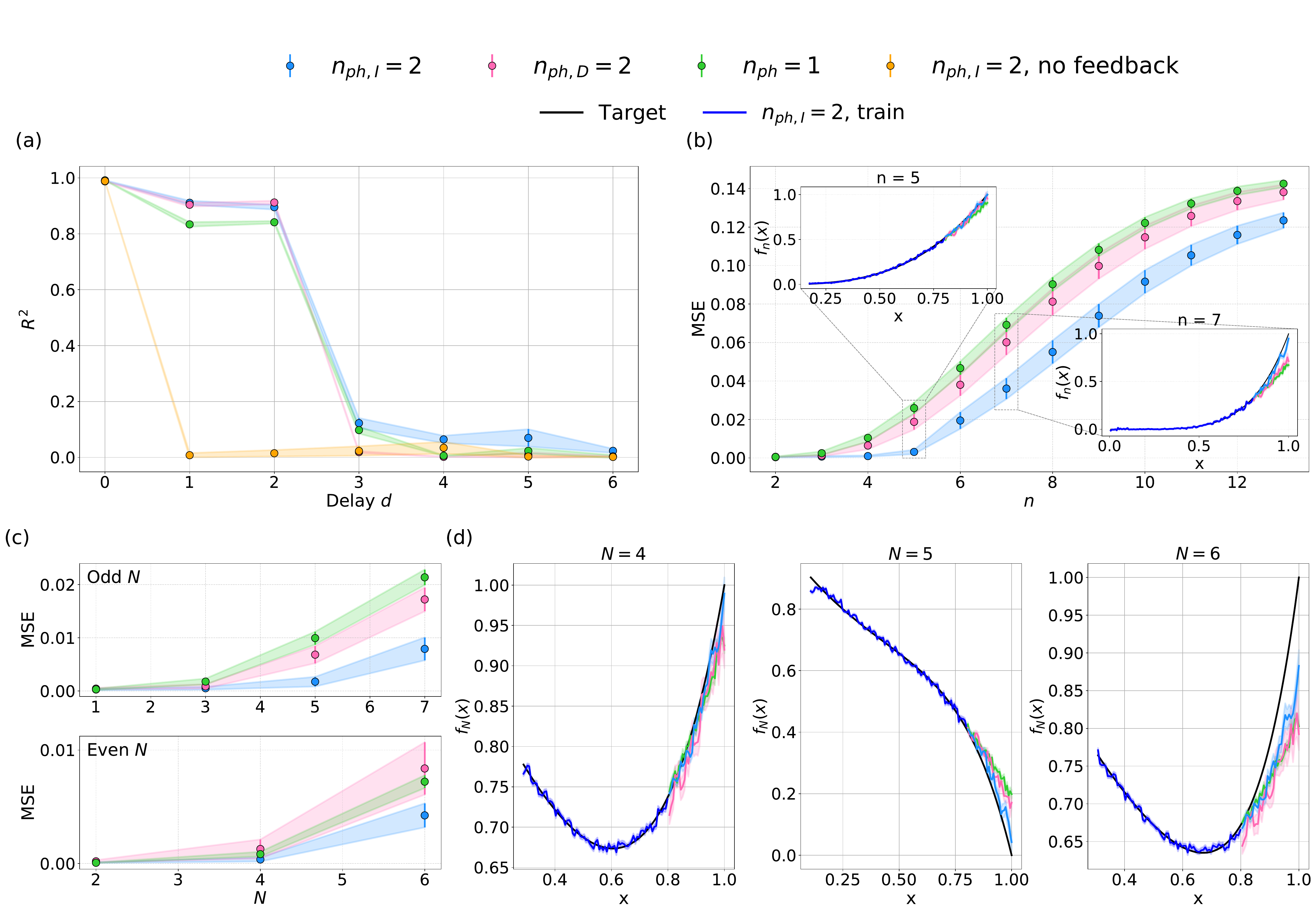}
    \caption{\textbf{Characterization of the quantum reservoir computing model: short-term memory and expressivity.} We report the quantum reservoir performance achieved for different photonic input states: two indistinguishable photons ($n_{\text{ph}, I} = 2$, light blue dots), two distinguishable photons ($n_{\text{ph}, D} = 2$, pink dots), and a single photon ($n_{\text{ph}} = 1$, green dots). (a) The short-term memory capacity is measured via the coefficient of determination $R^2_d$ as a function of the delay $d \in [0, 6]$. We report the results obtained when injecting into the device two indistinguishable photons but without feedback dynamics (yellow dots). The dataset contains 497 points in total.
    (b) The expressivity of the QRC model is measured by looking at the reservoir performance, in terms of the mean squared error (MSE) on the test set, in reconstructing $f_n(x) = x^n$ for $n = 2, \ldots, 13$. Insets display the predicted outputs for $n = 5$ and $n = 7$ for different photon-input configurations compared to the target (black lines). 
    It is also plotted the corresponding training output for $n_{\text{ph}, I} = 2$ (dark blue line). The datasets contain 150 points in total. (c) The expressivity of the model is further explored through the approximation of polynomial functions defined as $f_N(x)=\sum_{n=1}^N (-1)^n x^n$, with $N=1, \ldots, 7$. The MSE is plotted as a function of $N$. (d) Three examples of polynomial function approximations for selected values of $N=4, 5, \text{and } 6$. Error bars and shaded areas in the plot refer to the statistical fluctuations evaluated from 100 Monte Carlo extractions to account for the presence of Poissonian sampling noise.
    }
    \label{fig:Fig2}
\end{figure*}

%The whole experimental procedure for implementing the QRC protocol involves several steps. 
We start by choosing the configuration of phase shifters that maximizes the rank of the system's Gram matrix. This selection defines the transformation used to encode the input sequence and the ones of the two feedback loops applied during the protocol.
Consequently, the integrated device is calibrated to have full control over the selected phase-shifters.

Once the transformation is fixed, task-specific optimization can be performed by tuning the weights that map both the input sequence and the feedback signals, onto the corresponding optical phases. This corresponds to adjusting the input weight coefficient $a_{in}$, as well as the two feedback weights $a_{fb,D}$ and $a_{fb,4}$ which control the influence of the feedback on the optical phases $\phi_D$ and $\phi_4$, respectively. These parameters are treated as hyperparameters and are optimized considering the specific computational task under investigation.
To perform this optimization efficiently, we employ the Python \texttt{Optuna} package, a hyperparameter optimization framework based on Bayesian optimization algorithms \cite{akiba2019optuna}. In addition to tuning the weights, we also optimize the choice of which components $\mu'$ and $\mu''$ of the output probability vector are used to drive the two feedback loops. Indeed, depending on the input sequence, the informativeness of the output modes changes, making this choice task-dependent. These hyperparameter settings permit the system to adapt to the specific task demand: some tasks benefit from greater memory capacity to capture temporal correlations, while others require enhanced nonlinearity to improve the separation of complex input patterns\cite{stepney2024physical}. Further details are provided in the Supplementary Information S1, Supplementary Tabs. S1-2 and Supplementary Figs. S1-2.

We characterize our system performance in the regimes of single-photon input states ($n_{\text{ph}} = 1$) and two-photon input states. In particular, we consider both distinguishable ($n_{\text{ph},D} = 2$) and  indistinguishable ($n_{\text{ph},I} = 2$) two-photon inputs.

Starting with the study of short-term memory via the linear memory capacity task, we encode a randomly and uniformly sampled sequence of values $s_k \in [0,1]$ into the reservoir. The resulting reconstructed output probability vector is then used to train the linear regression model.
The results of the linear memory capacity task are reported in Fig. \ref{fig:Fig2}a. In this analysis, we compare the performance of the QRC system when exploiting the dynamics of the aforementioned different optical input states. 
These results highlight that the presence of feedback dynamics is the main factor that gives memory to the system. In contrast, quantum correlations alone do not appear to play a decisive role in this task, as the memory capacity is essentially identical for two-photon states in both distinguishable and indistinguishable configurations. Notably, when feedback is active, the QRC with a two-photon input state achieves slightly better performance compared to the single-photon case, as detailed in the Supplementary Information S2 and Supplementary Fig. S11.

The expressivity of the physical reservoir is evaluated by testing the ability of the QRC system to reconstruct monomial functions $f_n(x) = x^n$, with $x \in [0, 1]$ and $n$ ranging from 2 to 13. This benchmark is commonly used to quantify the nonlinear computational capacity of a reservoir, since higher-order monomials require more complex transformations of the input. 
The MSE in the monomial reconstruction on the test set is reported as a function of the exponent order $n$ in Fig.\ref{fig:Fig2}b for the different reservoir configurations investigated. We observe in this scenario that the configuration employing indistinguishable photon inputs in combination with feedback dynamics consistently achieves superior performance compared to other scenarios, resulting in lower MSE values, showing that quantum correlations can serve as a valuable resource for enhancing the system's expressivity. Such an enhancement is evident not only in improved prediction accuracy but also in the learning efficiency, which can be quantified either in terms of required training epochs for convergence \cite{siemaszko2023rapid} or by the number of training data to achieve a certain performance threshold. Specifically, as we show in the Supplementary Information S2 and Supplementary Figs. S3-9, the configuration using two indistinguishable photons reaches performance saturation faster than other input configurations, indicating that quantum interference enables the reservoir to construct richer feature representations with fewer training examples. 
Importantly, simply having access to joint classical probability distributions leads only to a modest improvement in terms of expressivity. This is evident in the configuration with two distinguishable photons, which slightly outperform the single-photon case, despite offering more degrees of freedom. These classical regimes tend to saturate at the same expressivity limit. In contrast, the use of indistinguishable photons provides a more evident enhancement in this nonlinear learning task.

The expressivity of the QRC is further tested by considering as a second benchmark the reconstruction of nonlinear polynomial functions of the form $f_N(x) = \sum_{n=1}^{N} (-1)^n x^n$.
Using the same dataset as in the monomial case, this task evaluates the reservoir’s ability to capture a richer combination of nonlinear terms, offering a stronger test of its computational capabilities.
The MSE in reconstructing these polynomial functions for increasing integer values of $N \in \{1, \ldots, 7\}$ is plotted in Fig. \ref{fig:Fig2}c. As in the monomial case, the configuration with indistinguishable photons achieves the lowest MSE, indicating superior expressivity. In contrast, configurations using distinguishable photons, as well as the single-photon input, exhibit significantly worse performance as $N$ increases.

\subsection{Experimental machine learning benchmarks} \label{sec:exp_ml}

\begin{figure*}[ht]
    \centering \includegraphics[width=0.99\textwidth]{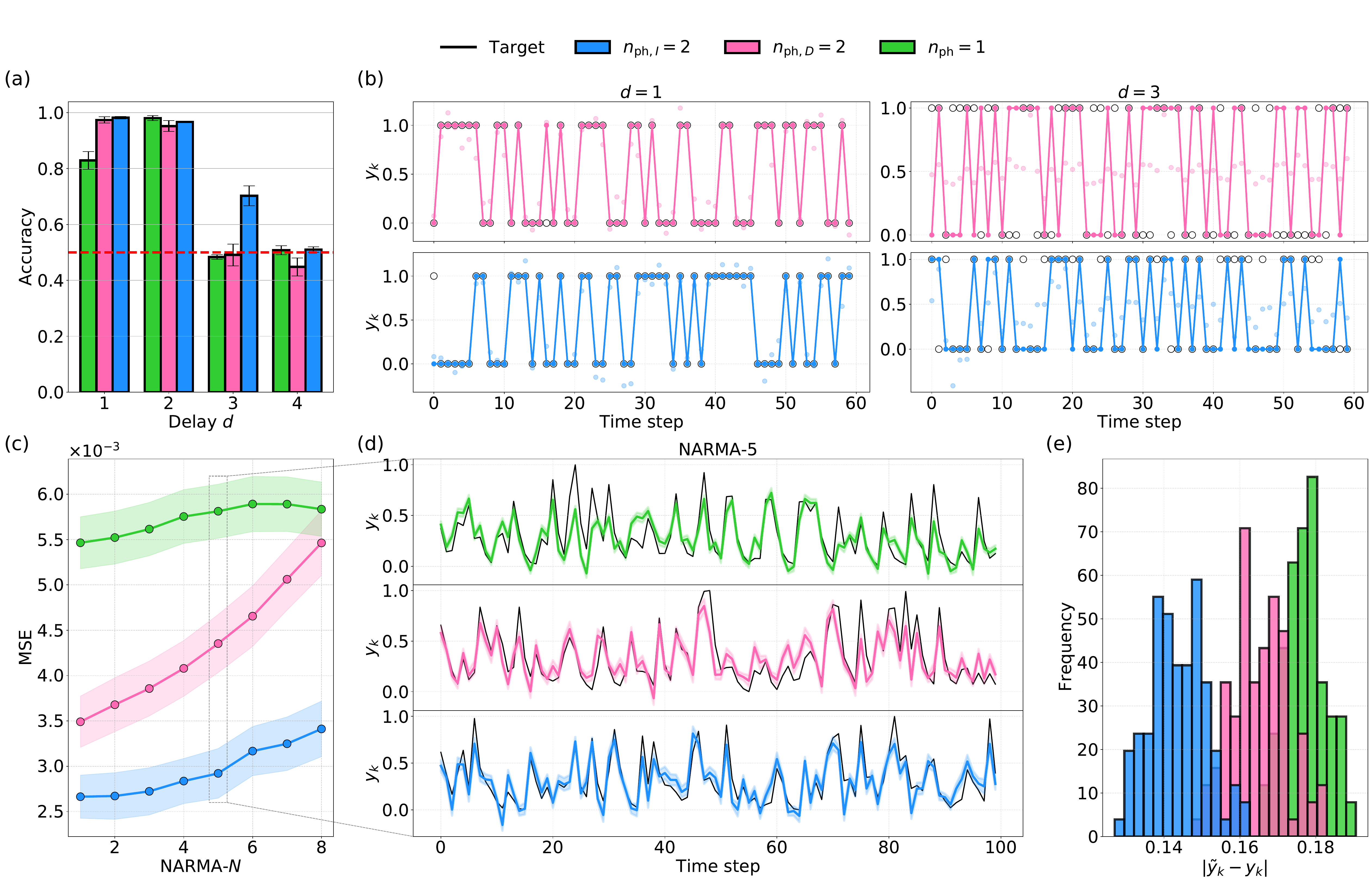} \caption{\textbf{Quantum reservoir performance on machine learning benchmarks: temporal XOR and NARMA-$N$.} The tasks are evaluated for different input configurations: two indistinguishable photons ($n_{\text{ph}, I} = 2$, light blue), two distinguishable photons ($n_{\text{ph}, D} = 2$, pink), and single photons ($n_\text{ph} = 1$, green). (a) The temporal XOR is defined as $s_k \bigoplus s_{k-d}$, where $s_k \in \{0, 1\}$ is the input sequence and $d$ is the temporal delay. The task is evaluated using classification accuracy for delays $d = 1,\ldots,4$.
    The datasets contain 300 points in total. 
    (b) The output predictions are compared with the targets for both $n_{\text{ph}, I} = 2$ and $n_{\text{ph}, D} = 2$ configurations, and for $d = 1$ and $d = 3$. While both perform well at $d = 1$, only the $n_{\text{ph}, I} = 2$ configuration successfully addresses the increased memory and nonlinearity demands at $d = 3$. (c) Performance on the test set for the NARMA-$N$ task \cite{atiya2000new}.
    Also this testbed shows the enhancement due to quantum correlations; indeed, the configuration with $n_{\text{ph}, I}=2$ always outperforms the one with $n_{\text{ph}, D}=2$ and $n_\text{ph}=1$. The datasets contain 500 points in total. (d) The test with respect to the target for the three configurations investigated is plotted for the function NARMA-$5$. Both the predicted and the target values $y_k$ are normalized to ensure a fair comparison. (e) The different normalized mean absolute errors between the predicted and the target values are compared in a histogram. Error bars and shaded areas in the plot refer to the statistical fluctuations evaluated from 100 Monte Carlo extractions to account for the presence of Poissonian sampling noise.}
    \label{fig:Fig3}
\end{figure*}

\begin{figure*}[ht]

    \centering \includegraphics[width=0.99\textwidth]{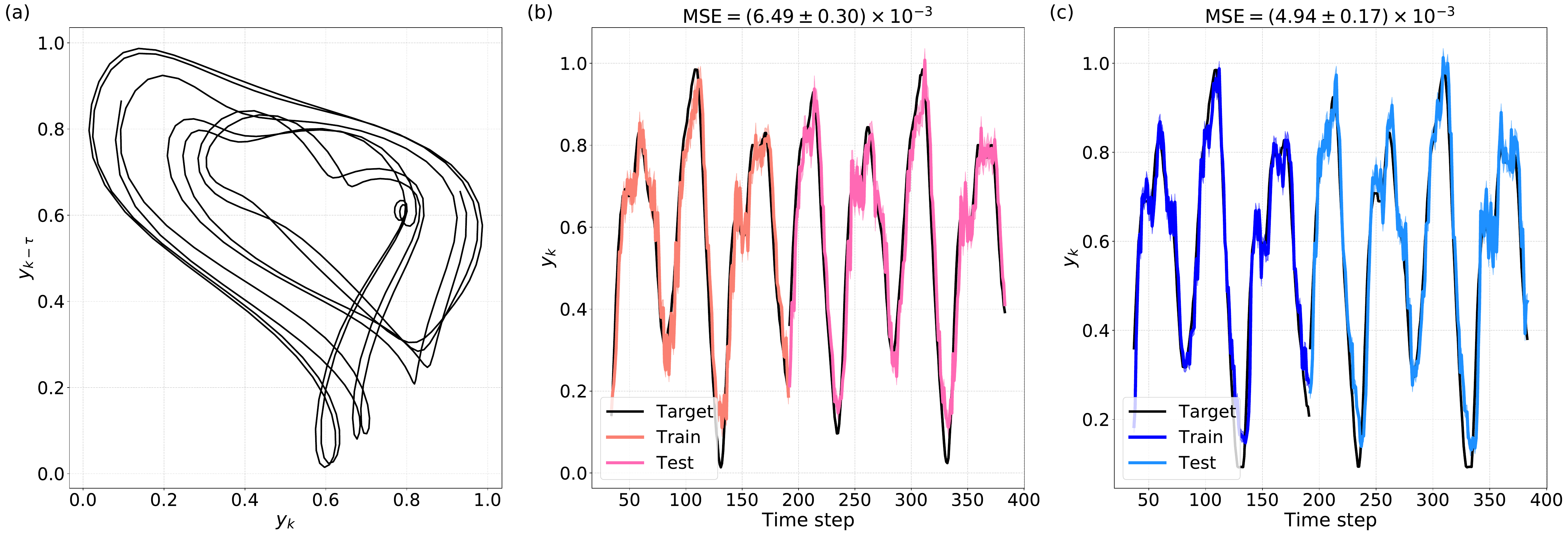} \caption{\textbf{Mackey-Glass time series forecasting.} (a) Chaotic attractor of the Mackey–Glass (MG) time-delay differential equation, simulated with parameters $\alpha = 0.2$, $\beta = 10$, $\gamma = 0.1$, and delay $\tau = 17$, which induces chaotic dynamics. The differential equation is numerically integrated, then time-discretized, normalized, and directly encoded as input to the quantum reservoir system. (b) and (c) Forecasting of future time steps with prediction horizon $t_f = 3$, comparing the target signal (black) with the predicted outputs during both training and testing phases. Two photon-input configurations are considered: (b) injection of two distinguishable photons ($n_{\text{ph}, D} = 2$, shown in salmon for training and pink for test), and  (c) injection of two indistinguishable photons ($n_{\text{ph}, I} = 2$, shown in dark blue for training and light blue for test). The achieved MSE is reported in the title of the corresponding plot. Each dataset contains 390 points. Shaded areas in the plot refer to the statistical fluctuations evaluated from 100 Monte Carlo extractions to account for the presence of Poissonian sampling noise.}
    \label{fig:Fig4}
\end{figure*}

The computational capabilities of the photonic QRC platform are finally evaluated on standard ML tasks. In particular, we benchmark its performance on tasks widely adopted in reservoir computing literature \cite{kobayashi2024feedback,garcia2024squeezing,garcia2024quantum,selimovic2025experimental,paparelle2025experimental}. These include the temporal XOR, the nonlinear autoregressive moving average family (NARMA-$N$), and the Mackey-Glass (MG) time-series forecasting task. Each of them is designed to probe different aspects of the system's computational performance. In particular, these testbeds allow us to analyze how the physical platform processes information and to investigate the role of quantum correlations as a computational resource when applied to real classical datasets. %\fromVale{questo non è vero alla fine mi pare no:Here, the feedback-loop used is the same for the linear memory capacity in order to enhance the memory capacity of the reservoir.}

The temporal XOR is a binary task where the target output is $y_k = s_k \bigoplus s_{k-d}$, with an input sequence of binary values $s_k \in \{0,1\}$. This task requires the system to store and nonlinearly combine temporally delayed inputs. Since the performance of the model is evaluated using the accuracy on a test sequence of new data, the raw outputs of the linear regression are rounded in a successive step to the binary values 0 and 1. 

The achieved results are shown in Fig. \ref{fig:Fig3}a-b, where it is studied how the accuracy performance changes varying the delay $d \in \{1, \ldots, 4\}$. Specifically, the bar plot in Fig. \ref{fig:Fig3}a shows the accuracy for the different delays $d$, for the different input states: $n_{\text{ph}} = 1$, $n_{\text{ph},D} = 2$ and $n_{\text{ph},I} = 2$. As expected, the accuracy decreases with increasing delay, reflecting the growing memory demand of the task. For $d = 1$ and $d = 2$, all the configurations perform nearly perfectly, with accuracies approaching $1$. However, as the delay increases to $d = 3$, the accuracy of the configuration with $n_{\text{ph},I} = 2$ continues to learn effectively, while the one with either $n_{\text{ph}} = 1$ or $n_{\text{ph},D} = 2$ drops significantly, approaching the random guess level of 50\%. %\fromRosario{fatto} \fromVale{così sembra che non ci sia alcun vantaggio e anche per d=3 va male, invece va sottolineato Proprio che per d=3 indistinguibili ancora impara qualcosa distinguibili già non ce la fa più}
Fig. \ref{fig:Fig3}b shows the output for the input state configurations with $n_{\text{ph},I} = 2$ and $n_{\text{ph},D} = 2$, considering $d = 1$ and $d = 3$. In each subplot, the predicted output $y_k$ is plotted over the 60 time steps representing the test set. At $d = 1$, both systems track the target sequence closely, while for $d = 3$, although the predictions become noisier and deviate more often from the target, the configuration with $n_{\text{ph},I} = 2$ still reaches an accuracy of $0.70 \pm 0.03$. In contrast, the configuration with $n_{\text{ph},D} = 2$ is no longer able to reconstruct the output, resulting in an accuracy equal to random guess. %, similarly to the case without a feedback loop. 
In contrast to the short-term memory study, these results show that exploiting quantum correlations in the input state enables the system to better handle more complex tasks. This is evidenced by the marked improvement in performance observed for the input state $n_{\text{ph},I} = 2$ compared to $n_{\text{ph},D} = 2$ configuration. 

Next, the indistinguishability advantage is investigated via the NARMA sequence task, which tests both short- and long-term memory combined with nonlinear transformations. The input sequence is randomly and uniformly sampled in $[0,1]$ and the target is:

\begin{equation}
    y_k = \alpha y_{k-1} + \beta y_{k-1} \sum_{j=k-N}^{k-1} y_j + \gamma \left( u_{k-1}^3 + u_{k-1}^5 \right) + \delta,
\end{equation}
where $u_k = \mu + \nu s_k$ with $\mu=0$ and $\nu=0.2$, while the other parameters are $\alpha=0.3$, $\beta=0.05$, $\gamma=50$ and $\delta=0.1$.
The evaluation of the task is illustrated in Fig. \ref{fig:Fig3}c, where performance are studied via the MSE as a function of the NARMA order $N$ \cite{atiya2000new}. The results show a similar behaviour to the one obtained for the temporal XOR task, indeed the system with $n_{\text{ph}, I} = 2$ outperforms both the $n_{\text{ph}} = 1$ and $n_{\text{ph},D} = 2$ cases across all the values of $N \in \{1, \ldots, 8\}$. The comparison for NARMA-$5$ is further emphasized in Fig. \ref{fig:Fig3}d showing how the configuration with indistinguishable photons tracks the target more closely than the distinguishable case. This supports the point that the presence of quantum correlations expands the accessible computational space in a meaningful way, enabling the reservoir to capture and reproduce complex temporal dependencies, especially when nonlinear dynamics are involved. Finally, the results in Fig. \ref{fig:Fig3}e summarizes this behavior through the mean absolute error among the different input states.

The last task examined is the forecasting capabilities of our QRC model for the Mackey-Glass time delay differential equation, described by complex continuous-time chaotic dynamics:

\begin{equation}
    \dot{s}(t) = \frac{\alpha s(t - \tau)}{1 + s(t - \tau)^\beta} - \gamma s(t),
\end{equation}
where the parameters are fixed to $\alpha = 0.2$, $\beta = 10$, $\gamma = 0.1$, and $\tau = 17$. The latter is responsible for the chaotic attractor dynamics shown in Fig. \ref{fig:Fig4}a. Then, solved numerically the differential equation, it is time discretized and encoded in the reservoir as $s_k$ normalized within the interval $[0,1]$. Differently from all the other tasks, in this scenario, we encode the nonlinear input $s_k$ directly in the input phase value. %\fromRosario{The dataset is then split evenly: 50\% is used to train the model, while the remaining 50\% is reserved for evaluating its performance on unseen data.} 
The aim of this task is to perform a time-series forecasting that generates predictions $t_f$ cycles into the future, i.e., trying to predict the target values $y_k = s_{k+t_f}$. 
In Fig. \ref{fig:Fig4}b-c, we evaluate the ability of the model to learn the Mackey-Glass time series,  comparing the forecasting for $t_f=3$ across the two configurations: $n_{\text{ph},D} = 2$ and $n_{\text{ph},I} = 2$.
While both scenarios involve two-input photons, the output spectra of the optical network show different numbers of frequency components. This difference arises from the total number of indistinguishable input photons, which affects the capacity of the model to represent complex signal structures \cite{gan2022fock}.
Despite this difference, both configurations succeed in capturing the fundamental frequency of the target signal. This is enabled by the direct encoding of the time-series and by the optimization of the input scaling hyperparameter $a_{in}$, which helps the output of the network to match its frequency. However, the configuration with two indistinguishable input photons also in this scenario slightly outperforms the reservoir with separable states in reconstructing the signal amplitude, showing an enhanced expressivity \cite{schuld2021effect}, thus expanding the set of functions accessible to the linear regression model. Additional details are provided in the Supplementary Information S2 and Supplementary Figs. S10-12.

\section{Discussion and conclusions}
The ambition to develop quantum neural networks has attracted growing interest recently. A widely adopted approach relies on variational strategies  \cite{cerezo2021variational}, where the parametrized quantum circuit is trained through a hybrid quantum-classical feedback loop. 
While successful in a range of applications \cite{peruzzo2014variational, agresti2024demonstration, hoch2025variational, cimini2024variational}, such methods face intrinsic scalability challenges, including the Barren plateau problem \cite{mcclean2018barren}, where gradients vanish exponentially with system size. 
This makes this procedure insufficient or infeasible when dealing with complex and high-dimensional systems that require extensive gradient optimization.
These limitations motivate the exploration of alternative models such as quantum reservoir computing.

In this work, we have experimentally implemented a photonic quantum reservoir protocol for temporal sequences forecasting using a reconfigurable integrated photonic platform, necessary for implementing real-time feedback loops. By injecting into the reservoir single-photon, two-photon distinguishable, and two-photon indistinguishable states, encoded in the path degree of freedom, we have systematically investigated how quantum correlations influence computational performance.

The successful forecasting of time-series data generally relies on two key capabilities of the computing platform: its capacity to memorize information about past inputs, to keep track of temporal correlations, and the ability to model nonlinear input-output relations. 
Our study demonstrates that the first capability can be engineered in integrated photonics systems through adaptive feedback mechanisms, with performance mainly governed by the number of active feedback loops and depending on the size of the accessible configuration space. These elements define the reservoir's memory. The second capability, linked to the expressivity of the system, can be improved by the use of multiphoton quantum states. When quantum correlations are introduced, exploiting photon indistinguishability, the system gains a clear advantage in reconstructing higher-order nonlinear functions, when fixing the number of physical resources. This enhancement directly translates into task-dependent improvements in accuracy on classical time series prediction benchmarks characterized by strong nonlinearities.
The observed performance gain can be attributed to the combination of multiphoton inputs and adaptive dynamical feedback, which increases the reservoir’s sensitivity to variations in the parameters encoding of the input sequence. Such increased sensitivity can be interpreted in terms of higher Fisher Information associated with indistinguishable two-photon states, which makes the output probabilities more responsive to changes in the encoded data. This has been largely proven in prior adaptive phase estimation experiments carried out on the same platform \cite{valeri2023experimental, cimini2023deep, cimini2024variational}, and investigating this connection further represents a promising direction for future research.

Overall, our results provide experimental evidence that quantum correlations and multiphoton states enrich the structure of the reservoir’s dynamics in a meaningful way, learning more effectively temporal dependencies when implementing adaptive feedback loops. Notably, this enhancement emerges already when employing two-photon states in a four-mode device, demonstrating the possibility to have an advantage even when operating in a relatively low-dimensional space.

\begin{figure*}[htb!]
    \centering \includegraphics[width=1\textwidth]{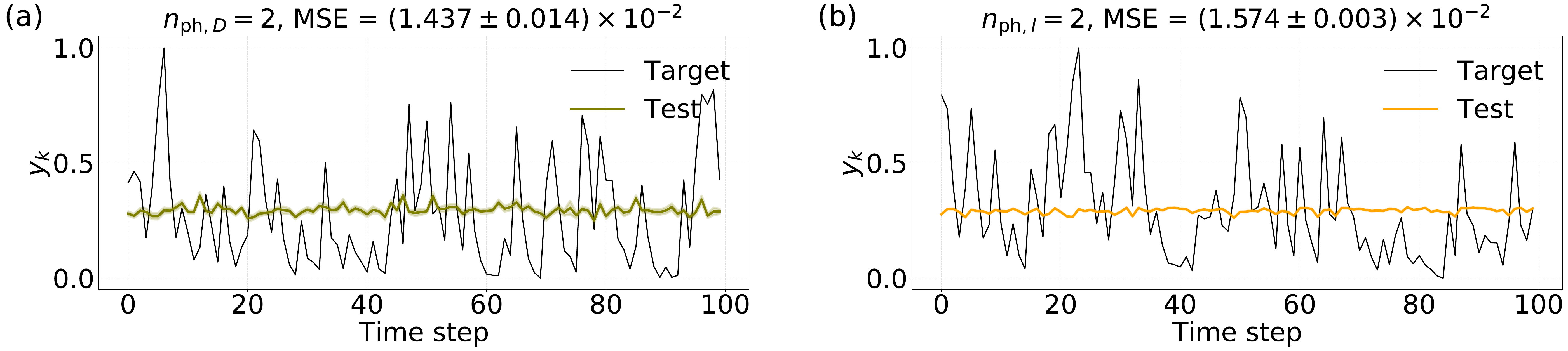} \caption{\textbf{Performance of NARMA-5 task using reservoir without feedback-loop mechanism.} (a) Output prediction (dark green) versus target (black) for the configuration with $n_{\text{ph}, D}=2$. (b) Output prediction (orange) versus target for the configuration with $n_{\text{ph}, I}=2$. In both cases, the datasets are reshuffled to eliminate any temporal correlations, and thus preventing the reservoir from retaining memory of past inputs. The resulting mismatch confirms that memory is essential for solving time-dependent tasks. Both the test and the target outputs $y_k$ are min–max normalized to the range $[0,1]$ to ensure a fair comparison between $n_{\text{ph}, D} = 2$ and $n_{\text{ph}, I} = 2$. Error bars and shaded areas in the plot refer to the statistical fluctuations evaluated from 100 Monte Carlo extractions to account for the presence of Poissonian sampling noise.}
    \label{fig:Fig5}
\end{figure*}

\section{Methods} \label{sec:methods}

\subsection{Digital linear layer} \label{sec:digital_linear_layer}

In quantum reservoir computing, a digital linear layer is the responsible of the training of the learning model. This layer takes as input the high-dimensional output features of the model, $\mathbf{X}$, and uses linear regression to adjust the weights of a matrix $W$ and approximate the target $\mathbf{y}$. Specifically, we use the linear model Ridge from the Python library \textit{sklearn} to perform linear regression efficiently.
In this framework, where the prediction function is defined as $f(\mathbf{p}_k, W) = \mathbf{p}_k^T W$ with $W$ being the vector of trainable weights, the training process consists in minimizing a regularized least-squares loss function:
\begin{equation} \label{eq:ridge}
    \min_{W} \sum_{k=1}^K \left(f(\mathbf{p}_k, W) - y_k \right)^2 + \alpha ||W||^2,
\end{equation}
with $\alpha$ the regularization strength and $||W||^2$ the squared $l_2$-norm of the weights.
Using the compact notation $\mathbf{X} = \bigl(\mathbf{p}_1, \ldots, \mathbf{p}_K\bigr)^T$ for the data matrix, and $\mathbf{y} = (y_1, \ldots, y_K)^T$ for the target vector, the least-squares error becomes $|\mathbf{X}W - \mathbf{y}|^2.$ When the regularization parameter $\alpha$ goes to zero the solution of such a minimization is $W = \mathbf{X}^+ \mathbf{y},$ where $\mathbf{X}^+$ is the Moore-Penrose pseudoinverse of $\mathbf{X}$, and, if the inverse exists, $\mathbf{X}^+ = (\mathbf{X}^T \mathbf{X})^{-1} \mathbf{X}^T.$ For $\alpha\not = 0$ it becomes: 
\begin{equation}
    W = (\mathbf{X}^T \mathbf{X} + \alpha I)^{-1} \mathbf{X}^T y,
\end{equation}
where $I$ is the identity matrix. Specifically, the role of the regularization term is to prevent overfitting by penalizing large weight values, with its strength controlled by the hyperparameter $\alpha \in [0, \infty)$.

Another hyperparameter considered in the training procedure is the \textit{washout}, which refers to the initial number of time steps during which the output of the reservoir is discarded and not used for training. This is done because, at the beginning of a sequence $\{s_k\}$, the reservoir state is more influenced by the initial conditions than by the input-driven dynamics. Discarding these initial steps exploits the fading memory property of the reservoir and allows the system to evolve into a dynamic regime that is independent of its initial state.
Consequently, the training for each task described in Sec. \ref{sec:experimental_results} is carried out by optimizing both $\alpha$ and the $washout$ using Optuna. The search ranges for the optimization are set to $\alpha \in [10^{-25}, 10^{-1}]$ and $washout \in [3, 50]$. The $\alpha$ range is chosen empirically to focus on regions associated with better performance, while the $washout$ range is selected based on the short-term memory analysis shown in Fig. \ref{fig:Fig2}a, ensuring that the amount of training data is not excessively reduced.

%\vspace{0.5\baselineskip} 

\subsection{Additional details on the feedback} \label{sec:feedback_details}
We explore two feedback configurations. In the first configuration, the feedback loops applied to the phases $\phi_D$ and $\phi_4$ depend only on the immediate past output, specifically at time step $k-1$. In the second configuration, the feedback incorporates a longer temporal history: the feedback applied to $\phi_D$ still depends on the output at time step $k-1$, while the feedback applied to $\phi_4$ is based on the output at time step $k-2$.
We have observed that, in the nonlinear regime, such as in the reconstruction of nonlinear functions, the expressivity of the system benefits from a feedback loop that relies only on the most recent step $k-1$. However, when the task places greater demands on temporal information processing, that is, when the model requires memory, performance improves by including information from earlier steps. In such cases, extending the feedback to incorporate additional past outputs enhances the system’s ability to capture temporal dependencies.

We further investigate the role of the feedback loop in the reservoir dynamics by evaluating performance when it is removed. As already shown in Fig. \ref{fig:Fig2}a, removing the feedback loop in a task that requires memory completely cancels the system’s memory capabilities. Here, we extend this analysis to a task requiring both expressivity and memory, that is the NARMA tasks. To this end, we use the same experimental datasets described in Sec. \ref{sec:exp_ml}, but apply a reshuffling of the input sequences.
The prediction results obtained using two reservoir configurations on the reshuffled datasets are shown in Fig. \ref{fig:Fig5}. In both cases, since the absence of any temporal structure, the reservoir is no longer able to reproduce the nonlinear dynamics required by the NARMA tasks, resulting in no agreement with the target signals. These results highlight the crucial role of feedback and memory in enabling the reservoir to model temporal functions.

\section{Funding}
This work is supported by the MUR PNRR project Spoke 4 and Spoke 7 (Grant No. PE0000023-NQSTI), the ERC Advanced Grant QU-BOSS (QUantum advantage via non-linear BOSon Sampling, Grant No. 884676), and the European Union's Horizon Europe research and innovation program under EPIQUE Project (Grant No. 101135288).
V.C. and F.S. acknowledge support from MUR FARE Ricerca in Italia QU-DICE (Grant No. R20TRHTSPA).

\section{Author contributions}
V. C., and F. S. conceived the experiment. R. Di B., V. C, and F.S. carried out the experiment and performed the data analysis. S.P., F.C., G.C., and R.O. fabricated the integrated photonic device and performed its initial characterization. All the authors discussed the results and contributed to the writing of the paper.

\section{Competing interests}
The authors declare no competing interests.

\section{Data availability}
All data and figures supporting the main conclusions of this study are available from the corresponding author upon reasonable request.

\section{Code availability}
The code used is available from the corresponding author upon reasonable request.

\section{Additional information}
Corresponding author: Valeria Cimini

\bibliography{QiSMG16.bib}

\end{document}

%\linenumbers
\title{\textit{Supplementary Information for}:\\Time-series forecasting with multiphoton quantum states and integrated photonics}

\let\comma,

\author{Rosario Di Bartolo}
\affiliation{Dipartimento di Fisica - Sapienza Università di Roma\comma{} P.le Aldo Moro 5\comma{} I-00185 Roma\comma{} Italy}

\author{Simone Piacentini}
\affiliation{Istituto di Fotonica e Nanotecnologie - Consiglio Nazionale delle Ricerche (IFN-CNR)\comma{} Piazza Leonardo da Vinci 32\comma{} I-20133 Milano\comma{} Italy}

\author{Francesco Ceccarelli}
\affiliation{Istituto di Fotonica e Nanotecnologie - Consiglio Nazionale delle Ricerche (IFN-CNR)\comma{} Piazza Leonardo da Vinci 32\comma{} I-20133 Milano\comma{} Italy}

\author{Giacomo Corrielli}
\affiliation{Istituto di Fotonica e Nanotecnologie - Consiglio Nazionale delle Ricerche (IFN-CNR)\comma{} Piazza Leonardo da Vinci 32\comma{} I-20133 Milano\comma{} Italy}

\author{Roberto Osellame}
\affiliation{Istituto di Fotonica e Nanotecnologie - Consiglio Nazionale delle Ricerche (IFN-CNR)\comma{} Piazza Leonardo da Vinci 32\comma{} I-20133 Milano\comma{} Italy}

\author{Valeria Cimini}
\email{valeria.cimini@uniroma1.it}
\affiliation{Dipartimento di Fisica - Sapienza Università di Roma\comma{} P.le Aldo Moro 5\comma{} I-00185 Roma\comma{} Italy}

\author{Fabio Sciarrino}
\affiliation{Dipartimento di Fisica - Sapienza Università di Roma\comma{} P.le Aldo Moro 5\comma{} I-00185 Roma\comma{} Italy}

\maketitle

\section{Experimental details}

\subsection{Quantum reservoir evolution}

In our photonic platform, the overall unitary transformation, that can be implemented via a four-arm integrated interferometer, is described by the unitary operator $\mathcal{U}$, given by:

\begin{equation}
\centering
\resizebox{0.94\textwidth}{!}{%
$\displaystyle
\mathcal{U} = \frac{1}{4} \begin{pmatrix}
(1 - e^{i\phi_B} + e^{i(\phi_D + \phi_4)} - e^{i\phi_4}) &
i(-1 + e^{i\phi_B} + e^{i(\phi_D + \phi_4)} - e^{i\phi_4}) &
i(-1 - e^{i\phi_B} + e^{i(\phi_D + \phi_4)} + e^{i\phi_4}) &
-(1 + e^{i\phi_B} + e^{i(\phi_D + \phi_4)} + e^{i\phi_4}) \\

i(-1 + e^{i\phi_B} + e^{i(\phi_D + \phi_4)} - e^{i\phi_4}) &
(-1 + e^{i\phi_B} - e^{i(\phi_D + \phi_4)} + e^{i\phi_4}) &
-(1 + e^{i\phi_B} + e^{i(\phi_D + \phi_4)} + e^{i\phi_4}) &
i(1 + e^{i\phi_B} - e^{i(\phi_D + \phi_4)} - e^{i\phi_4}) \\

i(- e^{i\phi_B} + e^{i\phi_D}) &
-(2 + e^{i\phi_B} + e^{i\phi_D}) &
(e^{i\phi_B} - e^{i\phi_D}) &
i(2 - e^{i\phi_B} - e^{i\phi_D}) \\

-(2 + e^{i\phi_B} + e^{i\phi_D}) &
i(e^{i\phi_B} - e^{i\phi_D}) &
i(2 - e^{i\phi_B} - e^{i\phi_D}) &
(- e^{i\phi_B} + e^{i\phi_D})
\end{pmatrix}
$%
}.
\end{equation}

This governs the transformation of the input $n_{\text{ph}}$-photon states through the interferometric circuit, driven by the effect of the internal phase settings $\phi_B, \phi_D, \phi_4$. In particular, $\phi_B$ is used to encode the input sequence while $\phi_D$ and $\phi_4$ are used to set the feedbacks.

The input state injected into the interferometer is described within a density matrix formalism that can be used for both single- and multi-photon configurations. In the multiphoton case, quantum interference effects, arising from the Hong-Ou-Mandel (HOM) effect \cite{hong1987measurement}, can lead to quantum-enhanced performance. The HOM effect, observable when two identical photons interfere at a beam splitter, manifests experimentally through bunching and anti-bunching behavior at the output of the chip. The degree of bunching depends on the indistinguishability of the photons, which can be tuned by manipulating a specific degree of freedom, such as the temporal delay, the frequency or the polarization. In our platform, multiphoton interference is achieved by operating a delay line that controls the injection time of the photons into the circuit. Thus, the indistinguishability is estimated by measuring the coincidence counts at the interferometer output as a function of temporal delay between photons, and it is quantified via the visibility $V$ of the HOM dip. A detailed characterization of our photonic circuit and the transition from distinguishable to indistinguishable photons is provided in the Supplementary Information of \cite{valeri2023experimental}. The actual input state $\rho(V)$ has to take into account the non-perfect photon indistinguishability, resulting in a visibility of the HOM dip lower than one, with $V \in (0,1)$, due to experimental imperfections. In the specific case of two-photon inputs, the state can be expressed as:

\begin{equation}
\rho(V) = V \left[ \left( \lvert \psi \rangle \langle \psi \rvert \right)_{2, I} \otimes \mathbb{1}_{2, D} \right]
+ (1 - V) \left[ \mathbb{1}_{2, I} \otimes \left( \lvert \psi \rangle \langle \psi \rvert \right)_{2, D} \right],
\end{equation}

where $I$ and $D$ are referred to respectively the indistinguishable and distinguishable photon contributions as reported in the main text. This formulation allows for a continuous interpolation between fully indistinguishable ($V=1$) and fully distinguishable ($V=0$) regimes. 

Finally, while the input state $\rho(V)$ is fixed, its evolution depends on the phase encoding implemented in the unitary operator $\mathcal{U}$. At each time step $k$, the state evolves according to:

\begin{equation}
    \rho_k(V) = \mathcal{U}_k \, \rho(V) \, \mathcal{U}_k^\dagger.
\end{equation}

The output is then obtained by performing a measurement described by a positive operator-valued measure (POVM) in the computational basis. The probability of observing the outcome $\mu$ is given by:

\begin{equation}
    p_{k}(\mu) = \mathrm{Tr} \left[ \Pi_\mu \, \rho_k(V) \right],
\end{equation}
where $\{\Pi_\mu\}$ are the POVM elements corresponding to the $\mu$ measurement outcomes.

In this quantum reservoir computing (QRC) protocol, while the input state $\rho(V)$ remains fixed throughout the evolution of the reservoir, its state is updated dynamically at each discrete time step $k$, depending on the phase configuration of the interferometer. Thus, the unitary $\mathcal{U}_k$ accounts for both the intrinsic transformation of the system and the effect of feedback conditioned on previous measurement results. This interplay leads to an effective nonlinear process, where each measurement outcome is not a merely passive readout, but actively shapes the subsequent system behavior.

\subsection{Experimental hyperparameters and photon counts}

The phase encoding scaling factor $a_{\text{in}}$, the two feedback phase weights $a_{\text{fb},D}$ and $a_{\text{fb},4}$, as well as the choice of which output probability $\mu$, and $\mu'$ drive the feedback operations, are treated all as hyperparameter of the system that can be optimized. This optimization is performed for each specific optical input state and is carried out using the \texttt{Optuna} package, which implements a Bayesian optimization strategy to efficiently explore the hyperparameter space and identify configurations that maximize performance on each task.
As a consequence, the resulting values of such hyperparameter optimization will depend both on the task and on the input photon state.
Specifically, distinct configurations were optimized separately for experiments using two indistinguishable photons ($n_{\text{ph},I} = 2$), two distinguishable photons ($n_{\text{ph},D} = 2$), and single-photon inputs ($n_{\text{ph}} = 1$), allowing the system to optimally adapt to different requirements depending on the selected configuration. The full set of optimized hyperparameter values, employed for acquiring the datasets in this work, is reported in Table \ref{tab:exp_hyp_2photons} for the two-photon inputs and in Table \ref{tab:exp_hyp_1photon} for single-photon states. The tables also include other experimental parameters such as the average coincidence counts used to reconstruct the output probability distributions, the dataset sizes $N_\text{samples}$ and the training/test splitting Tr/Ts. These parameters collectively characterize the operating conditions under which the performance of the quantum reservoir was assessed.

\begin{table}[H]
    \centering
    {\small
    \renewcommand{\arraystretch}{1.2}
    \setlength{\tabcolsep}{4pt}
    \begin{tabular}{@{}lcccccc@{\hskip 10pt}cccccc@{\hskip 10pt}cc@{}}
        \toprule
        & 
        \multicolumn{6}{c}{\textbf{Two indistinguishable photons ($n_{\text{ph,I}}=2$)}} &
        \multicolumn{6}{c}{\textbf{Two distinguishable photons ($n_{\text{ph,D}}=2$)}} &
        \multicolumn{2}{c}{\textbf{Dataset}} \\
        \cmidrule(lr){2-7} \cmidrule(lr){8-13} \cmidrule(lr){14-15}
        \textbf{Task}
        & $a_{in}$ & $a_{fb,D}$ & $a_{fb,4}$ & $\mu'$ & $\mu''$ & Counts [$10^4$]
        & $a_{in}$ & $a_{fb,D}$ & $a_{fb,4}$ & $\mu'$ & $\mu''$ & Counts [$10^4$]
        & $N_\text{samples}$ & Tr/Ts \\
        \midrule

        Linear memory
        & 0.17  & 5.94  & 2.33  & 5 & 8 & 1.61 (0.10)
        & 0.37  & 3.47  & 3.71  & 3 & 3 & 1.93 (0.08)
        & 497 & 80/20 \\

        Expressivity
        & -1.69 & 3.14  & 3.14  & 7 & 3 & 2.17 (0.14)
        & -1.18 & 2.81  & -2.42 & 5 & 6 & 2.36 (0.06)
        & 150 & 80/20 \\

        Temporal XOR
        & -1.82 & 0.77  & 3.14  & 3 & 3 & 1.23 (0.03)
        & -1.04 & 1.83  & 2.02  & 7 & 6 & 1.12 (0.13)
        & 300 & 80/20 \\

        NARMA
        & 1.32  & 1.16  & -0.42 & 3 & 8 & 1.32 (0.14)
        & 1.49  & 2.91  & -1.20 & 3 & 0 & 1.24 (0.12)
        & 500 & 80/20 \\

        Mackey-Glass
        & -2.65 & 1.98  & 2.59  & 3 & 6 & 1.13 (0.06)
        & -2.92 & 2.59  & 2.44  & 3 & 6 & 1.18 (0.08)
        & 390 & 50/50 \\

        \bottomrule
    \end{tabular}
    }
    \caption{Experimental hyperparameter settings for each task with two-photon input configurations: indistinguishable and distinguishable. Reported values include the input scaling weight $a_{in}$, the feedback weights $a_{fb,D}$ and $a_{fb,4}$, and the output probabilities $\mu$ and $\mu'$. The final two columns provide dataset information: the dataset size $N_{\text{samples}}$ and the training/test split Tr/Ts.}
    \label{tab:exp_hyp_2photons}
\end{table}

\begin{table}[H]
    \centering
    {\small
    \renewcommand{\arraystretch}{1.2}
    \setlength{\tabcolsep}{4pt}
    \begin{tabular}{@{}lcccccc@{\hskip 10pt}cc@{}}
        \toprule
        & 
        \multicolumn{6}{c}{\textbf{Single photon ($n_{\text{ph,I}}=2$)}} &
        \multicolumn{2}{c}{\textbf{Dataset}} \\
        \cmidrule(lr){2-7} \cmidrule(lr){8-9}
        \textbf{Task}
        & $a_{in}$ & $a_{fb,D}$ & $a_{fb,4}$ & $\mu'$ & $\mu''$ & Counts [$10^4$]
        & $N_\text{samples}$ & Tr/Ts \\
        \midrule

        Linear memory
        & 0.25  & 1.16  & 2.66  & 3  & 3  & 4.97 & 497 & 80/20 \\

        Expressivity
        & -2.87 & -3.03 & -0.23 & 2  & 0  & 4.12 & 150 & 80/20 \\

        Temporal XOR
        & -1.04 &  -1.15 & 3.14 & 2  & 2  & 3.00 & 300 & 80/20 \\

        NARMA
        & 0.96 & 1.67 & 2.35 & 2  & 3  & 5.00 & 500 & 80/20 \\

        \bottomrule
    \end{tabular}
    }
    \caption{
        Experimental hyperparameter settings for each task with a single-photon input configuration. Reported values include the input scaling weight $a_{in}$, the feedback weights $a_{fb,D}$ and $a_{fb,4}$, and the output probabilities $\mu$ and $\mu'$. The final two columns provide dataset information: the dataset size $N_{\text{samples}}$ and the training/test split Tr/Ts.
    }
    \label{tab:exp_hyp_1photon}
\end{table}

\begin{figure}[H]
  \centering
  \begin{subfigure}[t]{0.48\textwidth}
    \centering
    \includegraphics[width=\linewidth]{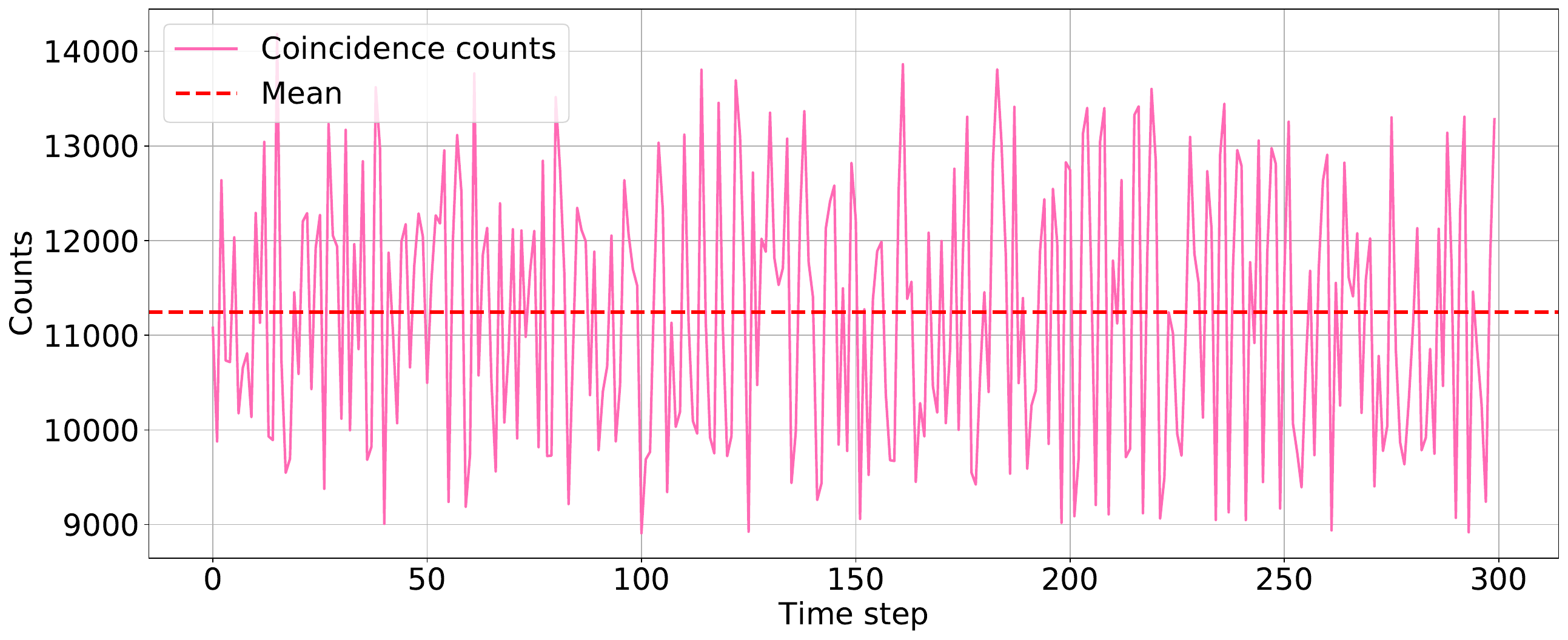}
    \caption{Distinguishable photons}
    \label{fig:distinguishable}
  \end{subfigure}%
  \hfill
  \begin{subfigure}[t]{0.48\textwidth}
    \centering
    \includegraphics[width=\linewidth]{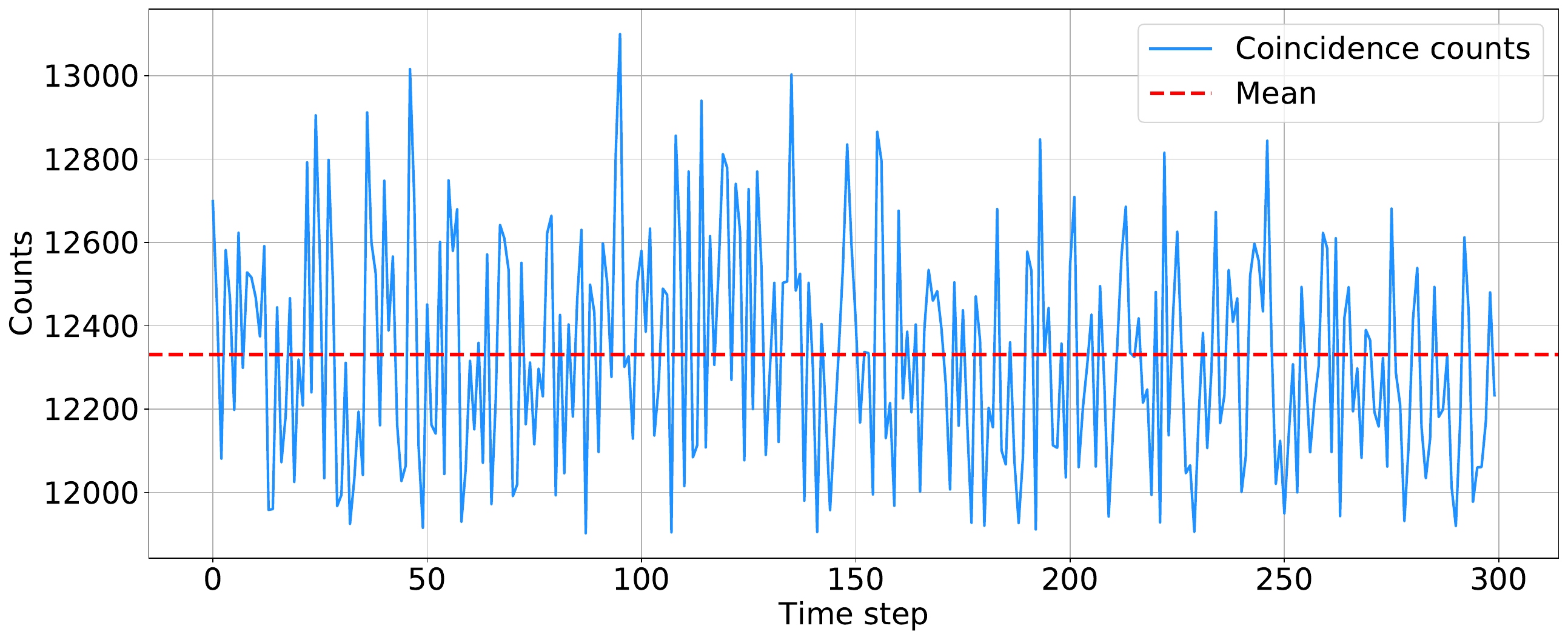}
    \caption{Indistinguishable photons}
    \label{fig:indistinguishable}
  \end{subfigure}
  \caption{Coincidence counts for the overall sequence used for evaluating the reservoir performance on the temporal XOR for (a) distinguishable and (b) indistinguishable two-photon inputs. The red dashed line represents the average number of coincidence counts, as detailed in the previous table.}
  \label{fig:photon_counts_memory}
\end{figure}

For each task, the output probability distribution is obtained by performing repeated measurements on the evolved quantum state. To ensure a fair comparison between different input configurations, we sample the output statistics using the same average number of coincidence events across all cases. As a representative example, in Fig. \ref{fig:photon_counts_memory}, we report the photon counting distributions obtained for the temporal XOR task.

\subsection{Feedback phase settings}

The phases used in the feedback loop are essential components of the system's dynamics, as they depend on past reservoir outputs and guide the future evolution of the quantum reservoir. As discussed in Sec. B of the Methods, the choice of feedback strategy depends on the specific computational task addressed. Specifically, for memory-based tasks, such as linear memory capacity, temporal XOR, NARMA \cite{atiya2000new}, and prediction of the Mackey-Glass time series, the feedback signal benefits from incorporating information over a longer temporal history. Therefore, in these cases we choose to encode the feedbacks depending on outputs from both the previous time steps $k-1$ and the earlier step $k-2$. Conversely, in tasks focused on nonlinear function reconstruction, such as monomials and polynomials, the feedback relies solely on the most recent output at time $k-1$.

\begin{figure}[H]
\centering

% Riga 1
\begin{subfigure}{0.48\textwidth}
  \centering
  \includegraphics[width=\linewidth]{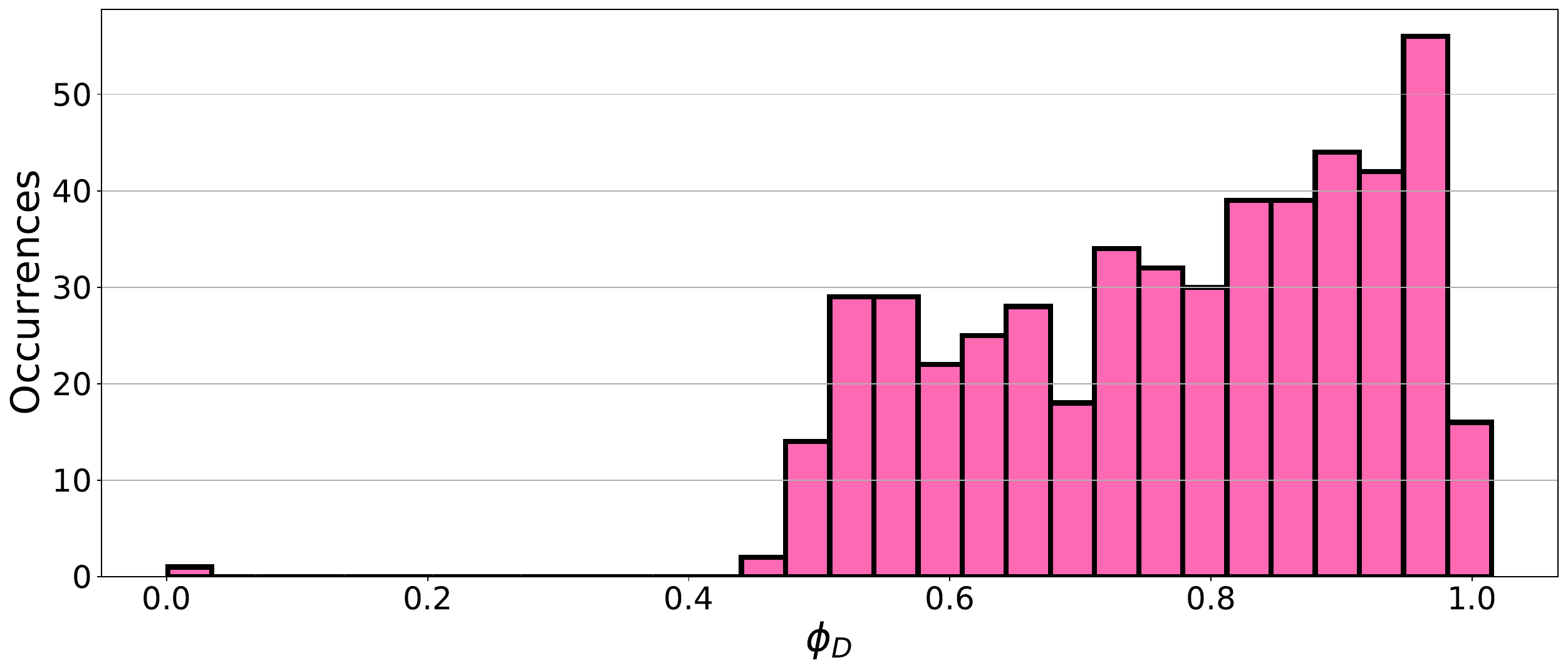}
  \caption{} \label{fig:a}
\end{subfigure}
\hfill
\begin{subfigure}{0.48\textwidth}
  \centering
  \includegraphics[width=\linewidth]{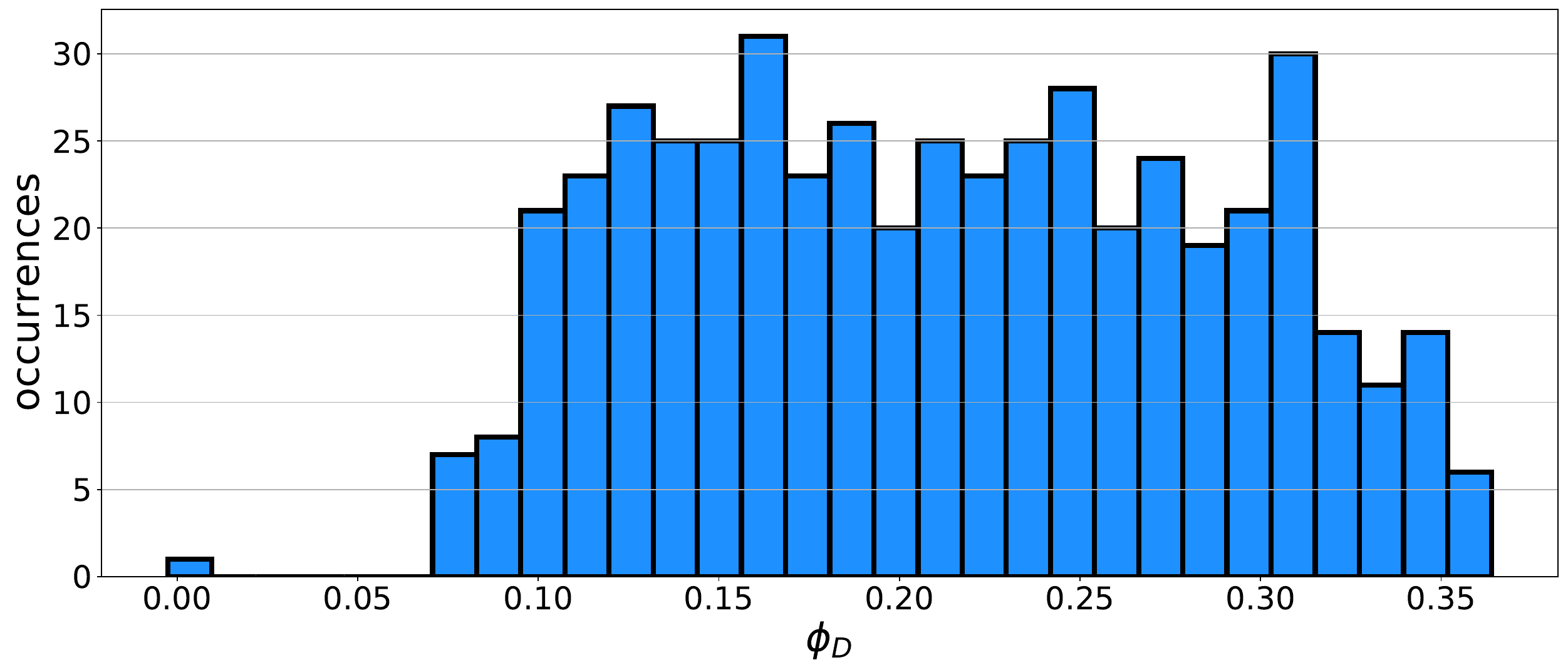}
  \caption{} \label{fig:b}
\end{subfigure}

\vspace{0.3cm}

% Riga 2
\begin{subfigure}{0.48\textwidth}
  \centering
  \includegraphics[width=\linewidth]{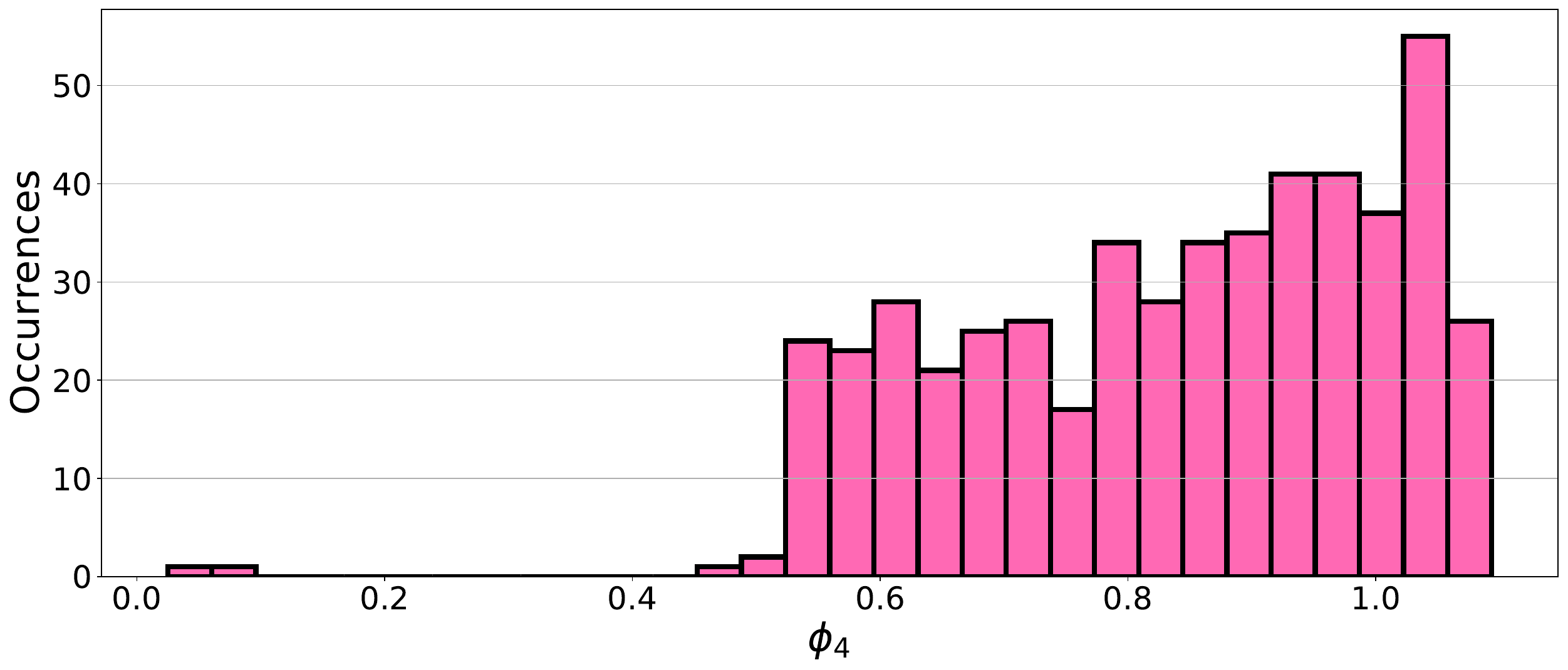}
  \caption{} \label{fig:c}
\end{subfigure}
\hfill
\begin{subfigure}{0.48\textwidth}
  \centering
  \includegraphics[width=\linewidth]{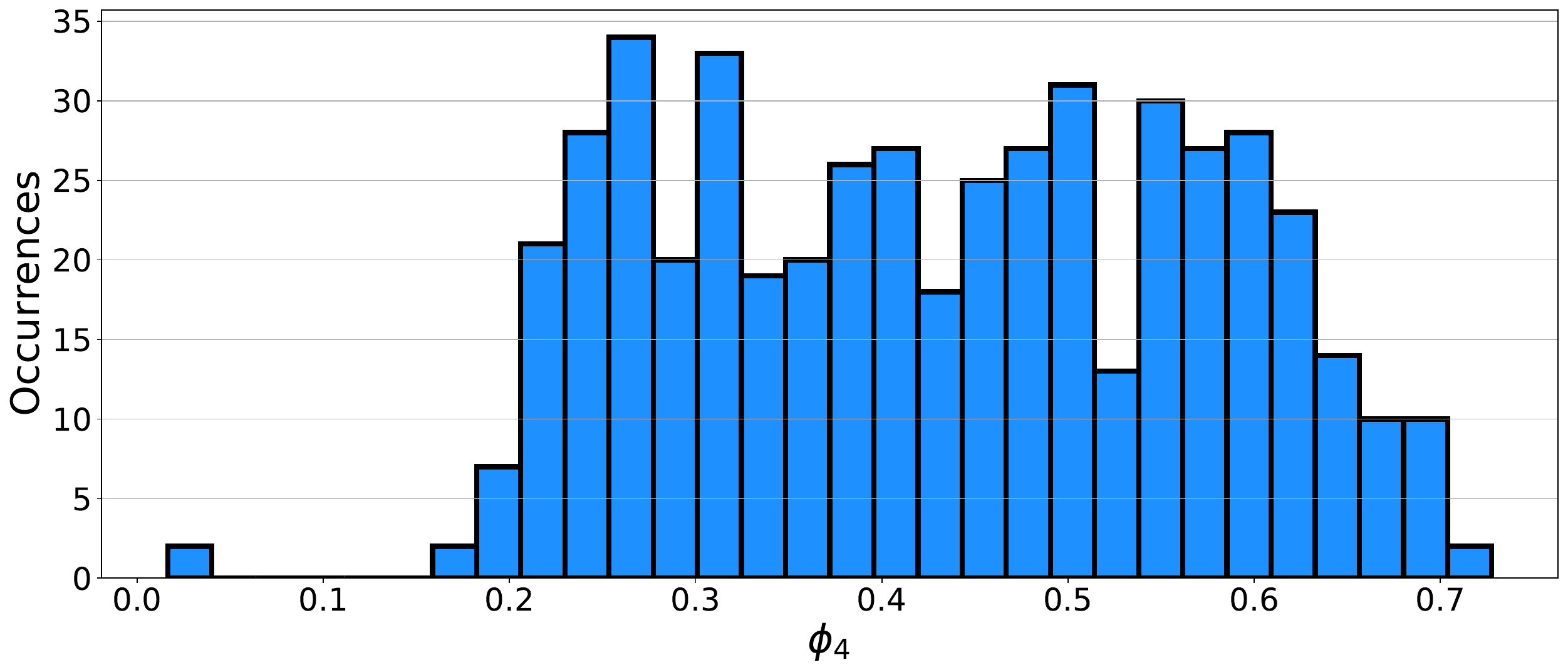}
  \caption{} \label{fig:d}
\end{subfigure}

\vspace{0.3cm}

% Riga 3
\begin{subfigure}{0.48\textwidth}
  \centering
  \includegraphics[width=\linewidth]{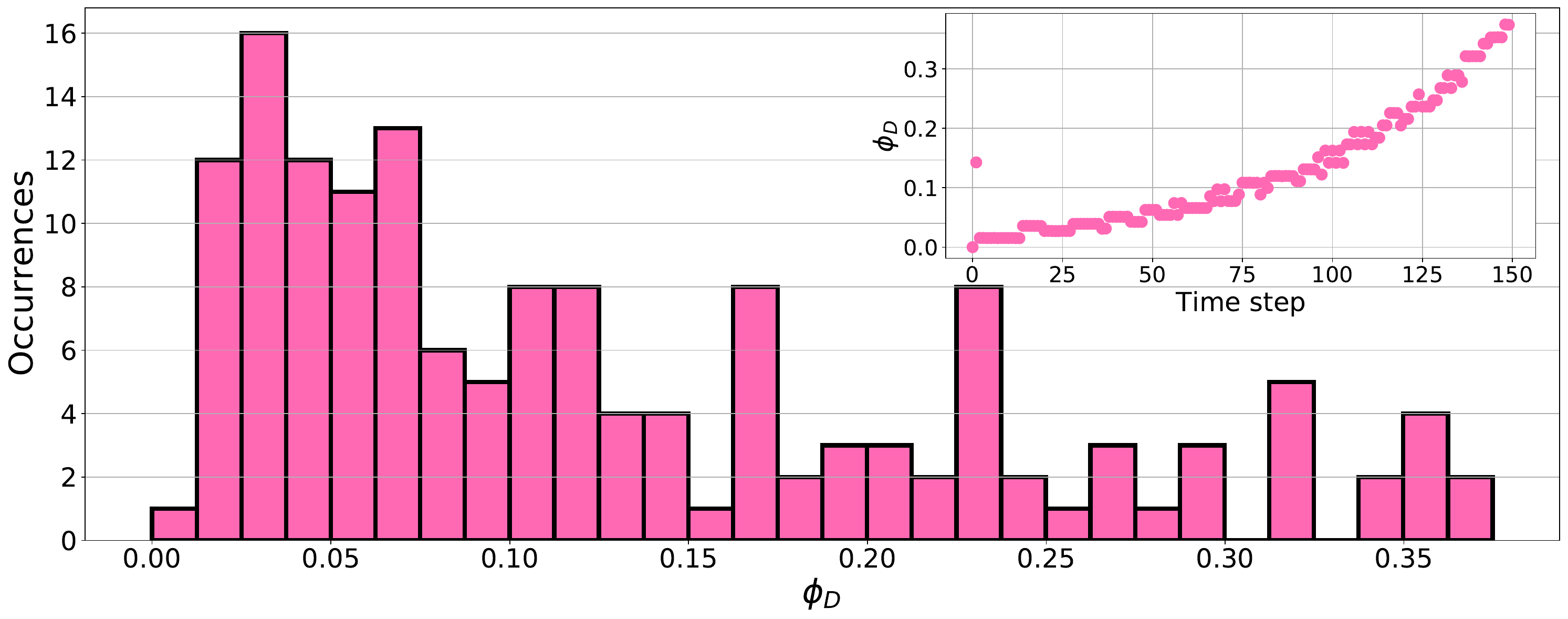}
  \caption{} \label{fig:e}
\end{subfigure}
\hfill
\begin{subfigure}{0.48\textwidth}
  \centering
  \includegraphics[width=\linewidth]{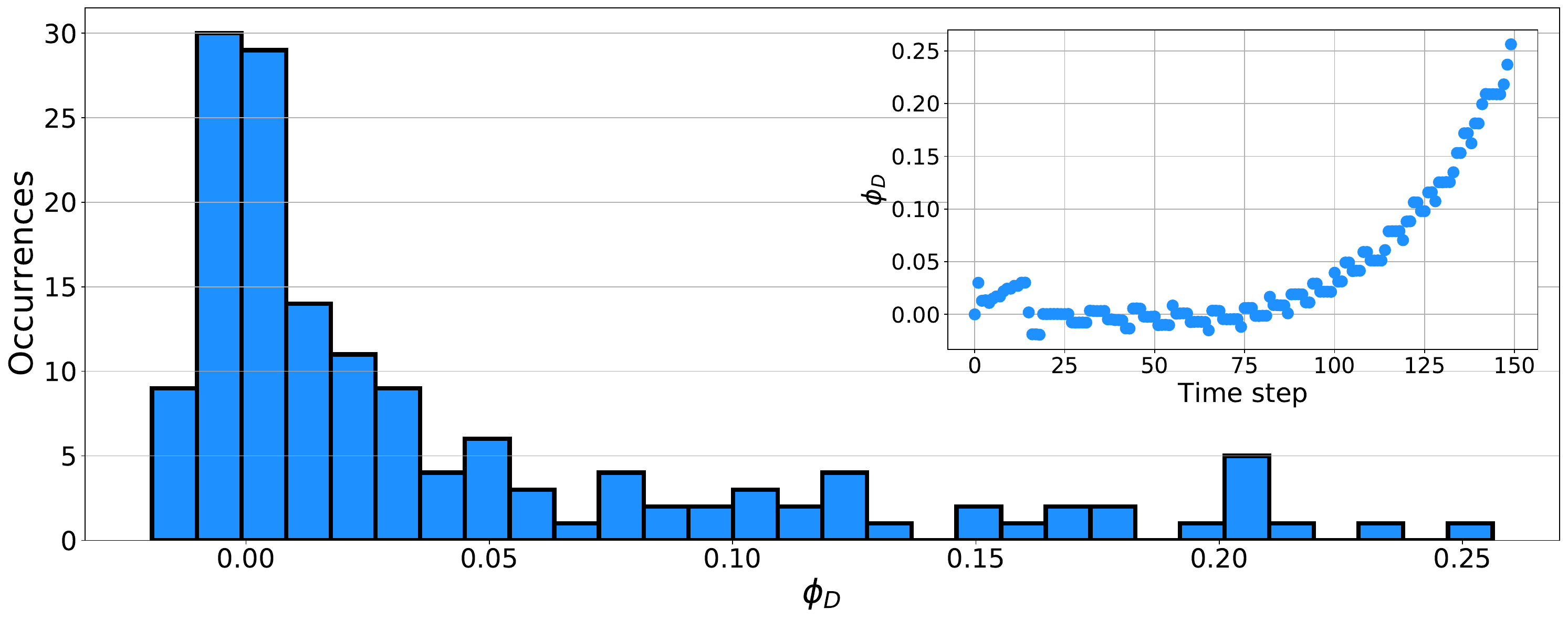}
  \caption{} \label{fig:f}
\end{subfigure}

\vspace{0.3cm}

% Riga 4
\begin{subfigure}{0.48\textwidth}
  \centering
  \includegraphics[width=\linewidth]{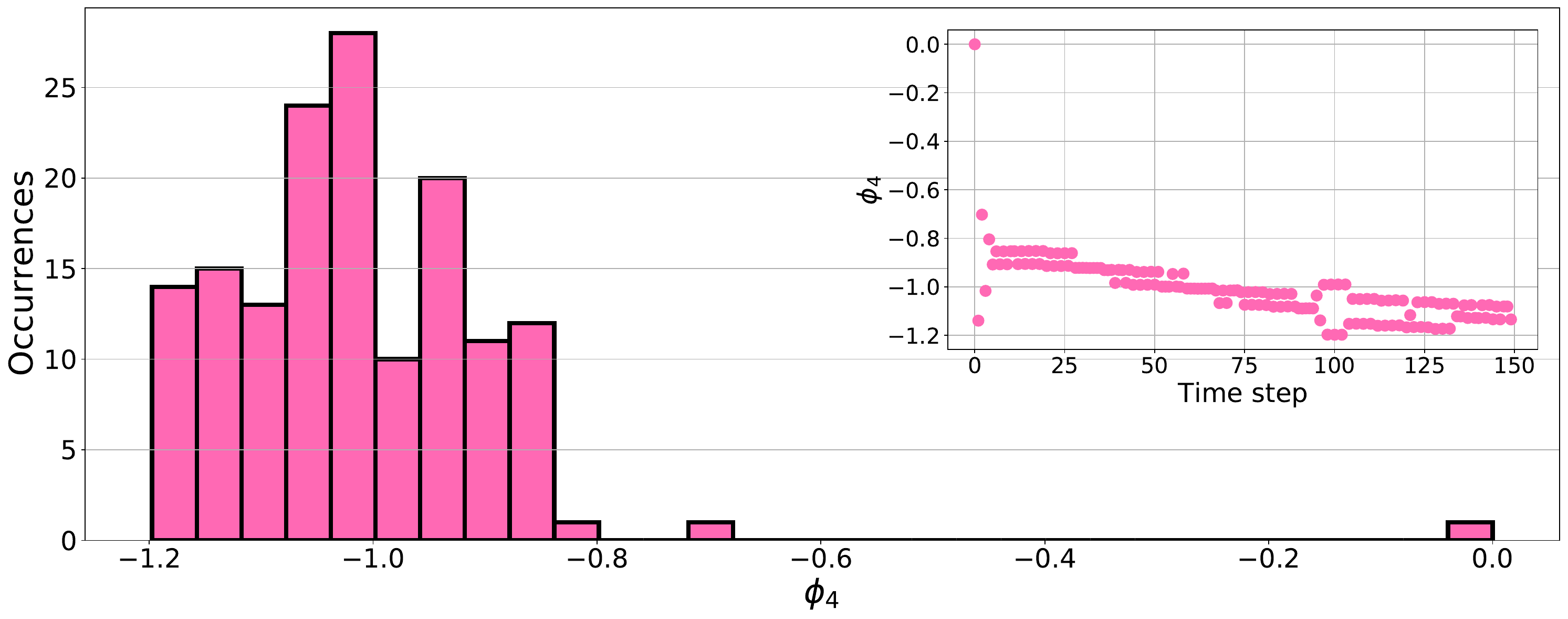}
  \caption{} \label{fig:g}
\end{subfigure}
\hfill
\begin{subfigure}{0.48\textwidth}
  \centering
  \includegraphics[width=\linewidth]{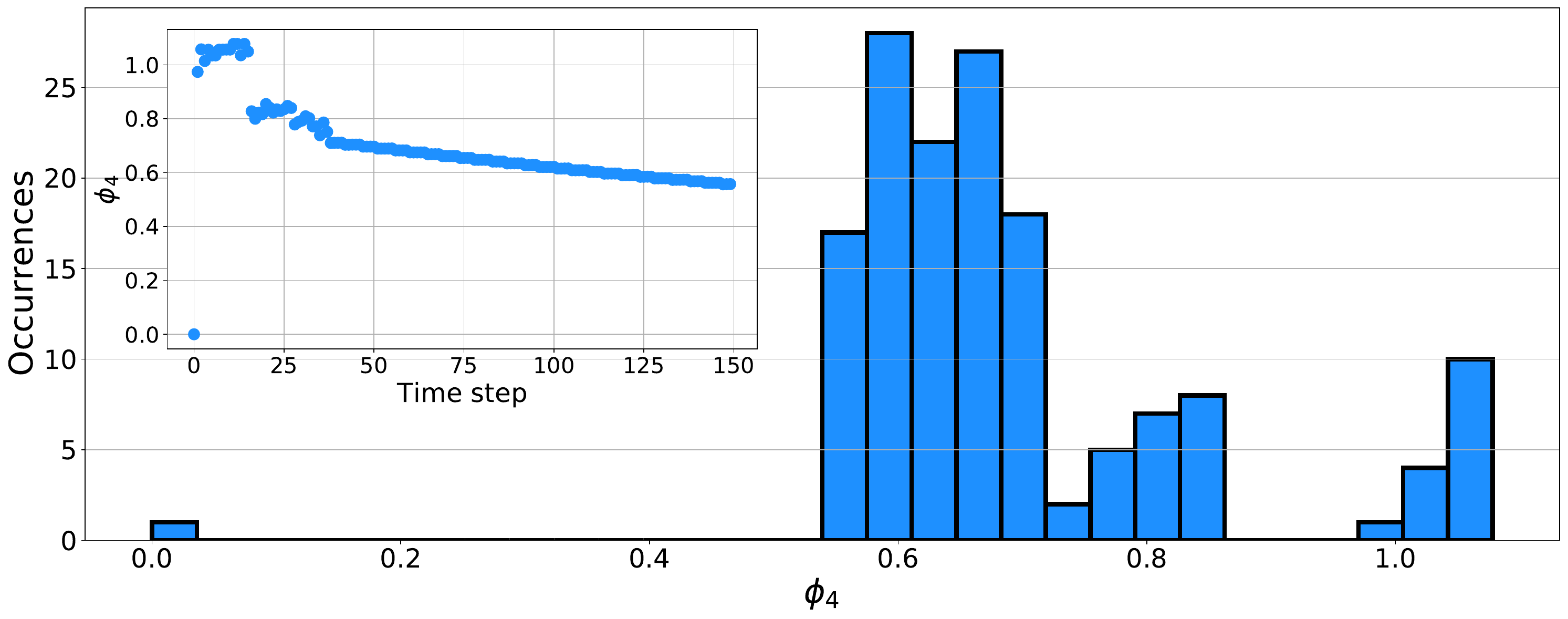}
  \caption{} \label{fig:h}
\end{subfigure}

\caption{Histogram of the feedback phase values $\phi_D$ and $\phi_4$ for task-independent characterization for two distinguishable photons input state (pink) and two indistinguishable photons input state (blue). (a - d): Linear memory task. (e - h): Nonlinear function reconstruction. The insets in panels (e - h) illustrate the temporal evolution of phase values set at each step of the input sequence.}
\label{fig:8panels}
\end{figure}

Figure \ref{fig:8panels} shows the statistical distributions of the feedback phase values that have been set experimentally across the linear memory and expressivity tasks. Histograms of phase occurrences are shown in pink for the two distinguishable photon configuration ($n_{\text{ph},D}=2$) and in blue for the two indistinguishable photon configuration ($n_{\text{ph},I}=2$). Panels \ref{fig:8panels}a-d correspond to the linear memory task, while panels \ref{fig:8panels}e-h refer to the expressivity task. 

In memory-related tasks, the phase distributions are broad and relatively uniform across the allowed range, with no strong concentration in specific regions of phase space. This reflects the input sequence, uniformly and randomly distributed in $[0,1]$, and suggests a need for diversity and a broad coverage of the state-space to effectively encode temporal dependencies.

By contrast, tasks focused on expressivity show more structured and localized phase distributions. The histograms are no longer uniformly distributed, and the insets reveal smooth and monotonic phase trajectories. These features reflect the regular and ordered nature of the input sequences. Regarding the phases reported in the insets, the set values arise from the discrete resolution of the thermo-optical phase shifters, which limits the available phase settings to a finite set of values. 

Note that these feedback phase dynamics are not externally imposed, but emerge from the autonomous evolution of the quantum reservoir in the feedback-driven regime. The internal reservoir state directly determines the feedback trajectory, resulting in phase dynamics that adapt naturally to the task. The contrast between the statistical and temporal behaviors observed across different computational tasks highlights the flexibility and task-dependent character of QRC.

\section{Additional results}

%\subsection{Effect of finite counting statistics}
\subsection{Effect of experimental imperfections}
\label{sec:poissonian_effect}

Here, we analyze the impact of finite counting statistics on the learning performance of the quantum reservoir. In our protocol, the output probabilities, which are used both to set the adaptive feedback phases and then for training the linear regression model, are reconstructed from a finite number of photon detection events. This constraint introduces statistical fluctuations that propagate through the reservoir's processing steps of the system, ultimately affecting the stability of its outputs. To systematically quantify these effects, we investigate the influence of limited sampling statistics through numerical simulations.

\begin{figure}[H]
\centering
\begin{subfigure}{0.5\textwidth}
  \centering
  \includegraphics[width=\linewidth]{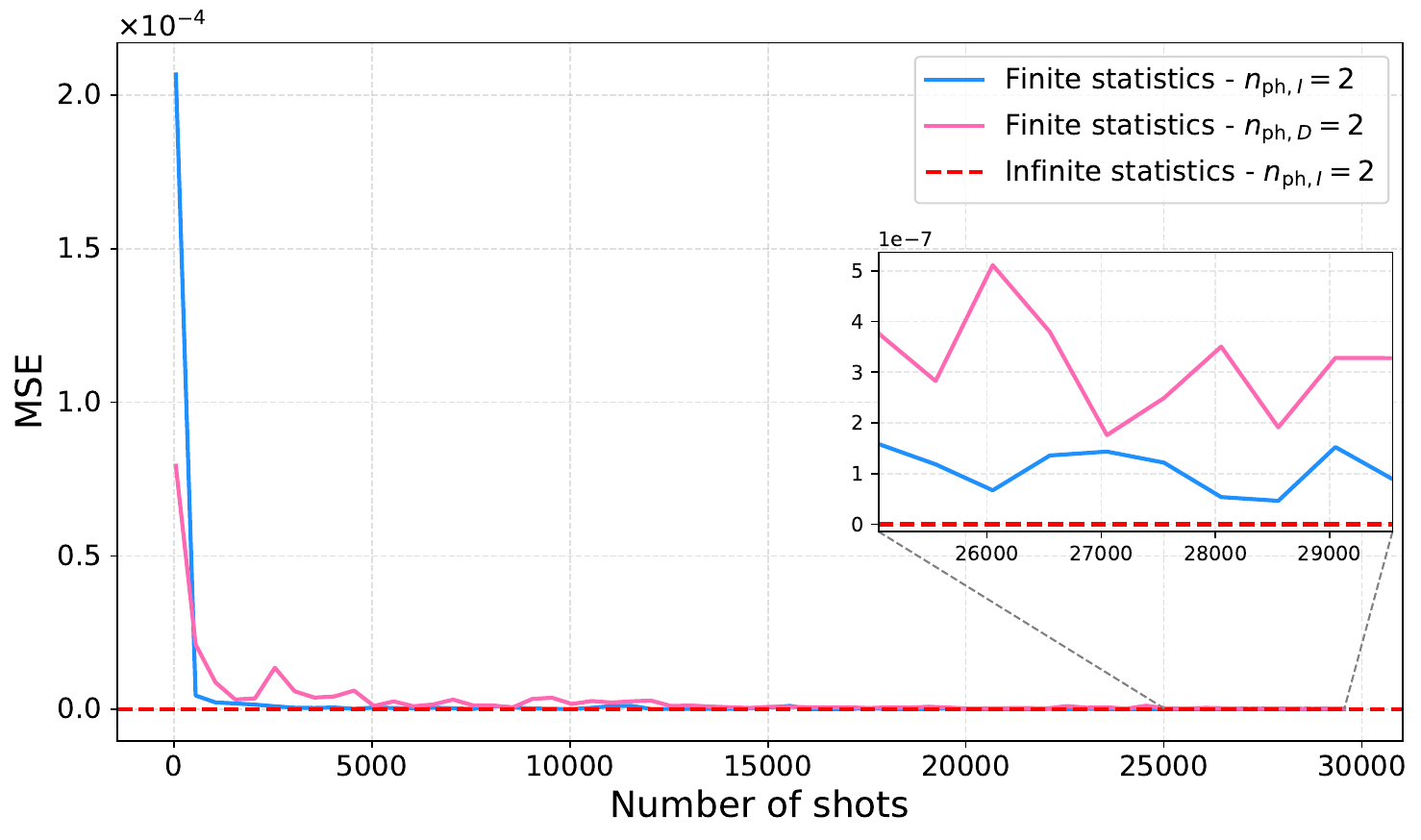}
  \caption{Continuous phase controls.} \label{fig:c}
\end{subfigure}
\hfill
\begin{subfigure}{0.48\textwidth}
  \centering
  \includegraphics[width=\linewidth]{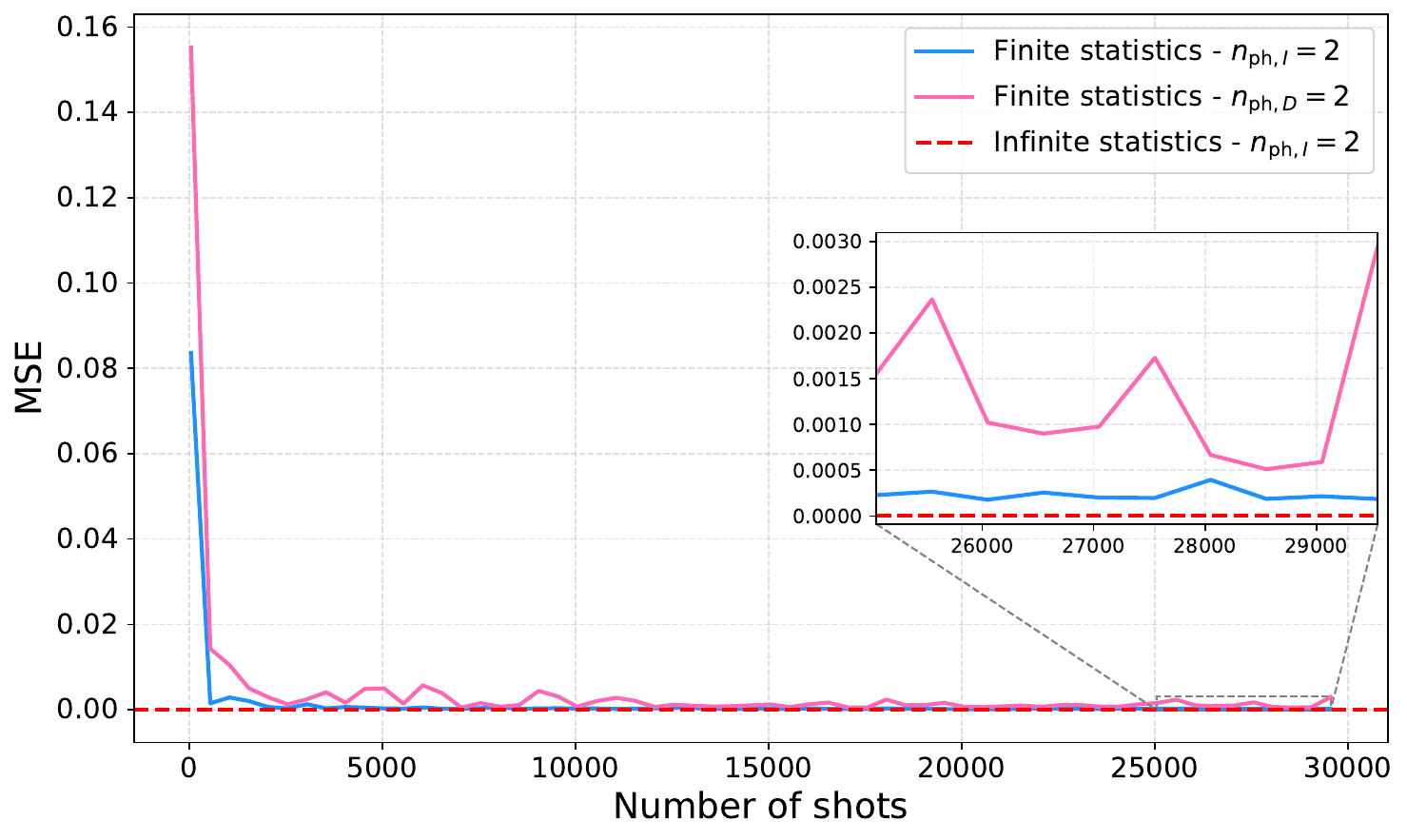}
  \caption{Discretized phase controls.}
\end{subfigure}
\caption{MSE on the test set in reconstructing $f_3(x)=x^3$ using the QRC simulated system. The results are shown for simulated output probabilities with $n_{ph,I} = 2$ (light blue) and $n_{ph,D} = 2$ (pink) while varying the sampling statistics. The red dashed line is the MSE achieved when simulating the system with the ideal (noisless) probabilities. Panel (a) shows the results obtained without any limitation on the phase settings, whereas panel (b) shows performance when discretized phase values are used for both the input sequence encoding and the feedback updates. Each dataset contains 150 points.}\label{fig:SI_counts}
\end{figure}

We study the system performance on a fixed test set when reconstructing a nonlinear target function, i.e. $f_n(x) = x^n$, as a function of the number of coincidence events used to reconstruct the output probabilities at each step of the protocol. The results here refer to the choice of $n = 3$. In the low statistics regime, the reconstruction is highly noisy and deviates from the target due to the large fluctuations in the estimated output distribution. The presence of poissonian noise, associated with the finite number of photon detection events, indeed plays a critical role in the learning process of the reservoir. As the number of shots increases, these fluctuations are progressively suppressed, and the model begins to accurately approximate the nonlinear target function, ultimately approaching the ideal scenario of infinite sampling statistics. This is shown in Fig. \ref{fig:SI_counts}a, where the performance of the model in terms of mean squared error (MSE) is reported as a function of the number of coincidences events used to reconstruct the output probabilities, both for the two distinguishable photons inputs and the indistinguishable state.

In addition to the finite-count effects, another significant source of experimental imperfection arises from the limited precision when setting the optical phase values. Specifically, phase controls in the photonic circuit are discretized, with only a finite set of phase settings available rather than a continuous range. This constraint further influences the reservoir’s performance, as it introduces additional errors due to such a discretization into the encoding and feedback mechanisms. 
The combined impact of finite counting statistics and discretized phase control is illustrated in \ref{fig:SI_counts}b, where we reproduced the simulations of the nonlinear function reconstruction as a function of the number of collected events per point while restricting the phase settings to those actually accessible in the experimental apparatus. The analysis of the convergence behavior allowed us to determine the level of measurement statistics required to carry out the experiment effectively. Beyond a certain threshold, the MSE saturates, and further increasing the number of detection events yields no appreciable improvement in reconstruction accuracy.

%Similarly, for distinguishable photons ($n_{\text{ph},D} = 2$), the model exhibits slower convergence but reaches approximately the same level of accuracy as the indistinguishable case.

\begin{figure}[H]
    \centering

    \subfloat[$N_{\text{shot}}=3.5\times 10^3$ and discretized phase controls.]{\label{a}\includegraphics[width=0.45\linewidth]{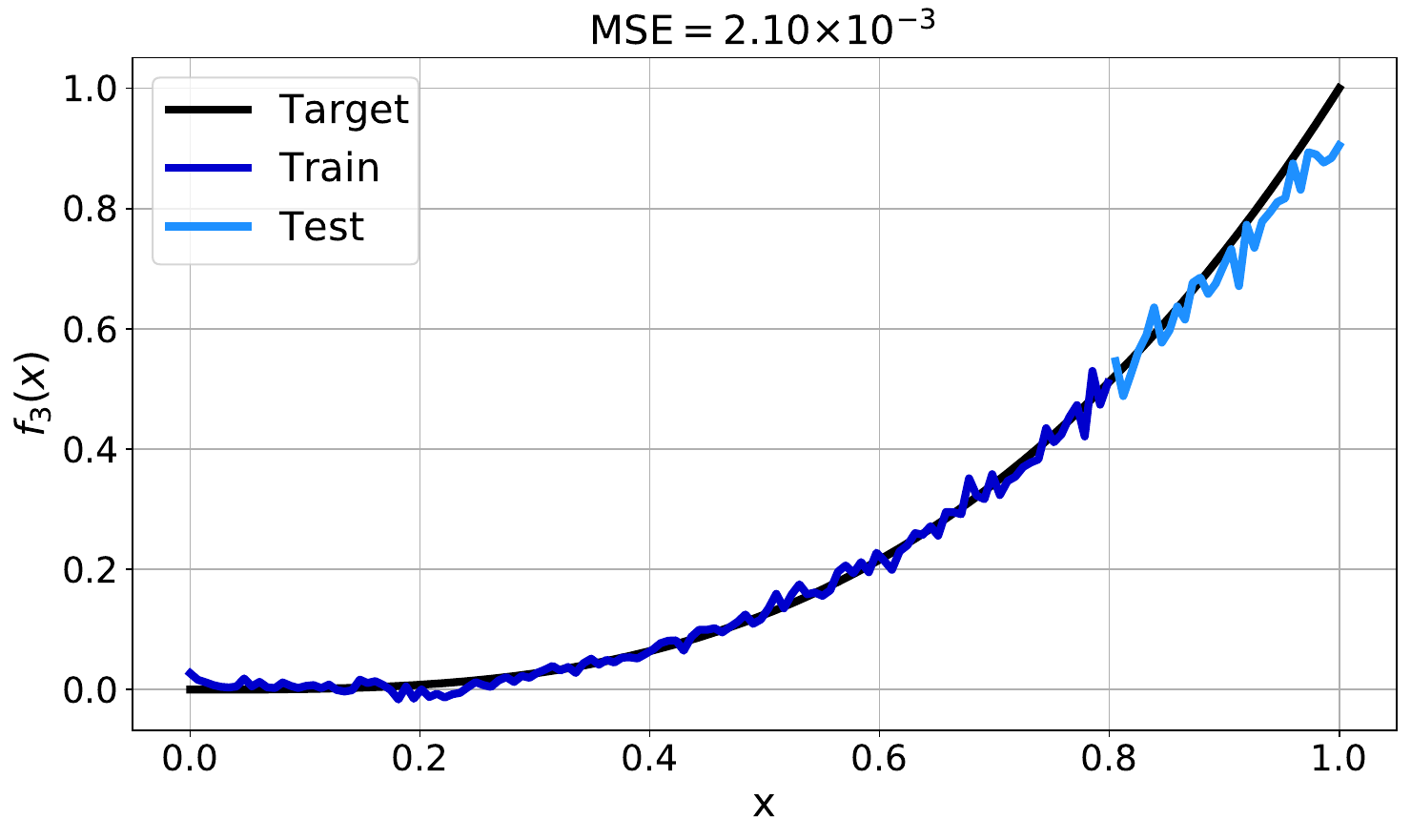}}\hfill
    \subfloat[Infinite $N_{\text{shot}}$ and discretized phase controls.]{\label{a}\includegraphics[width=0.45\linewidth]{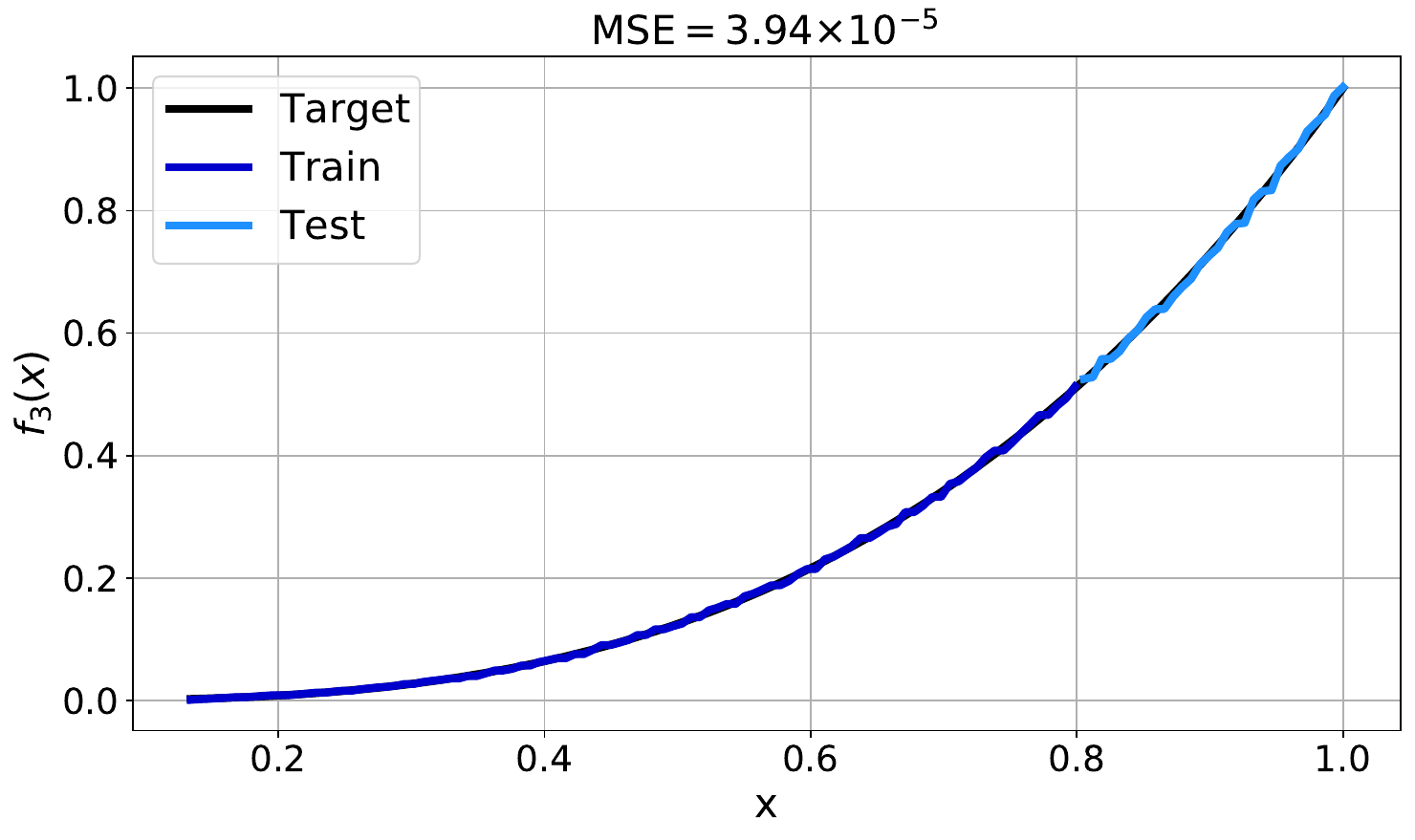}}\par\vspace{0.5em}
    
    \subfloat[$N_{\text{shot}}=3.5\times 10^3$ and continuous phase controls.]{\label{c}\includegraphics[width=0.45\linewidth]{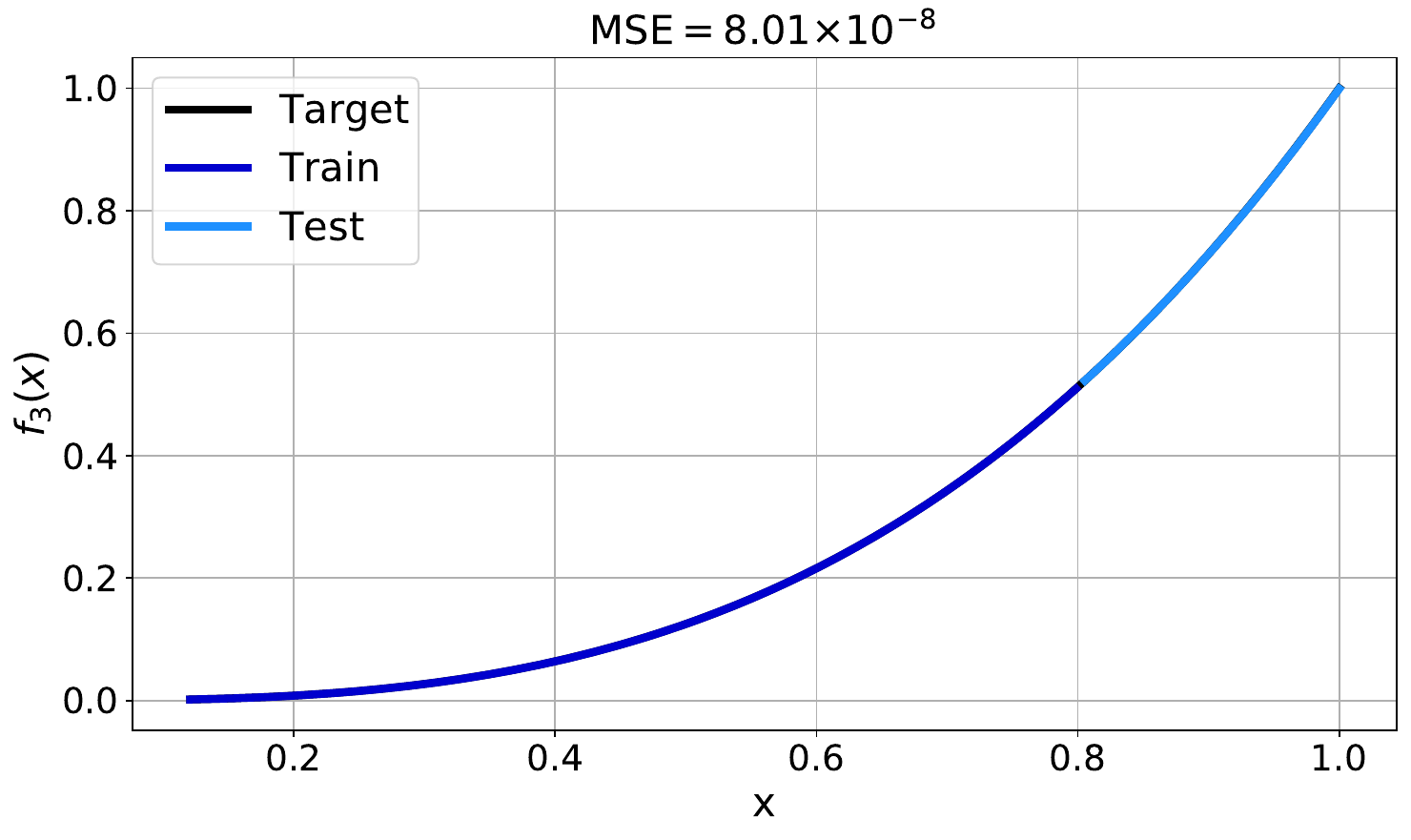}}\hfill
    \subfloat[Infinite $N_{\text{shot}}$ and continuous phase controls.]{\label{d}\includegraphics[width=0.45\linewidth]{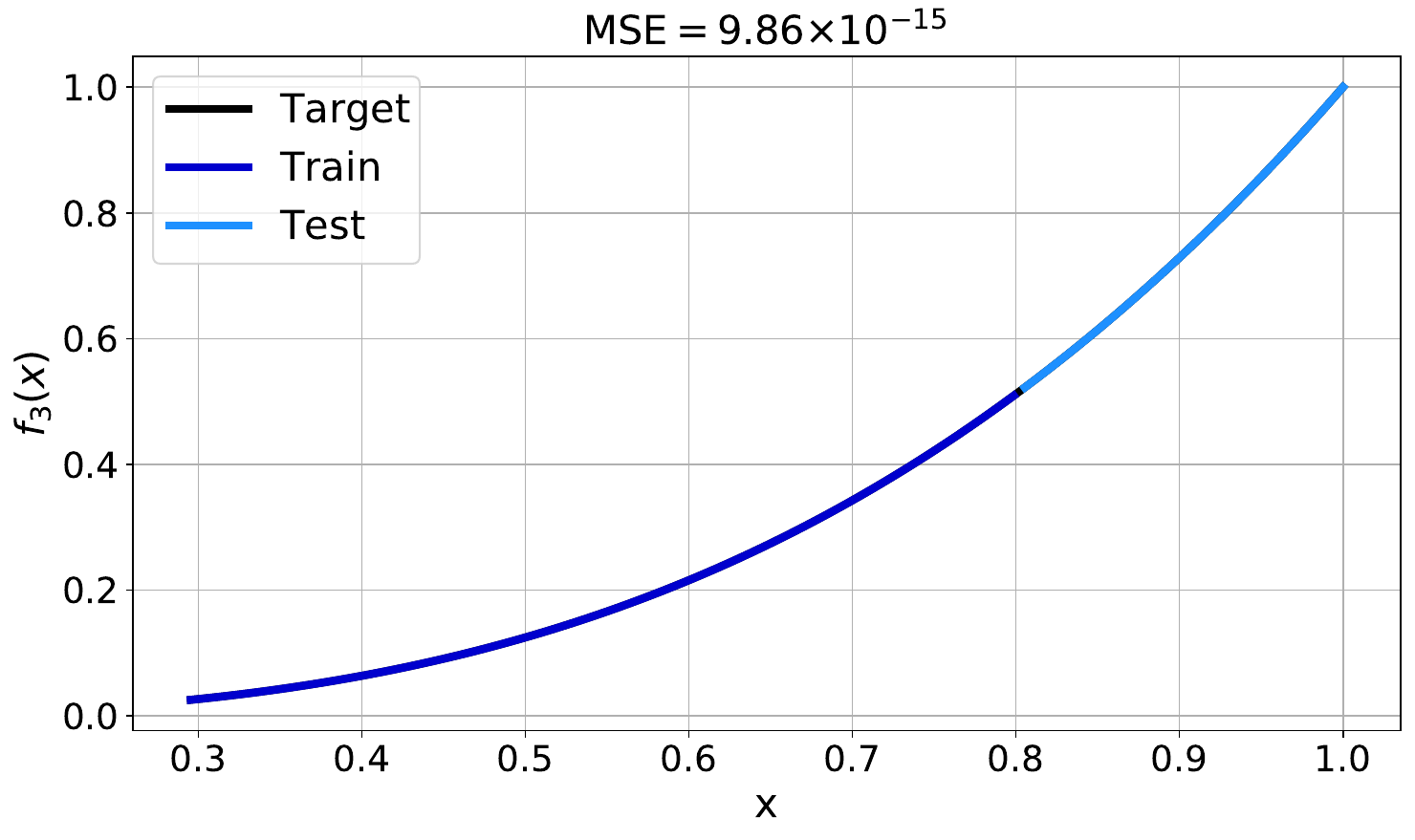}}
    
    \caption{Simulation of the reconstruction performance of the quantum reservoir computing model with two indistinguishable input photons for the target nonlinear function $f_3(x) = x^3$ under different experimental conditions. Panels (a) and (b) show the effect of increasing the number of shots from $N_{\text{shot}} = 3.5 \times 10^3$ to infinity using discretized phase controls. Panels (c) and (d), study the same effect with infinite shots. In the title of each panel is reported the MSE evaluated on the test set. Each dataset contains 150 points.}
    \label{fig:reconstruction_x357}
\end{figure}

The effect of experimental imperfections are further investigated in the case of $n_{ph,I} = 2$ by showing the reconstruction performance of the simulated QRC for the selected nonlinear target function, i.e. $f_3(x) = x^3$, for different statistics used to sample the output distribution and for discretized phase controls, as shown in Fig. \ref{fig:reconstruction_x357}. From the comparison between panels (a) and (b), where discretized phase controls are used but the number of shots increases from $N_{\text{shot}} = 3.5 \times 10^3$ to the ideal infinite case, 
%where the explicit probabilities are used to simulate the experiment, 
we observe a clear improvement of the function reconstruction capabilities. This suggests that increasing the number of measurements can partially correct errors, although it does not address all sources of inaccuracy. Indeed, comparing panels (b) and (d), both of which assume infinite sampling statistics, the difference in reconstruction accuracy highlights the dominant role of phase control resolution. The use of continuous phase controls in panel (d) allows for a perfect reconstruction of the target function. %, as compared to the discretized case in panel (b). 
Similarly, comparing panels (a) and (c), both with finite  $N_{\text{shot}}$, further confirms that the primary limitation comes from the discretization of the phase controls. The reconstruction in panel (c), with continuous controls, is notably smoother and more accurate than in panel (a), despite both being subject to statistical fluctuations due to finite sampling. Overall, since the negligible differences between (c) and (d), the figure demonstrates that while increasing the number of shots helps reduce statistical noise, the discretization of control parameters constitutes one of the main bottleneck to the model’s reconstruction performance.

%Starting from the low statistics of 500 shots, Fig. \ref{fig:x5_shots}a, the reconstruction is highly noisy and deviates from the target due to the large fluctuations in the estimated output distribution. As the number of shots increases to $3.5 \times 10^3$ and $2.1 \times 10^4$, respectively in Fig.s \ref{fig:x5_shots}b and \ref{fig:x5_shots}c, the fluctuations are progressively suppressed, and the model better approximates the nonlinear target function, approaching the infinite statistics scenario, shown in \ref{fig:x5_shots}d.

%\begin{figure}[htbp]
%    \centering
%    \subfloat[]{\label{a}\includegraphics[width=0.45\linewidth]{SI_Figures/SI_v1_500_mono_x3.pdf}}\hfill
%    \subfloat[]{\label{b}\includegraphics[width=0.45\linewidth]{SI_Figures/SI_v1_3k_mono_x3.pdf}}\par\vspace{0.5em}
%
%    \subfloat[]{\label{a}\includegraphics[width=0.45\linewidth]{SI_Figures/SI_v1_21k_mono_x3.pdf}}\hfill
%    \subfloat[]{\label{c}\includegraphics[width=0.45\linewidth]{SI_Figures/SI_mono_x3_perfetto_21k.pdf}}\par\vspace{0.5em}
%    
%    \subfloat[]{\label{e}\includegraphics[width=0.45\linewidth]{SI_Figures/SI_v1_inf_mono_x3.pdf}}\hfill
%    \subfloat[]{\label{f}\includegraphics[width=0.45\linewidth]{SI_Figures/SI_mono_x3_perfetto.pdf}}
%    
%    \caption{Descrizione generale della figura. \fromRosario{aggiungere 21k ideale}}
%    \label{fig:x5_shots}
%\end{figure}

%For completeness, Fig. \ref{fig:x5_shots}e shows the ideal case in which the phase shifters can be set with arbitrary precision and the output distributions are reconstructed using infinite statistics. This confirms that the stepped shape observed in the predicted curve is due to the finite phase-settings available. 

%\begin{figure}[H]
%\centering
%
%\begin{subfigure}{0.48\textwidth}
%  \centering
%  \includegraphics[width=\linewidth]{SI_Figures/SI_v1_inf_mono_x3.pdf}
%  \caption{} \label{fig:c}
%\end{subfigure}
%\hfill
%\begin{subfigure}{0.48\textwidth}
%  \centering
%  \includegraphics[width=\linewidth]{SI_Figures/SI_mono_x3_perfetto.pdf}
%  \caption{} \label{fig:d}
%\end{subfigure}
%
%\caption{}
%\label{fig:x5_shots_inf_ideal}
%\end{figure}

%Here, we observe that increasing the number of measurement shots not only improves the overall reconstruction accuracy but also reduces the amount of training data required to reach a given error threshold. For instance, with $2.1 \times 10^4$ shots, about 30 training points are sufficient to approach the asymptotic performance limit, whereas with only $5 \times 10^2$ shots, more than 60 training points are needed to reach similar accuracy. Interestingly, the experimental data appears to outperform the simulated finite-shot scenarios, particularly in the low-data regime. This behavior may be attributed to the presence of experimental noise in the quantum reservoir computer, which could act as a regularization or increase the expressivity of the model, effectively improving generalization.

\subsection{Additional experimental results}

\begin{figure}[H]
    \centering
    \includegraphics[width=1\linewidth]{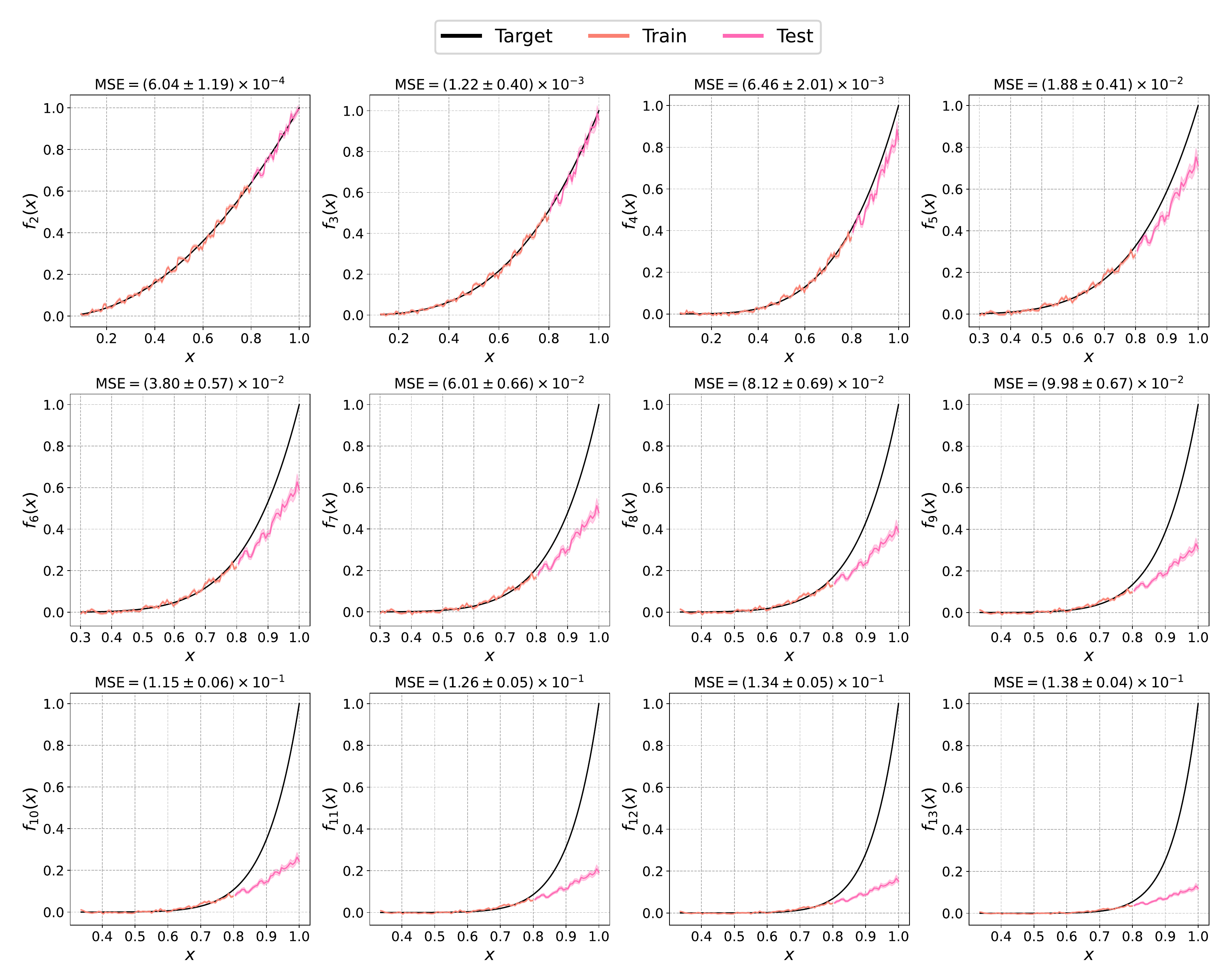}
    \caption{Reconstruction of the monomial target functions $x^n$ in the configuration with two distinguishable photons input. For each panel, the target function of order $n$ is shown as a black line. The reconstructed outputs are displayed for the training dataset (orange) and the test dataset (pink). The title of each plot reports the MSE computed on the test set. Each dataset contains 150 points. Error bars and shaded areas in the plot refer to the statistical fluctuations evaluated from 100 Monte Carlo extractions to account for the presence of Poissonian sampling noise.}
    \label{fig:monomial_v0}
\end{figure}

\begin{figure}[H]
    \centering
    \includegraphics[width=1\linewidth]{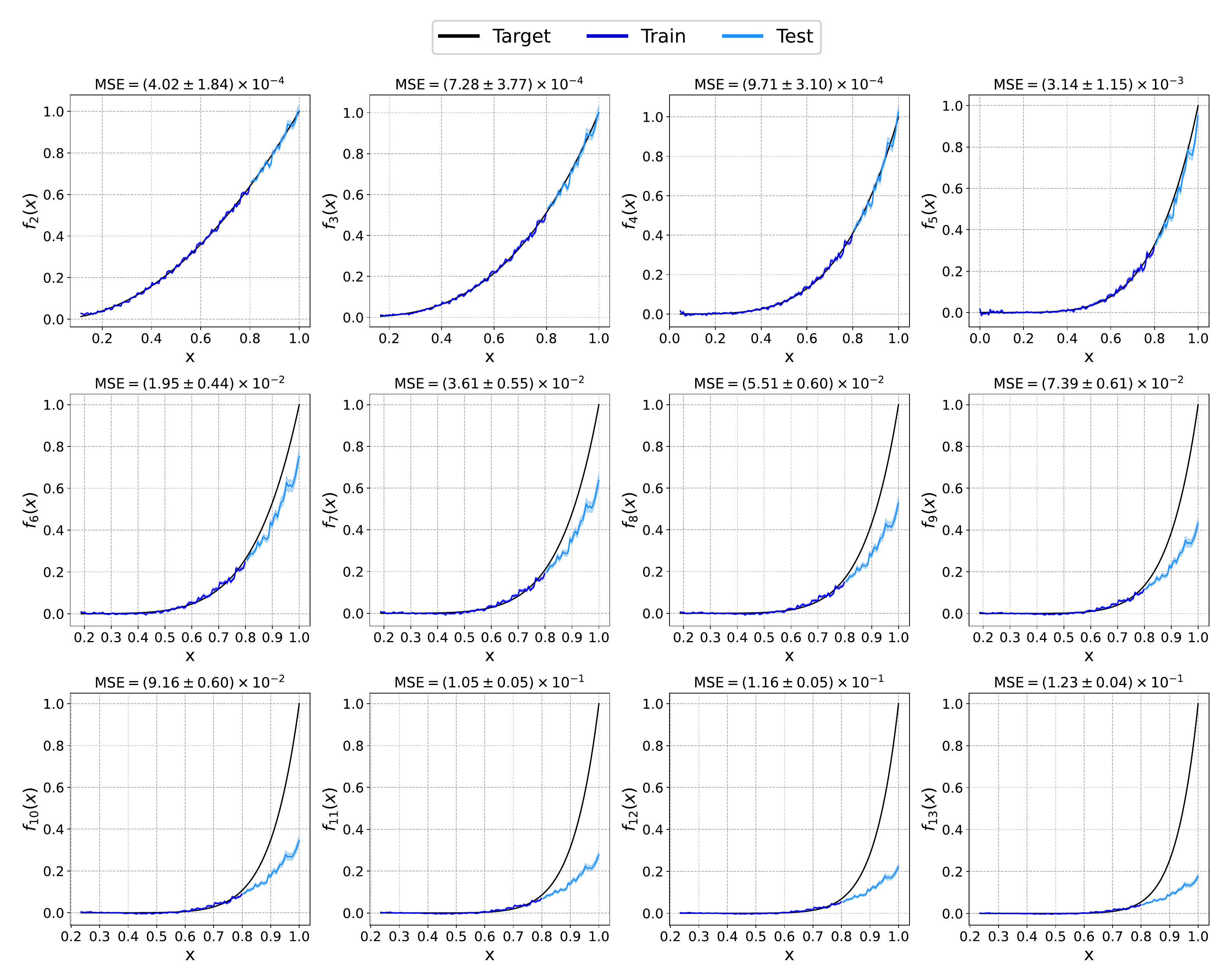}
    \caption{Reconstruction of the monomial target functions $x^n$ in the configuration with two indistinguishable photons input. For each panel, the target function of order $n$ is shown as a black line. The reconstructed outputs are displayed for the training dataset (dark blue) and the test dataset (light blue). The title of each plot reports the MSE computed on the test set. Each dataset contains 150 points. Error bars and shaded areas in the plot refer to the statistical fluctuations evaluated from 100 Monte Carlo extractions to account for the presence of Poissonian sampling noise.}
    \label{fig:monomial_v1}
\end{figure}

\begin{figure}[H]
    \centering
    \includegraphics[width=1\linewidth]{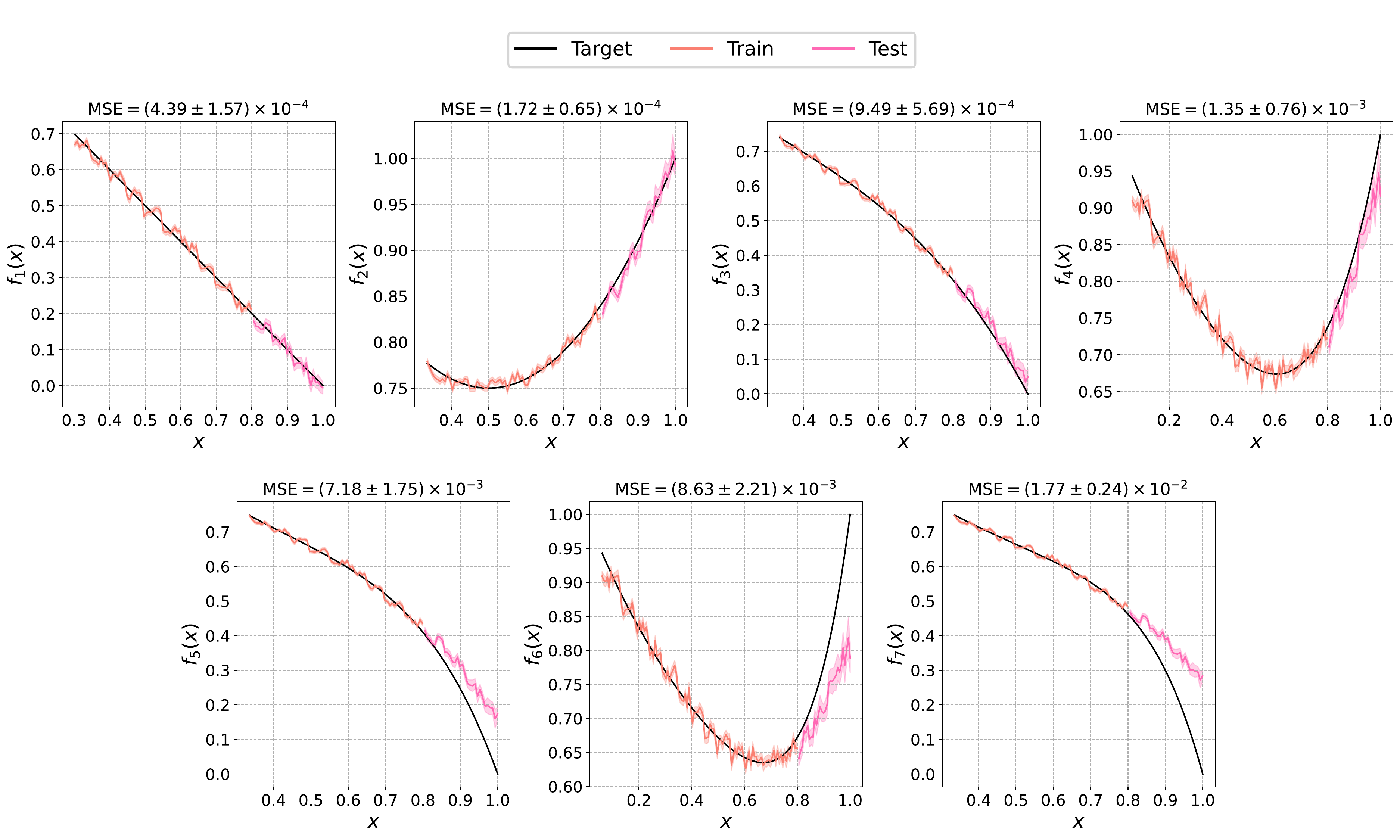}
    \caption{Reconstruction of the polynomial target functions of order $N$ in the configuration with two distinguishable photons input. For each panel, the target function is depicted as a black line. The reconstructed outputs are displayed for the training dataset (orange) and the test dataset (pink). The title of each plot reports the MSE computed on the test set. Each dataset contains 150 points. Error bars and shaded areas in the plot refer to the statistical fluctuations evaluated from 100 Monte Carlo extractions to account for the presence of Poissonian sampling noise.}
    \label{fig:polynomial_v0}
\end{figure}

\begin{figure}[H]
    \centering
    \includegraphics[width=1\linewidth]{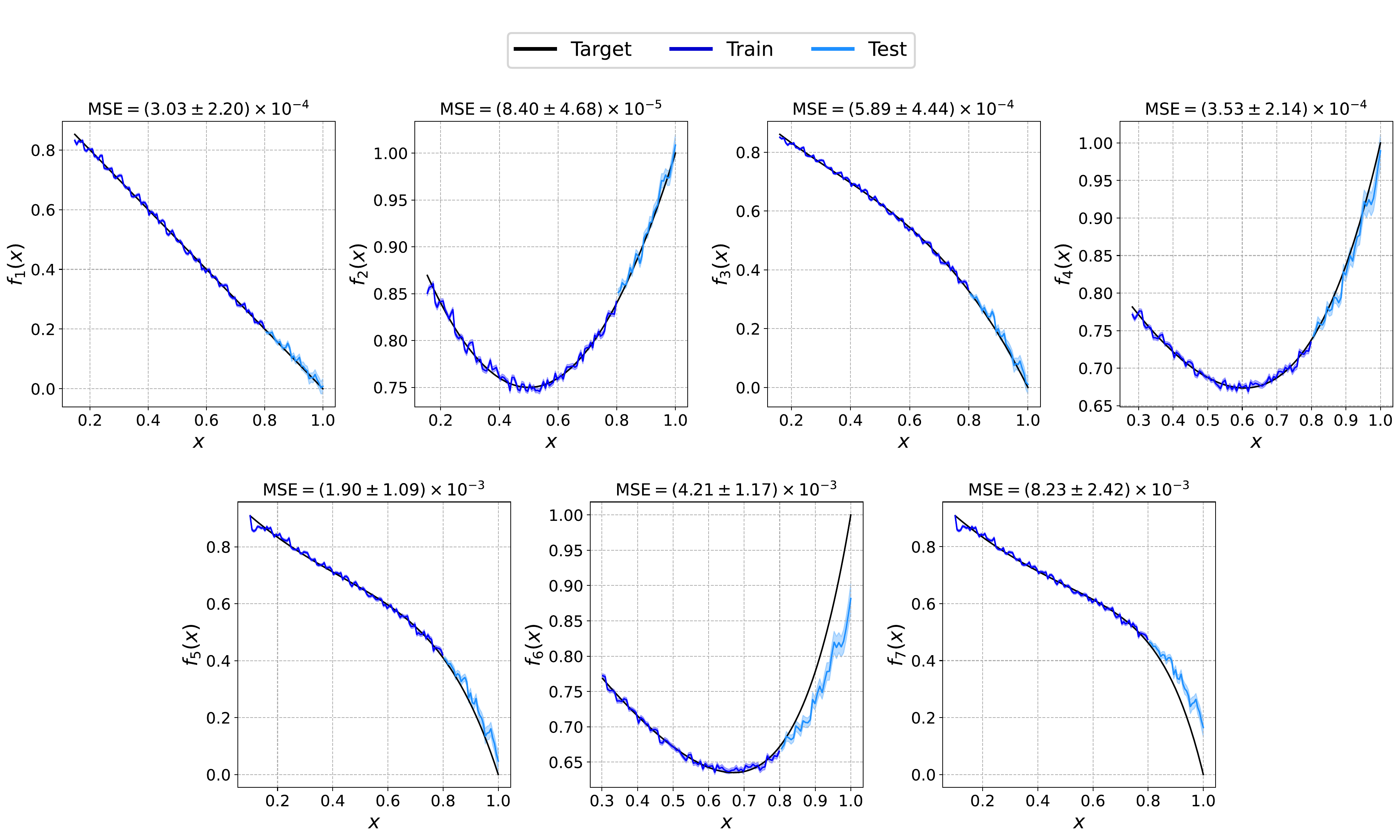}
    \caption{Reconstruction of the polynomial target functions of order $N$ in the configuration with two indistinguishable photons input. For each panel, the target function is depicted as a black line. The reconstructed outputs are displayed for the training dataset (dark blue) and the test dataset (light blue). The title of each plot reports the MSE computed on the test set. Each dataset contains 150 points. Error bars and shaded areas in the plot refer to the statistical fluctuations evaluated from 100 Monte Carlo extractions to account for the presence of Poissonian sampling noise.}
    \label{fig:polynomial_v1}
\end{figure}

Here, we focus on additional results obtained with the collected experimental data presented in the main text, with parameters, such as the number of shots, reported in Tab. \ref{tab:exp_hyp_2photons}. These results serve to further support the reported conclusions, showing that the observed outcomes are extended to a broader set of conditions. Specifically, we present the complete set of nonlinear functions reconstructed in our experiments, for which the MSE values were reported in the main text. This section includes reconstructions of all monomial functions of the form $f_n(x) = x^n$ for $n \in \{2, \ldots, 13\}$, as well as the investigated family of polynomials of the form $f_N(x) = \sum_{n=1}^{N} (-1)^n x^n$ with $N \in \{1, \ldots, 7\}$. For each function, we compare the reconstruction performance across different two-photon input configurations of the physical reservoir. These include scenarios employing distinguishable photons, shown in Figs. \ref{fig:monomial_v0} and \ref{fig:polynomial_v0}, and indistinguishable photons, shown in Figs. \ref{fig:monomial_v1} and \ref{fig:polynomial_v1}. The experimental reconstructions are compared with the target functions, and the uncertainty is evaluated through statistical noise estimated via repeated Monte Carlo simulations. In each case, these results provide an additional proof of the reservoir's ability to approximate increasingly nonlinear functions and, as already discussed in the main text, of the better performance obtained by employing indistinguishable photons. As expected, the reconstruction accuracy deteriorates with increasing nonlinearity, as higher-order monomials are more challenging to approximate. This trend is quantitatively reflected in the increasing MSE values with $n$, and analogously with $N$. Furthermore, the effect of the finite resolution in the phase control becomes apparent through the step-like behavior in the reconstructions, even in the regime where performance remains good. %, i.e., for monomials with $n < 8$ and for polynomials with $N < 5$. 
For the monomials with $n \geq 8$ and polynomials with $N \geq 5$, the results obtained with two indistinguishable photons exhibit greater robustness and better performance.

%The last effect examined is the difference in learning behavior between the indistinguishable and distinguishable photon cases. This analysis uses the same experimental datasets presented in the main text, with parameters such as the number of shots reported in Tab. \ref{tab:exp_hyp_2photons}, and thus employs discretized phase controls. The Fig. \ref{fig:lc_mono_3} presents the learning curves of the quantum reservoir computing model as a function of training set size. Each panel corresponds to a different monomial target function: (a) to $x^3$, (b) to $x^5$, (c) to $x^6$, and (d) to $x^7$. In all cases, the MSE on the test set decreases with increasing training size, indicating a progressive improvement in the model’s ability to approximate the target function. Across all panels, the curves consistently show that the use of indistinguishable photons leads to faster convergence.

\begin{figure}[H]\centering
\subfloat[$f_3(x)=x^3$]{\label{a}\includegraphics[width=.45\linewidth]{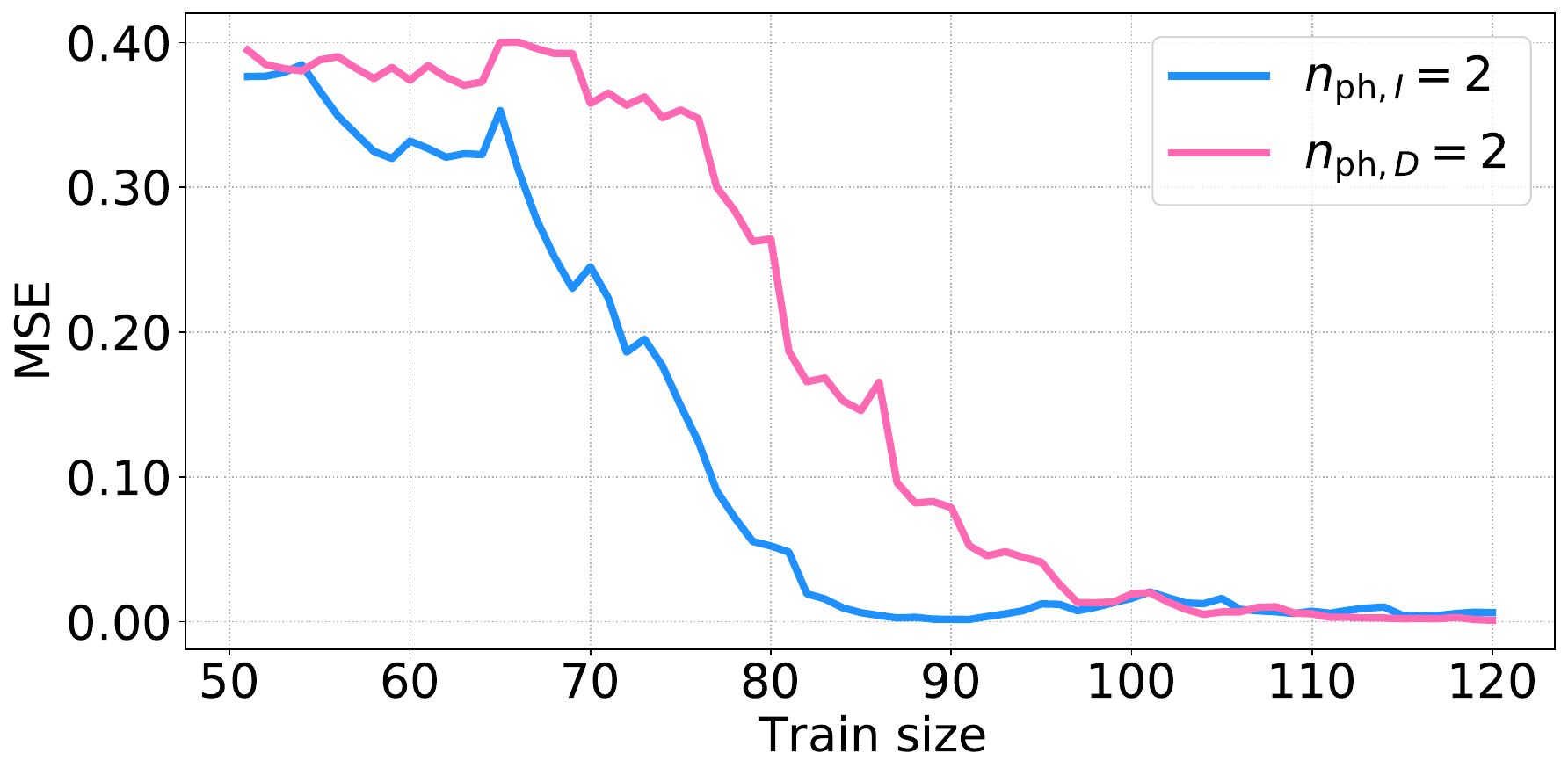}}\hfill
\subfloat[$f_5(x)=x^5$]{\label{b}\includegraphics[width=.45\linewidth]{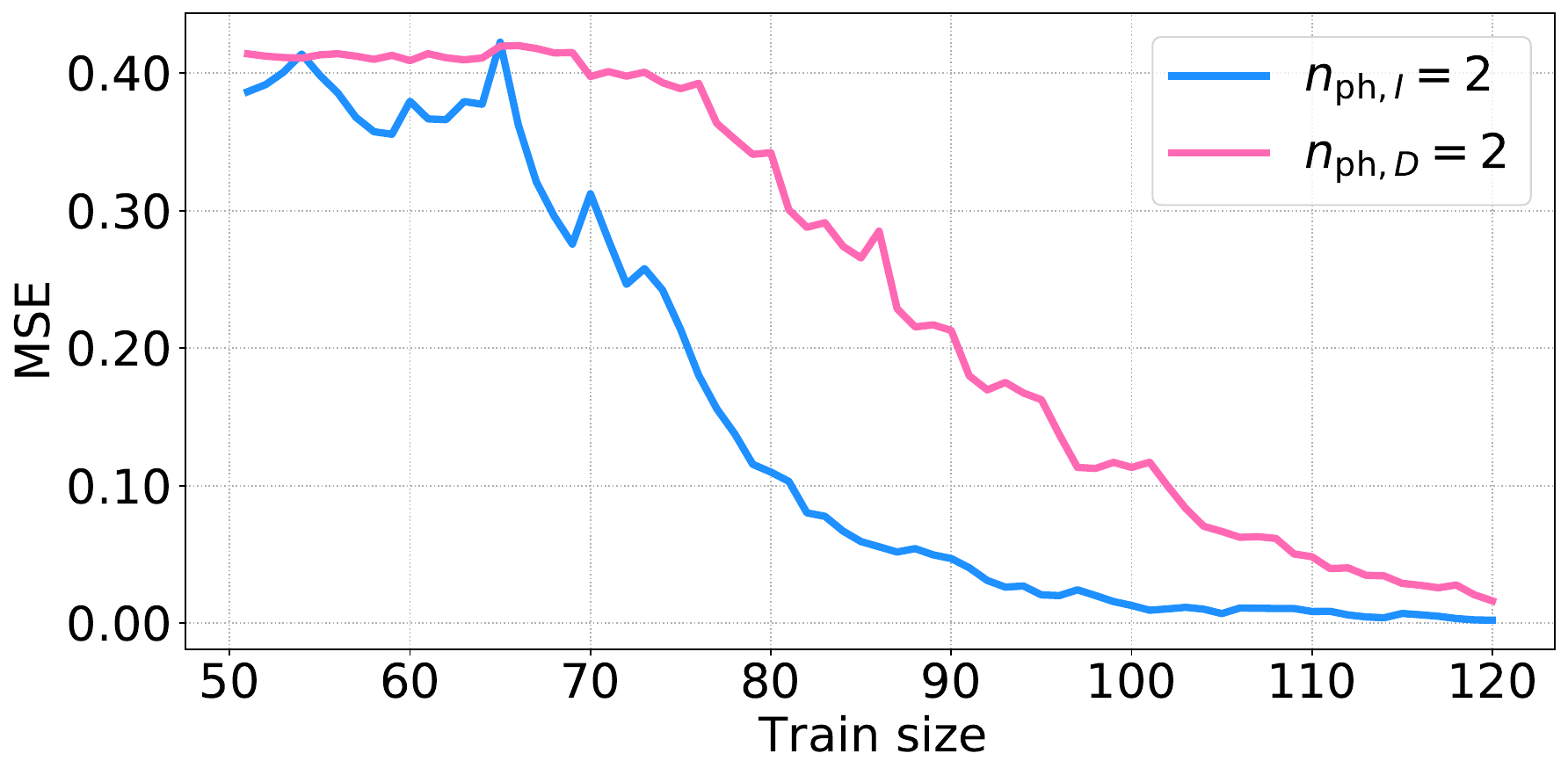}}\par 
\subfloat[$f_6(x)=x^6$]{\label{c}\includegraphics[width=.45\linewidth]{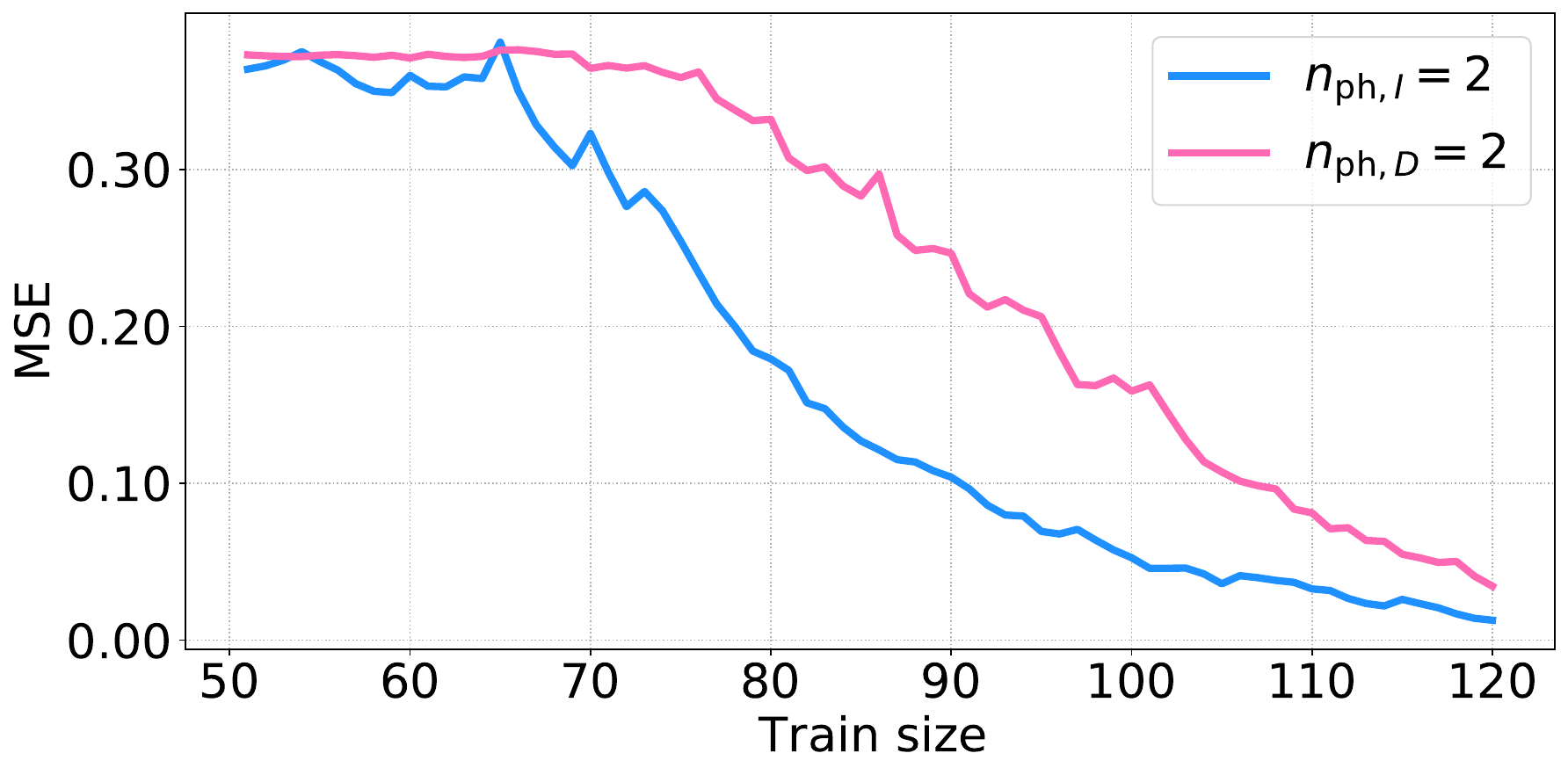}}\hfill
\subfloat[$f_7(x)=x^7$]{\label{c}\includegraphics[width=.45\linewidth]{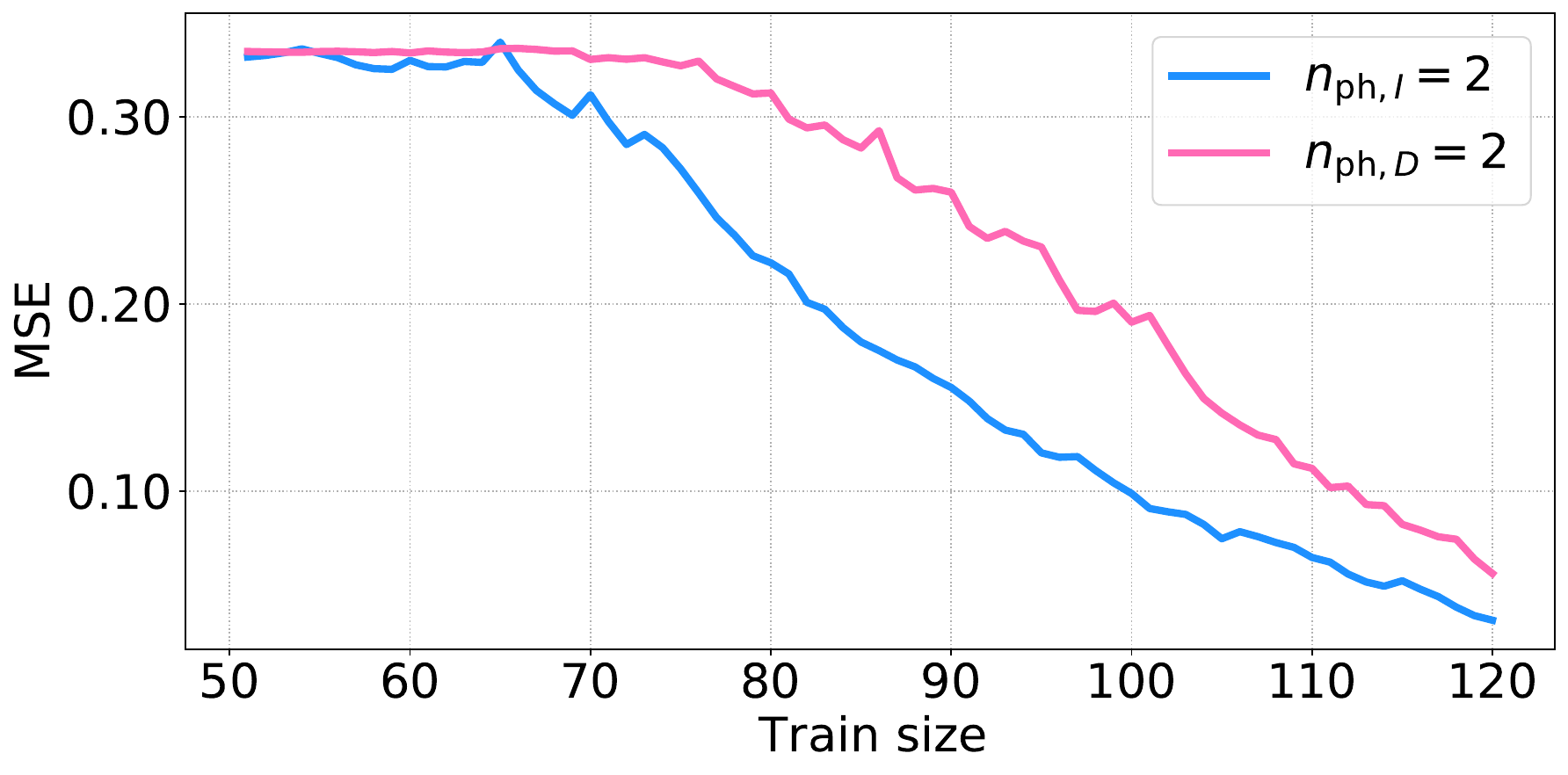}}

\caption{MSE on the test set as a function of the number of training samples, while keeping the test set fixed to the last 30 points, corresponding to $20\%$ of the total dataset. The blue line refers to experimental performance achieved with the two-indistinguishable-photon inputs, the pink line is relative to the two-distinguishable-photon input states. Panels a, b, c and d show the MSE obtained for the monomial reconstruction task for $x^3$, $x^5$, $x^6$, and $x^7$, respectively. Each dataset is the same experimental one used to present the results in the main text and contains 150 points.}
\label{fig:lc_mono_3}
\end{figure}

To further support this observation, we next examine the difference in learning behavior between the indistinguishable and distinguishable photon cases. The learning curves of the quantum reservoir computing model as a function of training set size are presented in Fig. \ref{fig:lc_mono_3}. Each panel corresponds to a different monomial target function $f_n(x)$: (a) $x^3$, (b) $x^5$, (c) $x^6$, and (d) $x^7$. In all cases, the MSE on the test set decreases with increasing training size, indicating a progressive improvement in the model’s ability to approximate the target function. Across all panels, the curves consistently show that the use of indistinguishable photons leads to faster convergence. This effect appears to be independent of the degree of the target polynomial, demonstrating that indistinguishability consistently enhances the ability of the reservoir to model complex nonlinear functions, both in scenarios where both configurations succeed, as in panel (a) with $f_3(x)$, and in those where neither fully succeeds, as in panel (d) with $f_7(x)$. This behavior shows that quantum interference during the reservoir evolution, enabled by the indistinguishable nature of the photons, increases the computational expressivity of the model.

%This effect is especially pronounced for training sizes between 60 and 90, where the MSE drops more rapidly compared to the distinguishable case. As the training set size increases, the performance gap between the two photon configurations gradually diminishes, suggesting that both cases eventually converge to similar levels of accuracy when sufficient data is available. However, in the low-data regime, the configuration with indistinguishable photons shows a clear advantage, likely due to enhanced expressivity or inherent noise robustness, ultimately resulting in better generalization performance.

\subsection{Simulations with ideal system}

%In this work, we have primarily investigated the role of photon indistinguishability within the framework of quantum reservoir computing. Here, through theoretical simulations, we explore how tuning the visibility impacts the expressivity of the system. In particular, we analyze how changes in visibility affect the principal component analysis of the reservoir states, a technique that offers insight into the effective dimensionality of the space explored by the reservoir.

%varoamo la strenght delle correlzzione qauntum 

%\fromRosario{citare Faccio} \cite{nerenberg2025photon}

In this work, we have primarily investigated the role of single and multiphoton states within the framework of quantum reservoir computing. Here, through theoretical simulations, further analysis is conducted to examine how the performance of the system, as the expressive power and the linear memory capacity of the system, varies with the scaling of the input-photon states and the tuning of the visibility $V$, which quantifies the degree of photon indistinguishability. In particular, under idealized conditions, i.e., perfect circuit transformations, infinite statistics, and continuous phase control, it is analyzed the impact of these parameters on the tasks and benchmarks considered in the main text.

\begin{figure}[H]
\begin{subfigure}{.5\textwidth}
  \centering
  \includegraphics[width=1\linewidth]{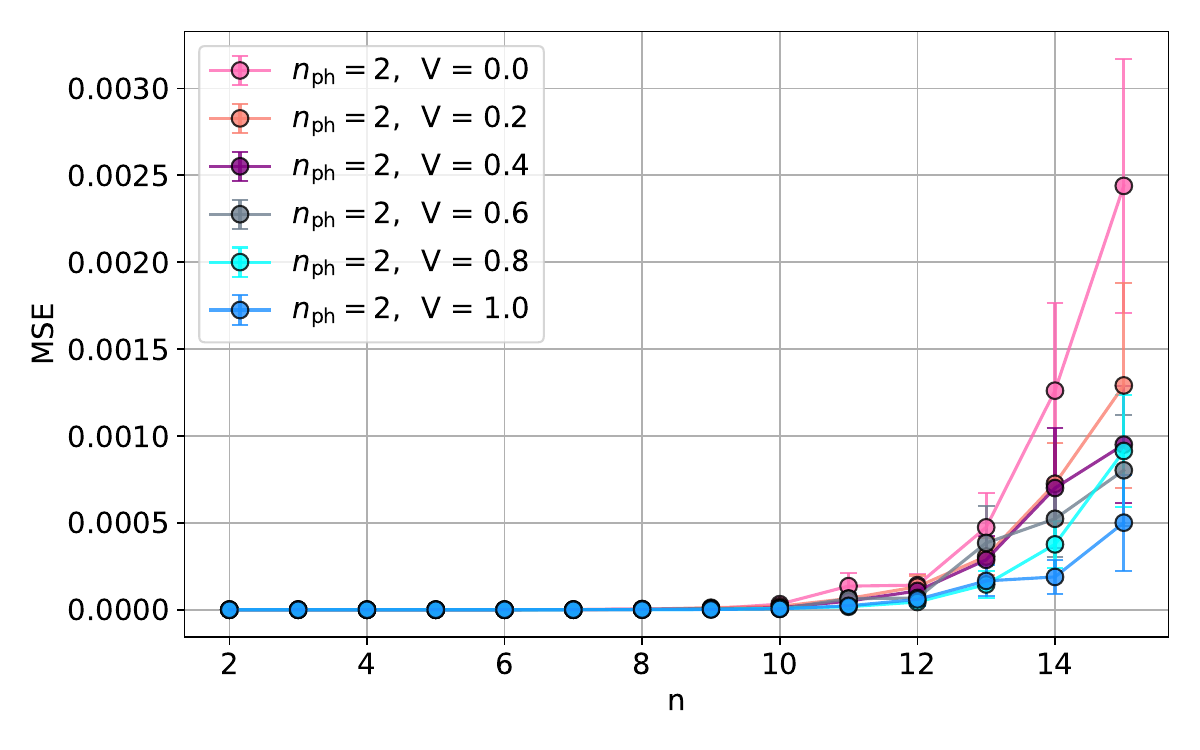}
  \caption{} \label{}
\end{subfigure}
\begin{subfigure}{.5\textwidth}
  \includegraphics[width=1\linewidth]{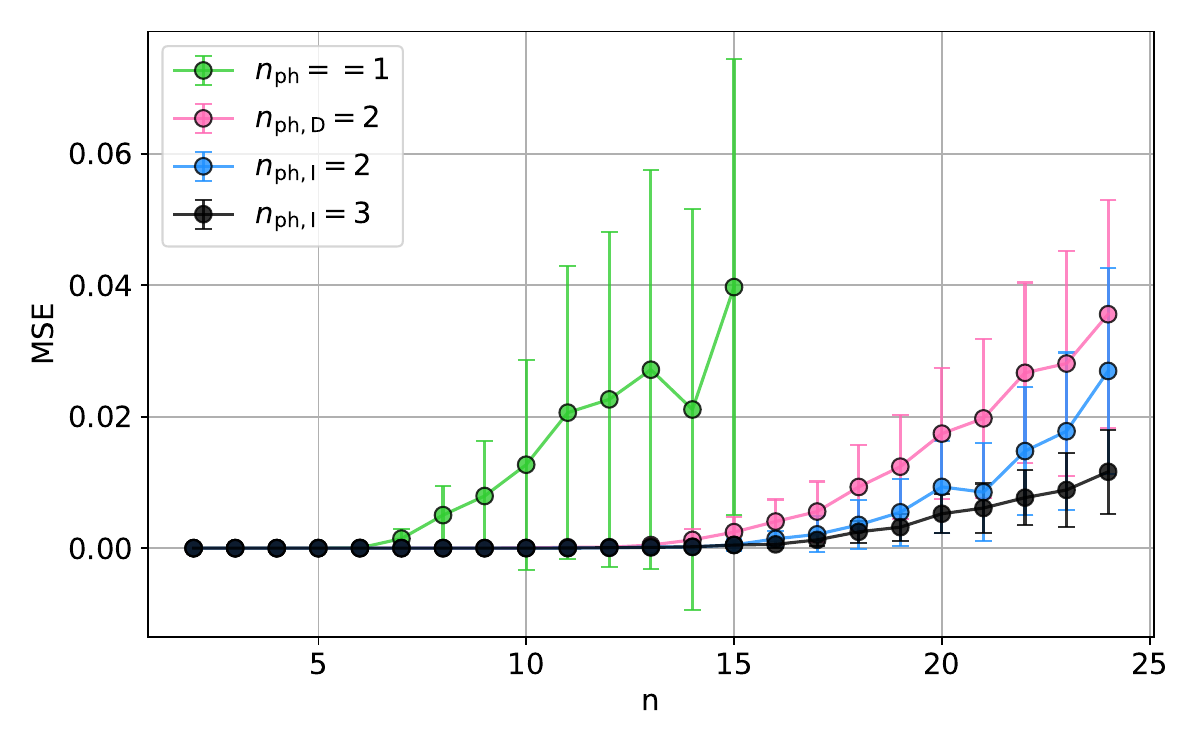}
\caption{}\label{}
\end{subfigure}
\caption{Theoretical simulation results illustrating the effect of photon indistinguishability and the number of photons in the input state on the reconstruction of monomial functions $f_n(x) = x^n$ in quantum reservoir computing. (a) Mean squared error versus the monomial degree $n$ for various values of visibility $V$, with two-input photons ($n_{\text{ph}} = 2$). (b) Comparison of MSE for: single-photon injection ($n_{\text{ph}} = 1$), two distinguishable photons ($n_{\text{ph,D}} = 2$), two indistinguishable photons ($n_{\text{ph,I}} = 2$), and three indistinguishable photons ($n_{\text{ph,I}} = 3$). The dataset contains 200 points and reports the mean MSE achieved over 30 different optimizations of the reservoir's hyperparameters; the uncertainties are the standard deviations.} \label{fig:monomials_theo_simulations_visibility}
\end{figure}

First of all, it is analyzed the role of visibility in the model’s ability to reconstruct nonlinear functions, specifically the monomials $f_n(x) = x^n$. As shown in Fig. \ref{fig:monomials_theo_simulations_visibility}a, increasing the degree of photon indistinguishability leads to significantly improved reconstruction performance. While for low $n$ all visibility values achieve comparably low MSE on the test set, differences become evident as $n$ increases. Indeed, for $n>10$ the MSE remains low even as the degree of the monomial $n$ increases when the two-photons are completely indistinguishable resulting in a HOM dip with visibility $V=1$. This indicates robust expressive power and reflects more capacity to learn nonlinear functions. Then, we focus on the role of photon number input states, as depicted in Fig. \ref{fig:monomials_theo_simulations_visibility}b. This provides a comparison across different numbers of input photon states $n_{\text{ph}}$, specifically considering a single photon, two photons with $V=0$ and $V=1$, and three indistinguishable photons. While all configurations perform similarly for $n<8$, differences in performance become increasingly pronounced as the power $n$ grows. For instance, at $n=8$ a clear performance gap emerges between the single-photon configuration and all other input states, highlighting that the limited amount of classical information carried by a single photon is not sufficient for processing more complex nonlinear information. Later, around $n=15$, the performance gap between indistinguishable and distinguishable two-photon states begins to widen significantly, indicating that while purely classical joint probabilities provide some advantage over the single-photon case by offering more degrees of freedom, they remain inferior to the enhancement achieved through quantum interference. As the degree increases further, the three-photon indistinguishable configuration clearly outperforms the others, maintaining high performance. These results confirm that the number of indistinguishable photons in the input state plays a critical role in enhancing the expressive power and accuracy of the quantum reservoir model. Moreover, adding a fourth photon does not yield further improvements in reconstruction performance, suggesting that the system has reached its maximum expressive capacity under the given conditions.

\begin{figure}[H]
\begin{subfigure}{.5\textwidth}
  \centering
  \includegraphics[width=1\linewidth]{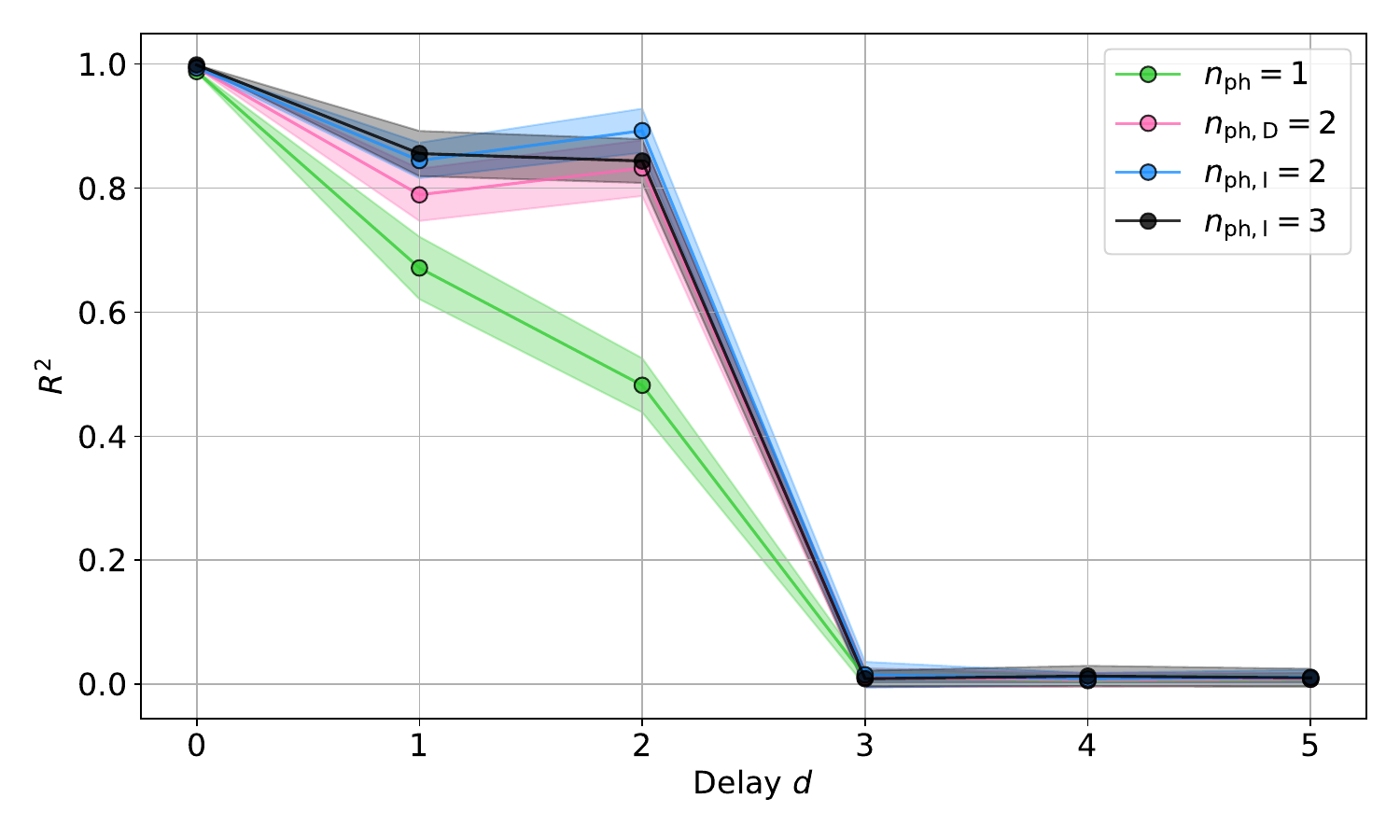}
  \caption{} \label{}
\end{subfigure}
\begin{subfigure}{.5\textwidth}
  \includegraphics[width=1\linewidth]{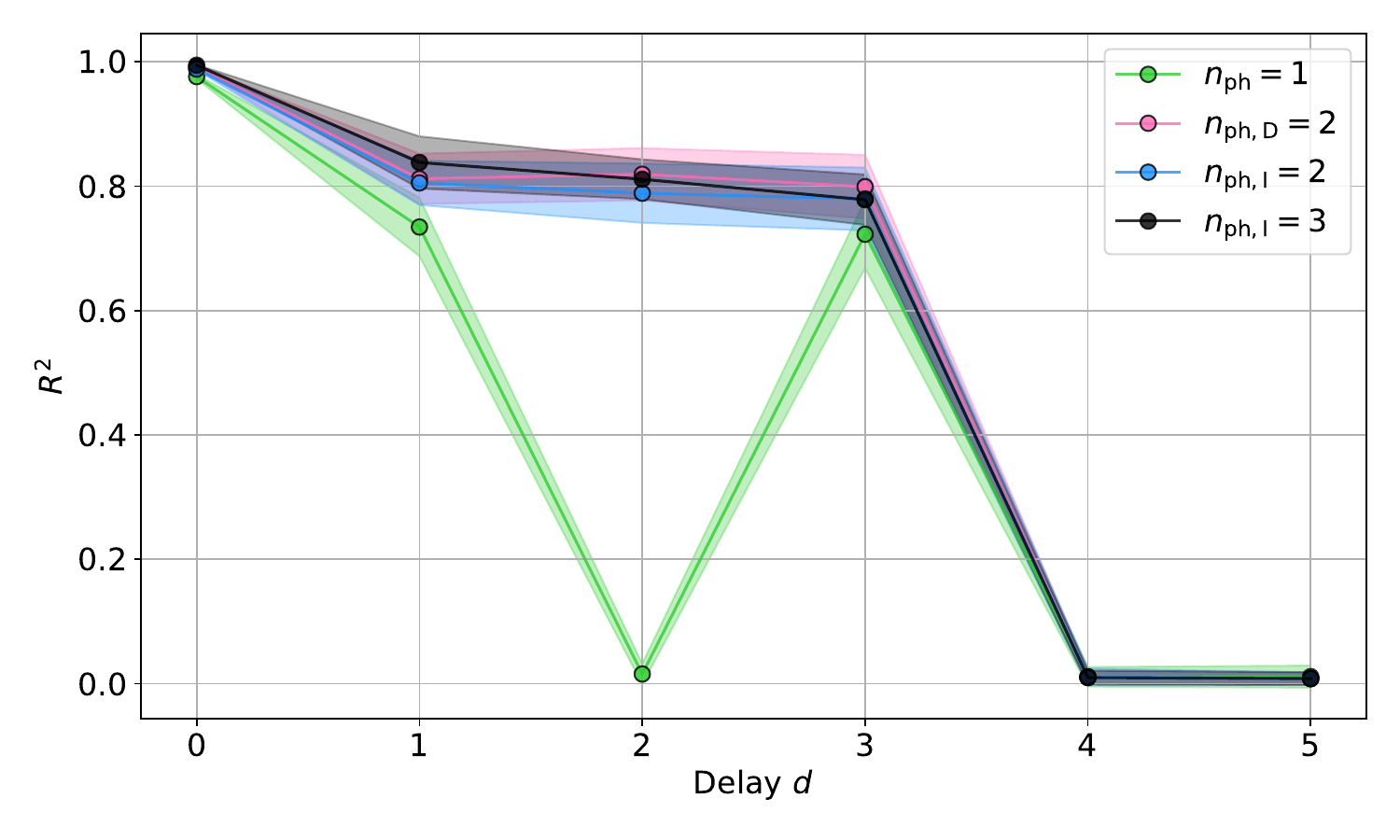}
\caption{}\label{}
\end{subfigure}
\caption{Theoretical simulation results illustrating the effect of photon indistinguishability and the number of photons in the input state on the reservoir's memory (a) The short-term memory is evaluated through the coefficient of correlation $R_d$ varying the temporal delay $d \in \{0,\ldots5\}$ and using in the dynamics two feedback loops. (b) The same task is evaluated with simulations using an additional feedback loop. The dataset contains 400 points and given a single optimization, the uncertainties are estimated by averaging over 30 iterations with different random sequences.} \label{fig:memory_photon_feedback}
\end{figure}

Then, we investigate how the memory capacity is influenced by both the number of photons in the input state and the number of active feedback loops in the reservoir dynamics, as shown in Fig. \ref{fig:memory_photon_feedback}. This analysis provides further insights into the mechanisms by which the system retains and retrieves temporal information throughout its evolution. In panel (a), we examine the short-term memory task with two feedback loops, as in the experiment, and evaluate the coefficient of determination 
$R_d^2$ as a function of the delay $d$. We observe that with two photons, both distinguishable and indistinguishable, the performance reaches a level comparable to what is achieved experimentally and shown in the main text. On the other hand, using a single photon yields a noticeably lower total memory capacity than the experimental counterpart. This discrepancy is likely due to the beneficial role of the experimental imperfections in QRC, such as the Poissonian noise, which introduces stochastic dynamics increasing the effective rank of the Gram matrix, and consequently the size of the accessible space \cite{nerenberg2025photon}. Interestingly, when using three indistinguishable photons, the performance saturates at the same level as with two photons, suggesting that in the absence of additional feedback loops, adding more photons does not further increase the system’s linear memory capacity. In panel (b), we include three feedback loops in the simulation. The overall memory capacity increases and reaches the delay $d=3$, showing an enhanced general performance. While the single-photon configuration still keeps memory only of 2 delays determined by the specific hyperparameter optimization that favored $d=3$ with respect to $d=2$. In contrast, the two- and three-photon configurations show sustained memory performance up to $d=3$, demonstrating the benefits of increasing both the photon number and the recurrence in the system's dynamics. These results support the idea that the enhancement in temporal information processing arises from the increased internal information recycling enabled by photon interference and recurrent feedback. However, the overall performance remains constrained by the size of the accessible configuration space, which is determined by the underlying photonic architecture.

\begin{figure}[H]
\begin{subfigure}{.5\textwidth}
  \centering
  \includegraphics[width=.92\textwidth]{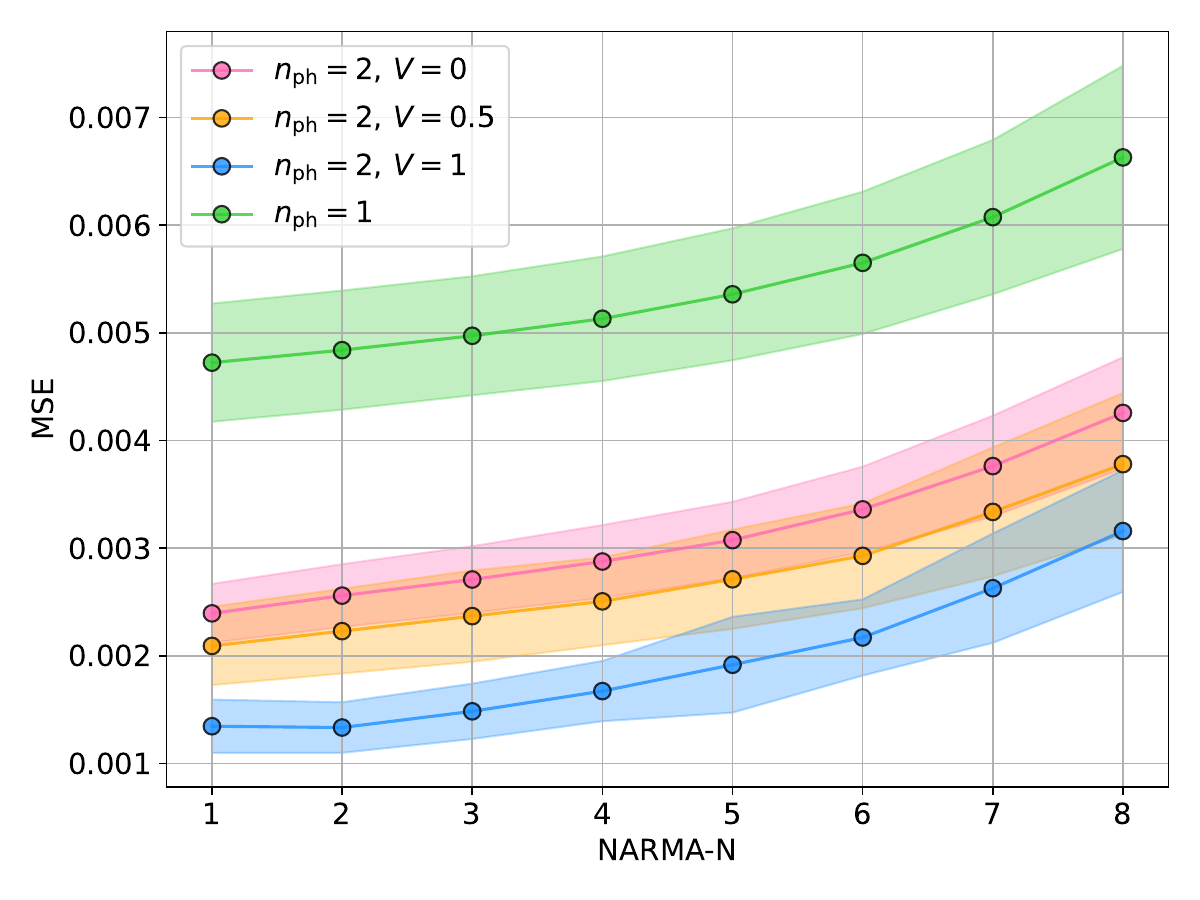}
  \caption{} \label{}
\end{subfigure}
\begin{subfigure}{.5\textwidth}
  \includegraphics[width=1\linewidth]{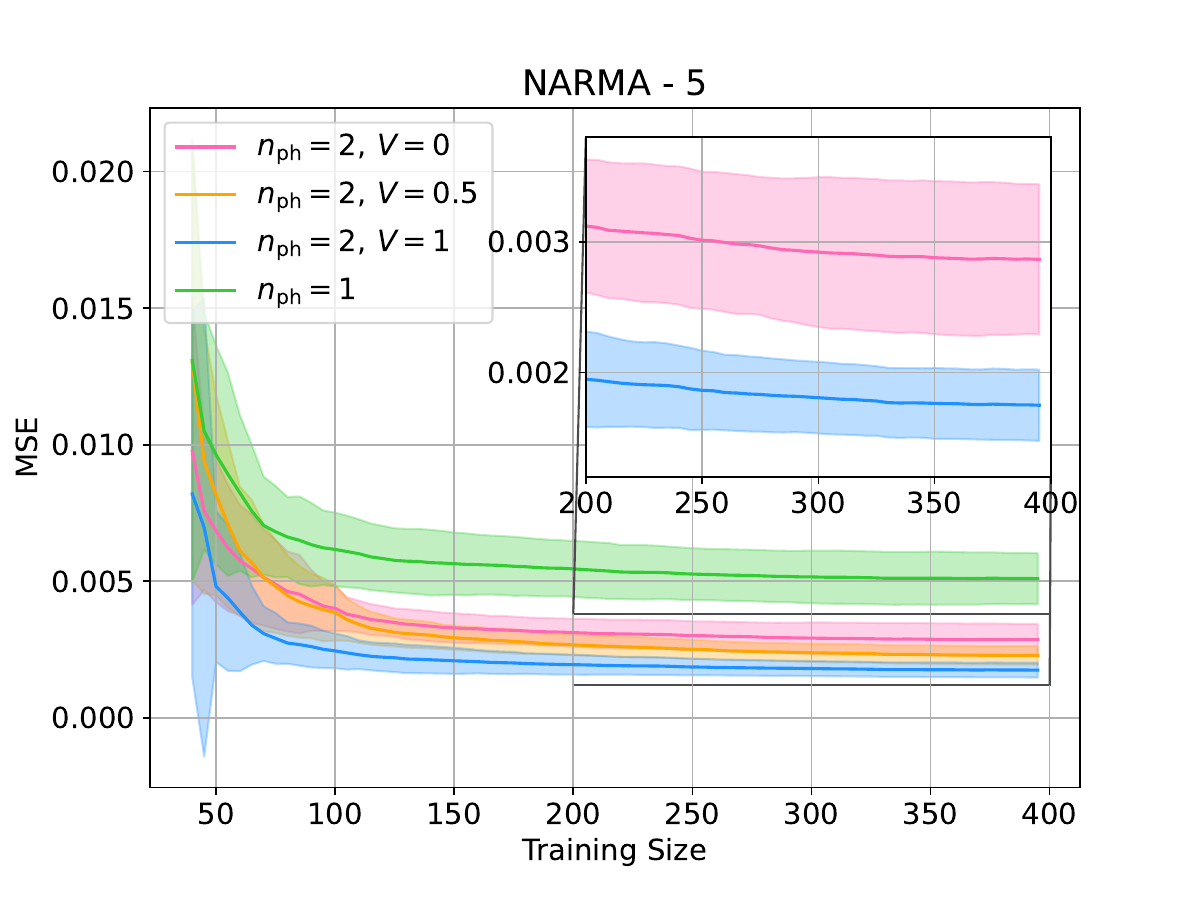}
\caption{}\label{}
\end{subfigure}
\begin{subfigure}{.5\textwidth}
  \includegraphics[width=1\linewidth]{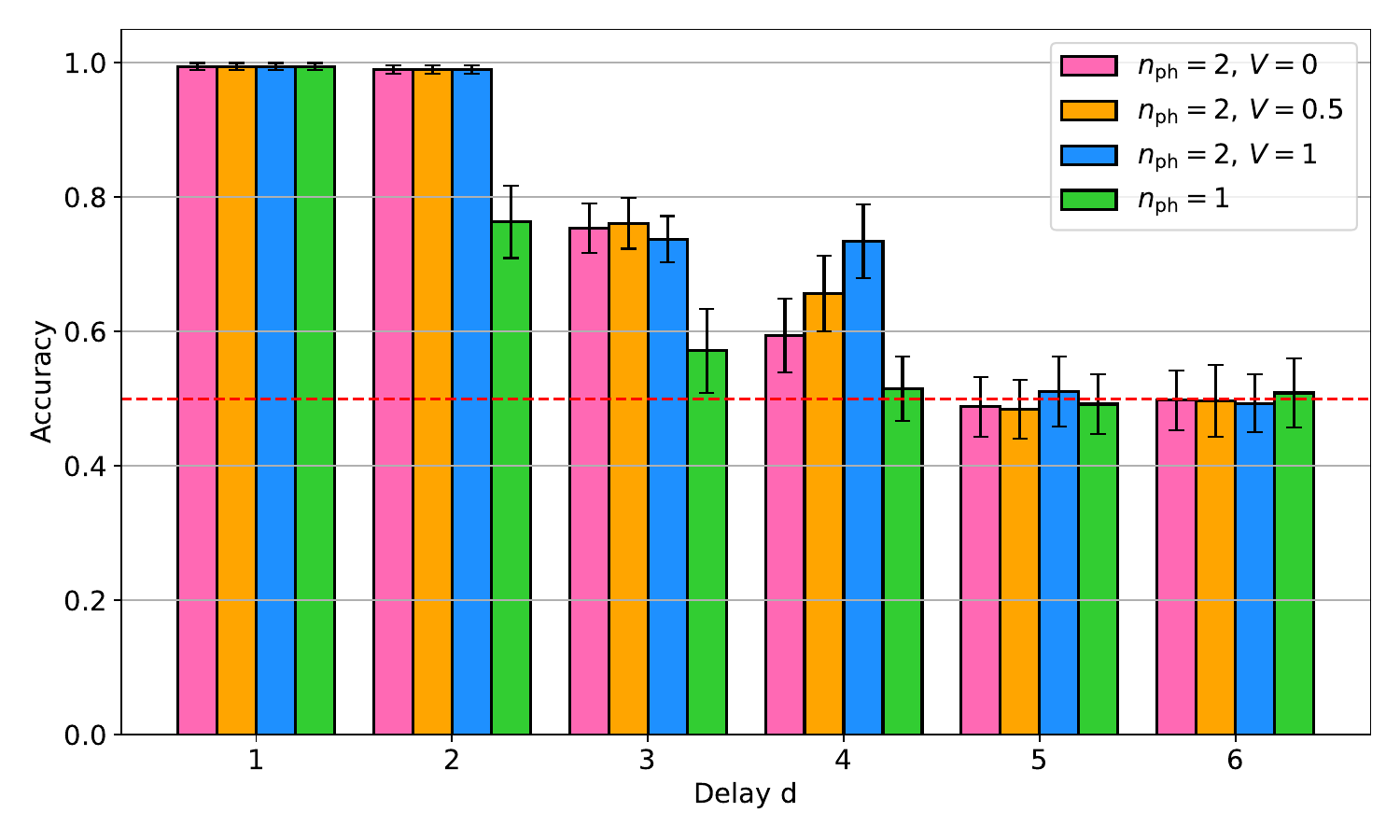}
\caption{}\label{}
\end{subfigure}
\begin{subfigure}{.5\textwidth}
  \includegraphics[width=.9\linewidth]{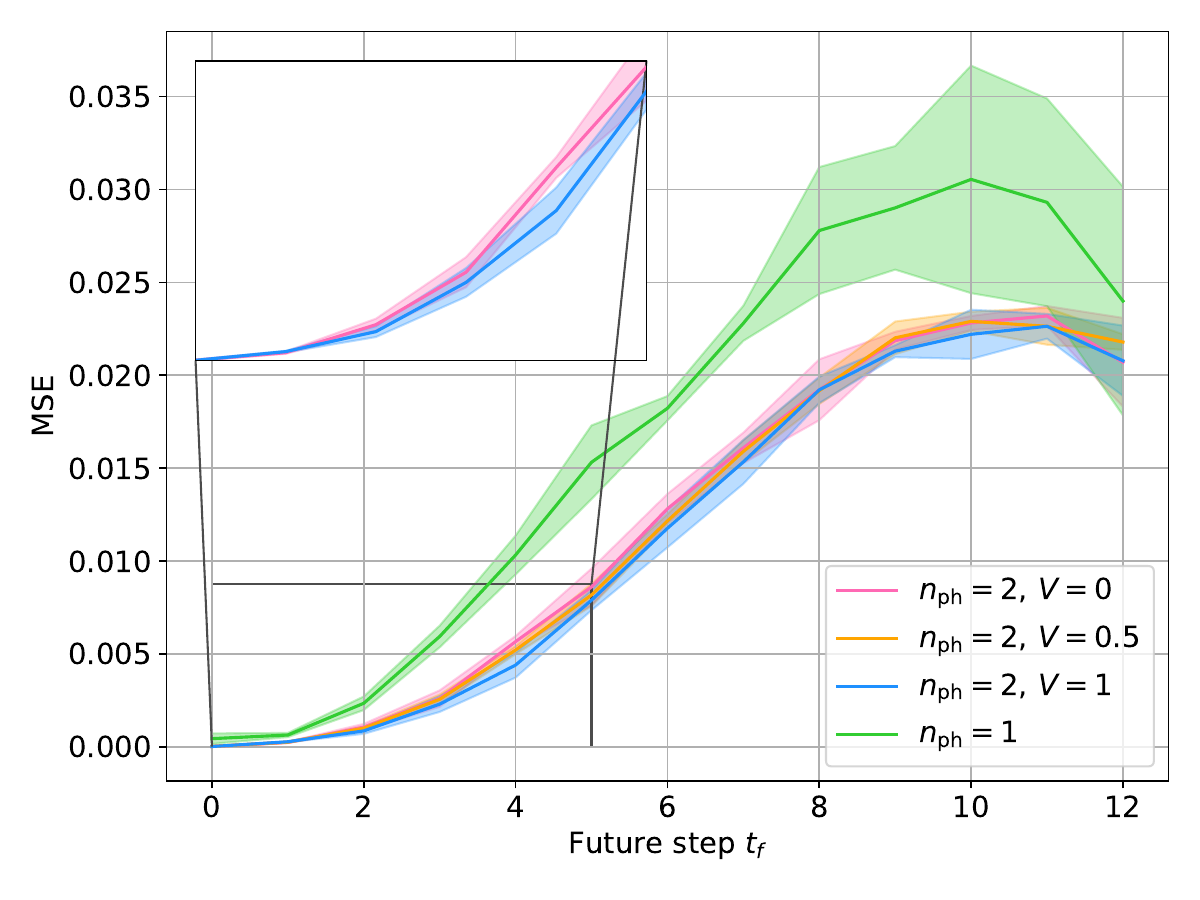}
\caption{}\label{}
\end{subfigure}
\caption{Theoretical simulation results illustrating the effect of single photon and two photons indistinguishability (with $V=0,\,0.5 \, \text{and} \, 1.0$) on NARMA sequence \cite{atiya2000new} and Mackey-Glass time-series forecasting tasks. (a) NARMA is evaluated through the mean squared error varying its order $N \in \{1, \ldots, 8\}$. The dataset contains 500 points. (b) Considering the sequence NARMA-5 it is studied the learning curve varying the training size from and testing on the last $100$ points. (c) The temporal XOR is evaluated through the accuracy varying the temporal delay $d \in \{1,\ldots6\}$. The dataset contains 500 points. (d) Mackey-Glass time-series forecasting is evaluated through the mean squared error varying the future step of the prediction $t_f \in \{0,\ldots,12\}$. The dataset contains 200 points. In these tasks, when the input sequence is random, as for NARMA and the temporal XOR, the uncertainties are estimated by averaging over 30 iterations with different random sequences, while regarding Mackey-Glass, since the input sequence is fixed, they are evaluated by averaging over 10 hyperparameters' optimization.}
\label{fig:narma_mg_theo_sim}
\end{figure}

Next, we investigate the performance of the ideal quantum reservoir across the benchmark tasks studied experimentally, focusing on the influence of the input photon number and the degree of indistinguishability. The obtained simulated results are reported in Fig. \ref{fig:narma_mg_theo_sim} and include the NARMA sequence with increasing nonlinearity order \cite{atiya2000new}, the temporal XOR task, and the Mackey-Glass chaotic time-series forecasting task. 

In panel (a), we evaluate the ability of the quantum reservoir to model increasingly nonlinear systems by applying the discussed protocol to the NARMA task with varying order $N \in \{1, \ldots, 8\}$. The performance is quantified computing the MSE on the test set. As the nonlinearity order increases, the difference between configurations becomes more evident. The two-photon indistinguishable setup ($n_{\text{ph}} = 2, V = 1$) outperforms both the distinguishable counterpart with $V = 0$, the partially distinguishable one with $V=0.5$, and the single-photon configuration, demonstrating lower error across up to the order $N=4$. For $N \ge 5$, the performance with two-input photons become all compatible with each other within one standard deviation. To better understand the advantage given by quantum correlations, panel (b) shows the learning curve for the NARMA-5 task, where the training size is varied up to 400, and the testing is performed on the final 100 points of the sequence. The results reveal that indistinguishability improves generalization performance. The inset highlights that while all configurations tend to converge as training size increases, the indistinguishable photon case reaches a lower asymptotic error, indicating more efficient use of training data.

The temporal XOR task is examined in panel (c). Here, we assess the prediction accuracy as a function of the delay $d \in \{1, \ldots, 6\}$. This task probes both the system’s short-term memory and its ability to capture nonlinear dependencies in binary sequences. The two-photon indistinguishable configuration consistently outperforms the others, maintaining higher accuracy across all delays. The single-photon setup shows a drop in performance for delays beyond $d = 2$, whereas the two-photon setups degrade more slowly. This is in agreement with the results obtained for the expressivity and the short-term memory in Fig.s \ref{fig:monomials_theo_simulations_visibility} and \ref{fig:memory_photon_feedback}, respectively. Moreover, the two indistinguishable photons input state maintains better performance for longer, highlighting the expressivity advantage gained by quantum correlations.  

Finally, panel (d) presents the results for the Mackey-Glass forecasting task, where the goal is to predict the future value of a chaotic time series as a function of the prediction horizon $t_f \in \{0, \ldots, 12\}$. The MSE is used as the evaluation metric. We underline that only in this case, the nonlinear input is directly encoded in the phase value, therefore, the main advantage of exploiting indistinguishable photons to gain in expressivity here is attenuated by the specific encoding. However, the indistinguishable two-photon configuration yields a slightly lower forecasting error, especially for short- to medium-term predictions, i.e., $t_f < 6$. Increasing the horizon, all configurations experience a degradation in performance. The inset plot highlights early steps where the benefit of photon indistinguishability is most evident. These results confirm that quantum interference improves the model’s ability to capture and forecast complex temporal dynamics.

Across almost all tasks, the simulations show that indistinguishable multiphoton input states enhance the computational performance of quantum reservoirs, particularly in terms of expressive power, and consequently in time-series forecasting requiring non-linearity. However, performance saturates with increasing photon number beyond $n_{\text{ph}} = 2$ or 3, reinforcing the idea that architectural constraints, such as the dimension of the accessible Hilbert space and the depth of internal transformations, ultimately bound the system’s capabilities.

\bibliography{QiSMG16}